\documentclass[oldversion]{aa}  
\usepackage{graphicx}
\usepackage{lscape}
\usepackage{natbib} 
\usepackage{siunitx}
\usepackage{url}
\usepackage[varg]{txfonts}
\usepackage{longtable}
\usepackage{supertabular}
\usepackage{cases}
\usepackage{dcolumn}
\usepackage{xcolor}
\usepackage{txfonts}

\bibpunct{(}{)}{;}{a}{}{,} % citation like A&A style
\def\changed{}
\begin{document}

\title{Testing massive star evolution, star formation history, and feedback at low metallicity}
\subtitle{Spectroscopic analysis of OB stars in the SMC Wing\thanks{Based on observations at the European Southern Observatory Very Large Telescope in program  086.D-0167(A)} \thanks{ Table\,B.1 will be only available in electronic form at the CDS via anonymous ftp to cdsarc.u-strasbg.fr (130.79.128.5) or via http://cdsweb.u-strasbg.fr/cgi-bin/qcat?J/A+A/}}
%\title{ The dichotomy in massive star evolution at low metallicity might be independent of rotation}
%\title{ Anomalies in massive star evolution, star-formation and  feedback  at low metallicity } 
 %\subtitle{Spectroscopic analysis of OB stars in the Wing of the SMC \thanks{Based on observations at the European Southern Observatory Very Large Telescope in program  086.D-0167(A)}}
   \author{Varsha Ramachandran\inst{1}
          \and W.-R. Hamann\inst{1}
          \and L. M. Oskinova\inst{1,2}
         \and J. S. Gallagher\inst{3} 
         \and R. Hainich \inst{1}        
          \and T. Shenar\inst{4}
          \and A. A. C. Sander\inst{5}
          \and H. Todt\inst{1}
          \and L. Fulmer\inst{6,7}
          %\and V. H{\'e}nault-Brunet
          }

   \institute{Institut f\"ur Physik und Astronomie,
              Universit\"at Potsdam,
              Karl-Liebknecht-Str. 24/25, D-14476 Potsdam, Germany \\
              \email{varsha@astro.physik.uni-potsdam.de}
\and Kazan Federal University, Kremlevskaya Ul. 18, Kazan, Russia     
\and Department of Astronomy, University of Wisconsin - Madison, WI, USA              
\and Institute of Astrophysics, KU Leuven, Celestijnenlaan 200 D, 3001, Leuven, Belgium
\and Armagh Observatory and Planetarium, College Hill, Armagh, BT61 9DG, Northern Ireland      
\and National Optical Astronomy Observatory, 950 North Cherry Ave. Tucson, AZ 85719, USA
\and University of Washington, Physics-Astronomy Bldg. 3910 15th Ave NE Rm. C319 Seattle, WA 98195, USA  
}       
   \date{Received <date> / Accepted <date>}

%-------------------  Abstract --------------------

\abstract
{
Stars that start their lives with spectral types O and early B are the
progenitors of core-collapse supernovae, long gamma-ray bursts, neutron
stars, and black holes. These massive stars are the primary sources of
stellar feedback in star-forming galaxies. At low metallicities, the
properties of massive stars and their evolution are not yet fully
explored. Here we report a spectroscopic study of 320 massive
stars of spectral types O (23 stars) and B (297 stars) 
in the Wing of the Small Magellanic Cloud
(SMC). 
The spectra, which we obtained with the ESO Very Large Telescope, were
analyzed using state-of-the-art stellar atmosphere models, and the
stellar parameters were determined.  We find that the stellar winds of
our sample stars are generally much weaker than theoretically expected. The stellar rotation rates show  broad, tentatively bi-modal distributions. The upper Hertzsprung-Russell diagram (HRD) is well populated 
by the stars of our sample from a specific field in the SMC Wing. A few
very luminous O stars are found close to the main sequence, while all
other, slightly evolved stars  obey a strict luminosity limit.
Considering additional  massive stars in evolved stages, with
published parameters and located all over the SMC, essentially confirms
this picture. The comparison with single-star evolutionary tracks 
suggests a dichotomy in the fate of massive stars in the SMC. Only
stars with an initial mass below $\sim 30\,M_\odot$ seem to evolve from
the main sequence to the cool side of the HRD to become a red
supergiant and to explode as type II-P supernova. In contrast, stars 
with initially more than $\sim 30\,M_\odot$ appear to stay always hot
and might evolve quasi chemically homogeneously, finally collapsing to
relatively massive black holes. However, we find no indication that
chemical mixing is correlated with rapid rotation. 

We measured the key parameters of
stellar feedback and established the links between  the rates of star
formation and supernovae.  Our study demonstrates that in metal-poor
environments stellar feedback is dominated by core-collapse
supernovae in combination with winds and ionizing radiation supplied by
a few of the most massive stars. We found indications of the stochastic
mode of massive star formation, where the resulting stellar population 
is fully capable of producing large-scale structures such as the
supergiant shell SMC-SGS\,1 in the Wing. The low level of feedback in
metal-poor stellar populations allows star formation episodes to persist
over long timescales. 
}

\titlerunning{ Massive star evolution, star-formation history and feedback at low metallicity}
\maketitle

%________________________________________________________________
\section{Introduction}
\label{sect:intro}

Massive stars ($M_{\rm init}\gtrsim 8\,M_{\odot}$) are cosmic engines that play a pivotal role in the physical and chemical evolution of the interstellar medium (ISM) and galaxies via UV radiation, stellar winds, and supernovae (SNe).  These stars are among the main sources of  reionization of the universe  \citep{Barkana2006Sci}. Massive stars are the progenitors of neutron stars and black holes, formed by their core-collapse. Detection of gravitational waves from coalescing black holes further highlighted a need for better understanding of massive stars, especially at low metallicity.

The evolution of a star is mainly decided by its mass, mass-loss
history, composition, rotation rate, and binary status. Mass-loss rates
($\dot{M}$) derived empirically are often lower compared to standard
prescriptions, especially for low-luminosity OB stars \citep[`weak wind
stars',][]{Bouret2003,Martins2005,Marcolino2009}. 
\changed{The variation of $\dot{M}$ with evolutionary phase and
metallicity is still under debate. Especially at highest masses, the
evolution of massive stars at low metallicity and their remnant mass at
collapse depends critically on the mass loss by winds and eruptions.
}
Stellar evolution models predict rotationally induced chemical mixing
in massive main sequence stars and homogeneous evolution channels,
predominantly at low metallicity. The presence of a companion can also
alter the evolutionary paths \citep{Sana2012}.

The quantitative spectroscopy of massive stars in metal-poor environments is a key to understand their properties, evolution, and feedback in detail. Up to now, the largest spectroscopically analyzed samples refer to the Large Magellanic Cloud (LMC) with metallicity $Z\sim0.5\,Z_{\odot}$ \citep{Schneider2018,Ramachandran2018b}. In this paper, we analyze 320 OB stars in the Small Magellanic Cloud (SMC), significantly enlarging the previous studies \citep{Hunter2008,Bouret2013,Castro2018_smc}.  Our sample stars belong to the Wing of the SMC, which is the nearest (DM =18.7\,mag) \citep{Cignoni2009} low density \citep{Stanimirovic2000} and \changed{low reddening region  at low metallicity.  The  chemical composition of B stars (C, N, O, Mg, Si and Fe ) obtained by \citet{Hunter2007} and \citet{Trundle2007}  compared to solar values \citep{Asplund2009} suggest the metallicity of SMC to be $\sim 1/7\,Z_{\odot}$}. The properties of the Wing match with the typical conditions for low-metallicity dwarf irregular galaxies, which are the most common among star-forming galaxies \citep{Gallagher1984}.

\begin{figure}
 \centering
 \includegraphics[width=8.5cm]{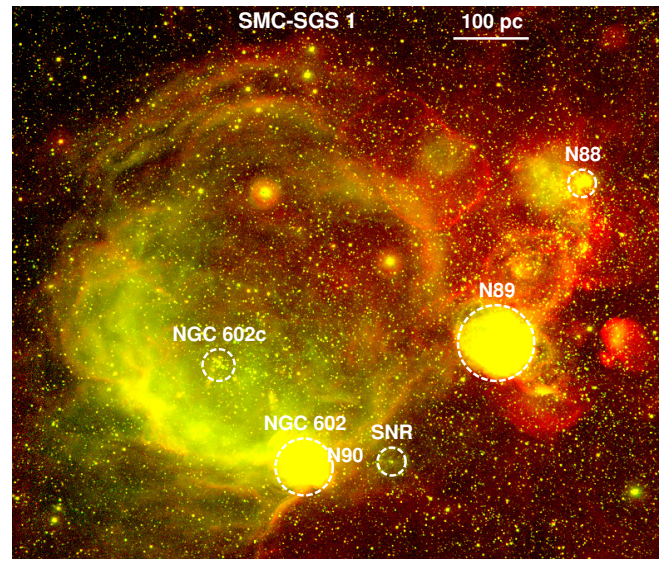}
 \caption{Supergiant shell SMC-SGS1 in the Wing of the SMC. The  \ion{H}{ii} regions (N\,88,89, and 90), clusters (NGC\,602 and NGC\,602c), and the supernova
remnant (SNR) in the region are indicated on the color composite (H$\alpha$, [\ion{O}{iii}]) image, taken from the Magellanic Cloud Emission-Line Survey \citep[MCELS, ][]{Winkler2005}.}
 \label{fig:sgsregion}
 \end{figure}

Among the spectacular manifestations of massive star feedback are the large-scale structures in the ISM, such as the H$\alpha$ ``supergiant shell'' (SGS) in the SMC \citep[SMC-SGS\,1,][]{Meaburn1980}, which contains the majority of our sample stars (Fig.\,\ref{fig:sgsregion}). It has a closed ring-like morphology (r $\sim 300$\,pc ) with a bright rim in the southeast. It is associated with a \ion{H}{i} supershell, which shows a central radial velocity of 173\,km\,s$^{-1}$ and expansion velocity of $\sim 10$\,km\,s$^{-1}$ \citep{Stanimirovic1999,Fulmer2019}.   However, the formation mechanism of such large structures is still a subject of debate.  One source responsible for the huge amount of energy needed for their formation could be the combined effect of ionizing radiation and winds from young clusters or OB associations and  SN explosions. Alternative mechanisms include the distortion of the ISM by $\gamma$-ray bursts \citep{Efremov1999}, collisions of high-velocity clouds (HVCs) \citep{Tenorio-Tagle1981,Tenorio-Tagle1986}, stochastic star formation propagation \citep{Seiden1979,Matteucci1983,Harris2008}, and the turbulent nature of the ISM \citep{Elmegreen1997}.  Quantitative feedback studies are necessary to unveil which of the possible mechanisms plays a key role.

The nebular complexes of the region were first identified by \citet{Davies1976} and classified into eight different emission regions DEM\,160-167.  Three \ion{H}{ii} regions, N\,88, N\,89, and N\,90, are located in the rim of the shell (see Fig.\,\ref{fig:sgsregion}).  A well-studied and prominent site of star formation in the SMC-SGS\,1 is the group of young clusters NGC\,602. The main cluster NGC 602a and the adjacent cluster NGC\,602b are immersed in the emission nebula N\,90, while NGC\,602c is located in the northeast \citep{Westerlund1964}.  The cluster hosts few of the most massive stars in the SMC , including a rare pre-SN star of WO type (Sk\,188) in NGC\,602c and an early-type O3 star (Sk\,183) in NGC\,602a \citep{Evans2012}. Many of the OB stars within the SGS do not appear to be bound to any of the clusters and it was suggested that they related to the same triggering event \citep{McCumber2005}.

 %---------------------------------------------------------------
%---------------------------------------------------------------

The quantitative spectroscopy of OB star populations in the SMC  allows us to test massive star evolution, their feedback, and star formation in low-metallicity galaxies. Through the analyses of OB stars using sophisticated model atmospheres, we provide the stellar and wind parameters of the individual stars and their energy feedback.  The spectroscopic observations and data reduction  are presented in Sect.\,\ref{sect:spec}. In Sect.\,\ref{sect:analysis}, we quantitatively analyze the OB star spectra using the Potsdam Wolf-Rayet (PoWR) model atmospheres. We present the main results and discussions in  Sect.\,\ref{sect:hrd} to Sect.\,\ref{sec:feedback}. The final Sect.\,\ref{sec:summary} provides a summary and general conclusions. The Appendices encompass additional data, figures and tables.

\section{Observations and data reduction}
\label{sect:spec}

We obtained the optical spectra of the OB stars associated with SMC-SGS\,1 on 2010 October 24-26 with the Fiber Large Array Multi-Element Spectrograph (FLAMES) on ESO's Very Large Telescope (VLT). Using the Medusa-fiber mode of FLAMES \citep{Pasquini2002}, spectra on 132 targets were simultaneously recorded, where each fiber has an aperture of $1.2\arcsec$ on the sky. 
Three of the standard settings of the Giraffe spectrograph LR02 (resolving power $R$\,=\,6000, 3960--4567\,\AA), LR03 ($R$\,=\,7500, 4501--5071\,\AA), and HR15N ($R$\,=\,19200, 6442--6817\,\AA) were used  for this survey. The higher resolution spectra of  H$ \alpha $ serve for the determination of the wind parameters and to distinguish nebular emission from the stellar component. 

\changed{Our observation constraint was a magnitude cut of $V \leq 17$\,mag \citep[based on][]{Massey2002}}, corresponding to unreddened O and early B-type stars at the distance of the SMC. We could not apply further constraints since no other photometry data were available at the time of observation. The exposure time for each pointing is 1800\,s  in all three spectrograph settings. We took three to six exposures for each target to get an S/N$>50$.

The ESO Common Pipeline Library (CPL)\footnote{http://www.eso.org/observing/cpl} FLAMES reduction routines  were executed for the standard data processing stages, i.e., bias subtraction, fiber location, summed extractions, division by a normalized flat-field, and wavelength calibration. All these spectra were then corrected to the heliocentric frame. A number of fibers was placed on the sky background. Each sky fiber was inspected and the contaminated sky fibers were rejected prior to the creation of a median sky spectrum for each observation. This median sky spectrum was then subtracted from all stellar spectra. However, because of the prominent nebular regions in the shell, the majority of the spectra are still contaminated by nebular emission lines such as H$\alpha$, [\ion{O}{iii}], [\ion{N}{ii}], and [\ion{S}{ii}].

To clean the extracted spectra of cosmic rays, we calculated the ratio of the two exposures for each spectrum. Since these exposures were consecutive, we assumed that any unexpected and significant deviations in the ratio are indicative of a cosmic ray. The pixels identified as suspect were rejected, then replaced with the value from the sister exposure, appropriately scaled by the ratio of the surrounding pixel.

The obtained spectra are not flux-calibrated. The spectra were rectified by fitting the stellar continuum with a cubic spline function. For each star, the LR02 and LR03 spectra were merged to form the medium resolution blue spectrum from 3960 to 5071\,\AA. We obtained spectra of 543 individual stars. Since we only applied a V magnitude cutoff, the sample also contains late-type stars, foreground stars, and objects with poor S/N. By inspection of the blue spectra, we eliminated them from our final catalog. We are only interested in hot blue stars in this study. The final catalog contains 320 OB stars. The survey covers a large part of the SGS with a total area of 0.28\,kpc$^{2}$.  We adopt a naming convention for all the objects in the survey starting with SMCSGS-FS (SMC supergiant shell FLAMES Survey) and a number in ascending order with their right ascension (1--320).   The full catalog of the survey targets are  available in Table\,B.1 in electronic form at the CDS with their names, positions, and spectral types.  
  
Flux-calibrated UV spectra are available in the Mikulski Archive for Space Telescopes (MAST\footnote{ http://archive.stsci.edu/}) for nine OB stars of our sample. All these stars (SMCSGS-FS\,96, 216, 231, 284, 287, 288, 292, 298, and 310) have an International Ultraviolet Explorer (IUE) spectrum in the wavelength range 1150--2000\,\AA. For SMCSGS-FS\,310, there exists an HST/Space Telescope Imaging Spectrograph (STIS) spectrum, a far-UV Far Ultraviolet Spectroscopic Explorer (FUSE), and a high-resolution Ultraviolet and Visual Echelle Spectrograph (UVES) spectrum. The HST spectrum was taken with the G130M grating in the wavelength range 1135--1429\,\AA, with an effective resolving power of $R$\,=\,18000. The FUSE spectrum was taken with a large aperture ($ 30\arcsec \times 30\arcsec$), and extends over the wavelength range 905--1187\,\AA. The UVES spectra cover wavelength ranges of 3750--5000\,\AA, 4600--6600\,\AA, and 6700--11000\,\AA\, with a resolving power of $R \sim 35000$.

In addition to the spectra, we adopted various photometric data from the VizieR archive, for constructing the spectral energy distributions (SEDs). Ultraviolet and optical ($U, B, V, $ and $ I$) photometry was taken from \citet{Massey2002}, \citet{Zaritsky2004}, \citet{Girard2011}, and \citet{Zacharias2012}. The infrared magnitudes ($JHK_{s}$ and \textit{Spitzer}-$IRAC$) of the OB stars are taken from the catalogs \citet{Kato2007}, \citet{Bonanos2009}, and \citet{Cutri2012}.

\subsection*{Spectral classification}
\label{sect:class}

 We visually compared the blue optical spectra of our sample stars to those of classified stars from \citet{Sota2011,Sota2014}, \citet{Walborn2014}, \citet{Evans2015}, and \citet{McEvoy2015}, while also taking into account the low metallicity of the SMC compared to the Galaxy and the LMC. Detailed descriptions of the subtype classification of OB stars are  given in  \citet{Ramachandran2018} and \citet{Ramachandran2018b}.  The spectral classifications of all stars are tabulated in Table\,\ref{table:App_coord}.

We identified 23 O stars and 297 B stars in the whole sample based on \ion{He}{i}/\ion{He}{ii} 
ionization equilibrium. In O stars, the \ion{He}{i} lines get weaker and \ion{He}{ii} lines get stronger as going toward the earliest subclasses. So, for the classification of early O subtypes, we used the strength and morphology of the optical nitrogen lines \ion{N}{iii\,$\lambda\lambda$4634--4642} (hereafter \ion{N}{iii}) and \ion{N}{iv\,$\lambda4058$} (hereafter \ion{N}{iv}) \citep{Walborn1972,Walborn2002}. We identified these nitrogen emission lines in four of our sample stars, known as Of stars. The nitrogen emission lines of these stars are weaker than the LMC stars of similar spectral subtype.

 The sample contains a very early-type O3 star, SMCSGS-FS\,231 (Sk\,183), which shows   
\ion{N}{iv} and \ion{N}{iii} emission lines in the spectra. Part of our team has studied this star in detail previously \citep[see][]{Evans2012}. We identified another early spectral subtype O6 star, SMCSGS-FS\,287, which shows weak \ion{N}{iii} emission lines and very strong \ion{He}{ii} absorption lines. Two O7 stars (SMCSGS-FS\,241 and 292) are present in the sample, where only SMCSGS-FS\,292 is identified as an Of star based on very weak \ion{N}{iii} emission lines (no available LR03 spectra for SMCSGS-FS\,241 to assign this classification). All these three Of stars have a prominent \ion{He}{ii}\,4686 absorption feature that is stronger than any other \ion{He}{} line in the blue-violet spectra. This characteristic represents the Vz luminosity class. Stars of this class are expected to be very young and may be near or on the zero-age main sequence (ZAMS) \citep{Walborn1973, walborn_2009}. However, analysis of O\,Vz  stars in 30 Doradus by \citet{Sabin-Sanjulian2014} revealed that they might be in more evolved phases because of  weaker O-star winds in the low-metallicity environment of the LMC compared to the Galaxy.
 
A peculiar Of supergiant, SMCSGS-FS\,310 (Sk\,190) is part of the sample. We classified this star as O7.5 based on its strong \ion{He}{ii} absorption lines and very weak \ion{N}{iii} emission lines. The \ion{He}{i} absorption lines are very weak in the blue optical spectrum. The weak \ion{He}{ii}\,4686 emission wings in the spectrum qualify its Onfp nature. This spectral class was introduced by \citet{Walborn1973} to describe  a composite emission + absorption profile at \ion{He}{ii}\,4686.  In this case ``f'' denotes \ion{He}{ii}\,4686 and \ion{N}{iii} emission, ``p'' refers to peculiar early-type spectra and the broadened absorption lines due to rapid rotation is indicated with ``n'' \citep{Walborn1973,Walborn2010}. Since the \ion{He}{ii}\,4686 absorption line is negligible or very weak  compared to the emission wings, the star might be an Of supergiant. Hence we assigned a spectral type of O7.5\,In(f)p to the star. Such evolved stars are not expected to rotate rapidly  because of strong mass loss. \citet{Walborn2010} discussed possible reasons for their rapid rotation such as binarity and stellar mergers. These types of stars are also considered as gamma-ray-burst progenitors.

All other 18 O stars are of late subtype O8--9.7. These are mostly main sequence stars since their 
 \ion{He}{ii}\,4686 absorption line is strong. One of these, SMCSGS-FS\,269, is identified to be a Vz luminosity class star. Because of the weak \ion{He}{ii}\,4686 absorption, SMCSGS-FS\,288 is classified as a giant.

\begin{figure}
\centering
\includegraphics[scale=0.48,trim={0cm 0 0 0cm}]{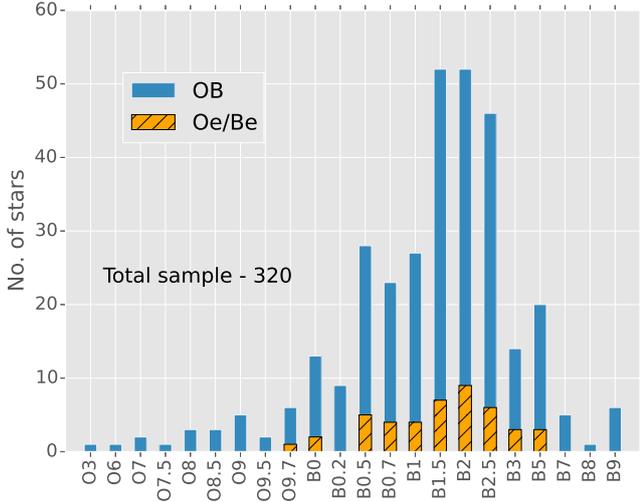}
\caption{Spectral subtype distribution of all OB stars in our sample. 
For each spectral subtype, the number of detected Oe/Be stars are represented  in orange with black lines.}
\label{fig:spec}
\end{figure}

We subclassified all sample B stars into B0--9 based on the ionization equilibrium of helium and silicon. The main diagnostics used for early B types (B0--2) are the \ion{Si}{iii\,$\lambda$4553}/\,\ion{Si}{iv\,$\lambda$4089} line ratio and the strength of \ion{He}{ii\,$\lambda4686$}, \ion{He}{ii\,$\lambda4542$}, and \ion{Mg}{ii\,$\lambda4481$}. Approximately two-thirds of the sample are in the early B-star category, as highlighted by the histogram shown in Fig.\,\ref{fig:spec}. The spectral subtype with highest number of sample stars  is B2. This is because, the sample is not complete in the case of late B stars (B2.5--9). We identified a total of 95 stars later than B2. The main diagnostic lines used to subclassify these late B-type stars are \ion{Si}{ii\,$\lambda\lambda$4128-4132}, \ion{He}{i\,$\lambda4144$}, \ion{He}{i\,$\lambda4471$}, and \ion{Mg}{ii\,$\lambda4481$}.

The main criteria used for determining the luminosity classes of B stars are the width of the Balmer lines and the intensity of the \ion{Si}{iv} and \ion{Si}{iii} absorption lines \citep{Evans2015}. We identified five B supergiants in the sample, which show rich absorption-line spectra. Among these, four are late B9 supergiants and one is a B2 supergiant. There are also 20 giants / bright giants present in the sample and the rest are main sequence B stars.  \cite{Ramachandran2018b} and \citet{Evans2015}  provide a detailed description of spectral and luminosity classifications of B stars.

%%---------------------------------------------------------------

%---------------------------------------------------------------

We identified 44 stars with strong emission lines among the whole sample. The optical spectra of these non-supergiant stars exhibit broad Balmer emission lines, especially H$\alpha$. These are characteristic features of Be (or Oe) stars, which emerge from their circumstellar decretion disk. The high-resolution H$\alpha$ observations (HR15N) helped us to distinguish disk emission profiles. In some cases, we observed a double-horn profile in H$\alpha$ and H$\beta$. Some spectra are contaminated by nebular emission.

The fraction of Be stars in the sample is about 15\%. This is slightly higher than that of our LMC sample described in  \cite{Ramachandran2018b}. This fraction is just a lower limit since we have only single epoch spectroscopic observations of our sample stars. We might have missed a fraction of Be stars owing to their transient nature and variability in the emission line profiles \citep [about one-third to one-half,][]{McSwain2008}.
Among this Be sample, 15 stars are identified as giants or bright giants, while the rest are dwarfs. These Be stars have spectral subtypes ranging from B0 to B5 (see Fig.\,\ref{fig:spec}). Interestingly, 15 out of 25 B-type giants are Be stars.   
 
\bigskip
 The statistics of spectral subtypes of all investigated OB stars is shown in Fig.\,\ref{fig:spec}. Emission line stars in each subtype are also illustrated. We can see a gradual increase in the number of stars with spectral subtypes starting from O3, until B2. Stars with spectral subtype B2 are the most common, while this spectral class also encompasses more Be stars than any other. The number of stars declines toward late B subtypes most probably because of our brightness limit.

\section{Quantitative analysis}
\label{sect:analysis}
We performed the spectral analysis of all 320 sample OB stars in SMC-SGS\,1 using the  PoWR model atmosphere code.  By fitting the model spectra to the observational data, we derive stellar and wind parameters of the individual OB stars.

\subsection{Models}
\label{sect:models}

PoWR is a state-of-the-art stellar atmosphere code suitable for the spectroscopic analysis of any hot stars with and without winds, across a broad range of metallicities \citep{Hainich2014,Hainich2015,Oskinova2011}. The PoWR models have already been used to analyze the complete sample of Wolf-Rayet (WR) stars and binaries in the SMC \citep{Hainich2015,Shenar2016}.  The PoWR code solves the radiative transfer equation for a spherically expanding atmosphere and the statistical equilibrium equations simultaneously under the constraint of energy conservation. Stellar parameters were determined iteratively.  Details of the PoWR code are described in \citet{Graefener2002}, \citet{Hamann2004}, \citet{Sander2015}, and \citet{Todt2015}.

A PoWR model is specified by the star's luminosity $L$,  stellar temperature $T_\ast$,  surface gravity $g_\ast$, and  mass-loss rate $ \dot{M} $ as the main parameters. The stellar temperature relates to $R_\ast$ and $L$ via the Stefan-Boltzmann law $L = 4 \pi \sigma_{\mathrm{SB}}\, R_\ast^2\, T_\ast^4$. In this case the ``stellar temperature'' $T_\ast$ is the effective temperature $T_\mathrm{eff}$ corresponding to the stellar radius $R_\ast$.  We place the latter at the Rosseland continuum optical depth of 20. In the case of our program stars, the winds are optically thin and the differences between $T_\ast$ and $T_\mathrm{eff}$ ($\tau = 2/3$) are negligible.  

In the subsonic region of the stellar atmosphere, a velocity field is defined such that a hydrostatic density stratification is approached \citep{Sander2015}. In the supersonic  region, the prespecified wind velocity field $\varv(r)$ is assumed to follow 
the so-called $\beta$ -law \citep{CAK1975} 
%\begin{equation}
%\varv(r) = \varv_\infty \left( 1- \frac{r_{0}}{r} \right)^{\beta}.
%\end{equation}
In this work, we adopt $\beta$=0.8, which is a typical value for O-type stars 
\citep{Kudritzki1989}. 

In the non-LTE iteration in the co-moving frame, the line opacity and emissivity  profiles  are  
Gaussians  with  a  constant Doppler width $\varv_{\mathrm{Dop}}$. We set this 
parameter to 30\,km\,s$ ^{-1} $ for our OB sample. In the formal integral for the calculation of the emergent spectrum, the Doppler velocity is split into the depth-dependent thermal velocity and a ``microturbulence velocity'' $\xi(r)$. The pressure broadening is also taken into account by means of microturbulent velocity. We adopt  $\xi(r) = \rm max(\xi_{min},\, 0.1\varv(\emph{r}))$ for OB models, where $\rm \xi_{min}= 20 $\,km\,s$ ^{-1} $ \citep{Shenar2016}. For main-sequence B stars,  $\rm \xi_{min}$ can be low as  $5 $\,km\,s$ ^{-1} $  \citep{Hunter2008,Dufton2006}. By constructing some models with lower microturbulence velocities, we found that this effect is within the uncertainty limits of our grid parameters.

Optically thin inhomogeneities in the model iterations are described by the 
``clumping factor''  $D$ by which the density in the clumps is enhanced compared 
 to a homogeneous wind of the same $\dot{M}$ \citep{HK98}.
 For all the OB stars in our study, we account for depth-dependent clumping assuming that clumping begins at the sonic point, increases outward, and reaches a density contrast of $D=10$ at a radius of $R_{\mathrm{D}} = \, 10\,R_\ast$ \citep{Runacres2002}.  We note that the empirical mass-loss rates when derived from H$\alpha$ emission scale with $D^{-1/2}$, since this line is mainly fed via recombination.

The models are calculated using complex atomic data of H, He, C, N, O, Si, Mg, S, P, and Fe group elements. The iron group elements are treated with the so-called superlevel approach  as described in \citet{Graefener2002}.  We adopt the following chemical abundance based on \citet{Trundle2007}\,: $X_{\rm H} = 0.7375$, $X_{\rm C} = 2.1 \times 10^{ -4}$ , $X_{\rm N} = 3.26 \times 10 ^{-5}$, $X_{\rm O} = 1.13 \times 10^{ -3}$, $X_{\rm Si} = 1.3 \times 10^{ -4}$, $X_{\rm Mg} = 9.9 \times 10^{-5}$, $X_{\rm S} = 4.42 \times 10^{ -5}$, $X_{\rm P} = 8.32 \times 10^{ -7}$, and $X_{\rm Fe} = 3.52 \times 10^{ -4}$.

\subsection{Spectral fitting}
\label{subsec:specfit}

%---------------------------------------------------------------
 \begin{figure*}
 \sidecaption
\centering
\includegraphics[width=12cm,trim={0 6cm 0 0}]{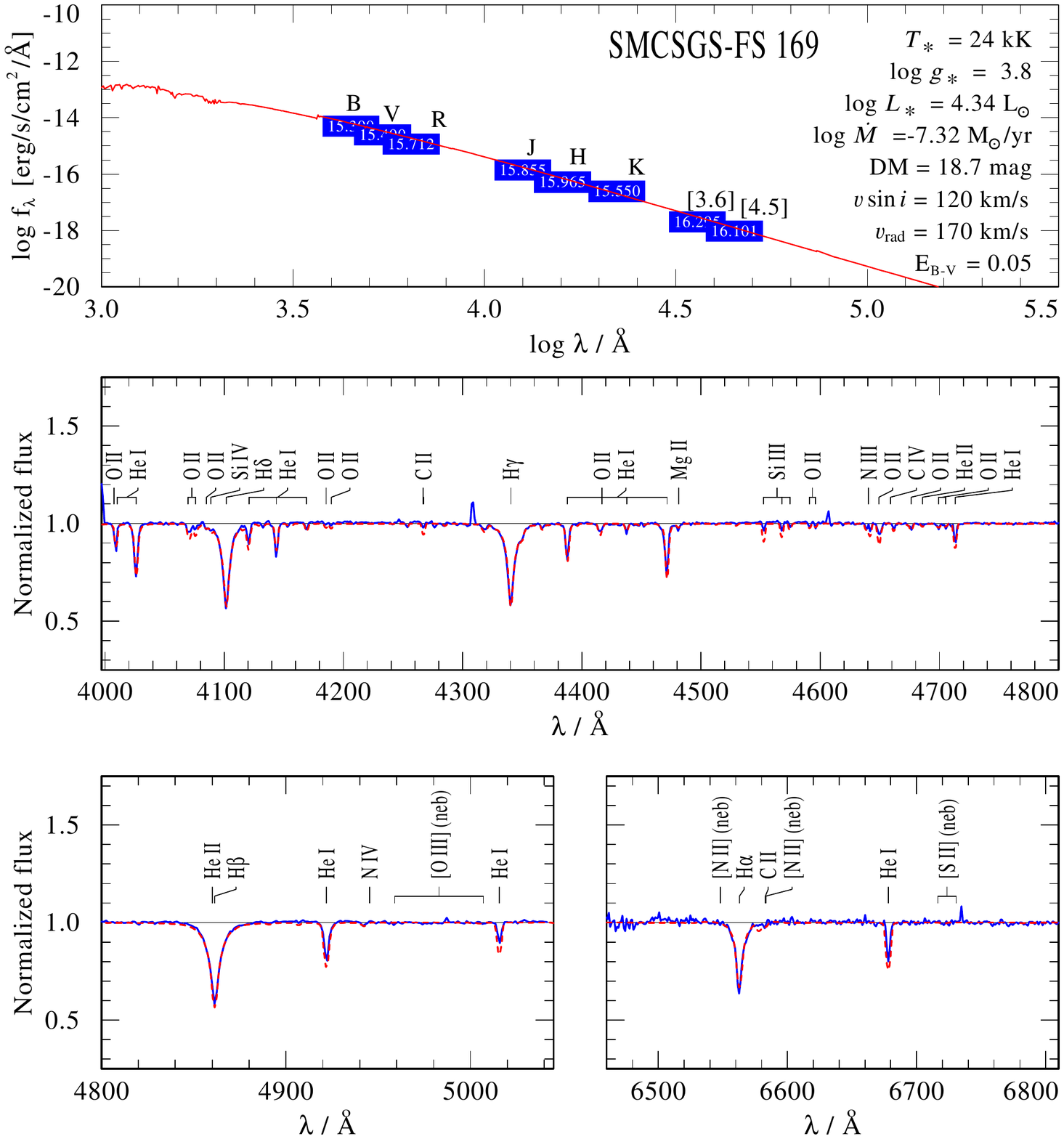}
\caption{Spectral fit for SMCSGS-FS\,169. Top panel: The model SED (red) fitted to the available photometry from optical ($UBV$ and $I$), and infrared ($JHK_{s}$ and $IRAC$ 3.6 \& 4.5 $\rm \mu m$) bands (blue boxes). Botton panels: The VLT-FLAMES spectra (blue solid line), overplotted with the PoWR model (red dashed line). The parameters of this best-fit model are given in Table\,\ref{table:App_stellarparameters}.}
\label{fig:optfit}
\end{figure*}
%---------------------------------------------------------------
%---------------------------------------------------------------
 \begin{figure*}
 \sidecaption
\centering
\includegraphics[width=12cm,trim={0cm 7.5cm 0 5.5cm}]{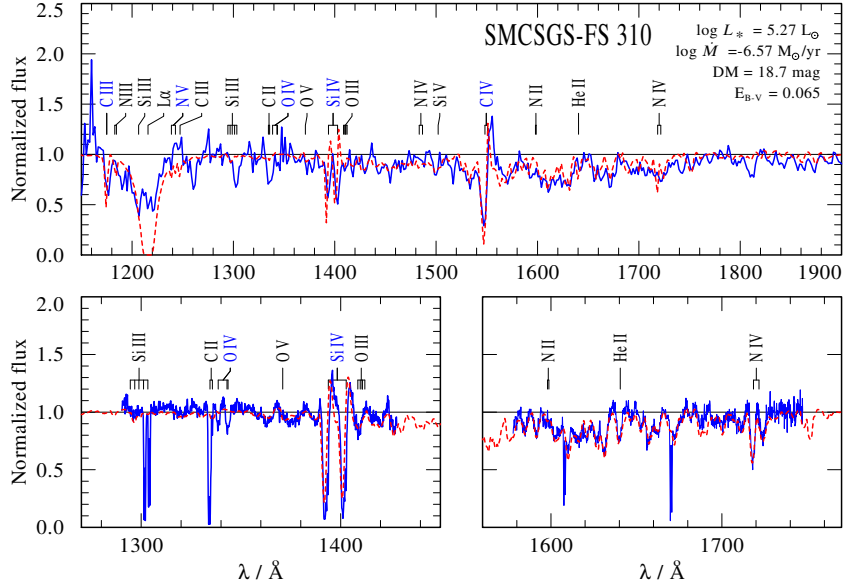}
  \caption{Spectral fit for SMCSGS-FS\,310. The panels show the normalized HST (blue solid line), overplotted with the PoWR model (red dashed line). The parameters of this best-fit model are given in
Table\,\ref{table:App_stellarparameters}.}
  \label{fig:310uv}
\end{figure*}
%---------------------------------------------------------------

For a systematic spectral analysis, we constructed grids of OB-star model atmospheres for SMC metallicity with the stellar temperature $T_\ast$ and the surface gravity $\log\,g_\ast$ as main parameters. The grid spans from $T_\ast$\,=\,10\,kK to 50\,kK with a spacing of 1\,kK, and $\log\,g_\ast$\,=\,2.0 to 4.4\,[cm\,s$^{-2}$] with a spacing of 0.2\,dex. For a given value of ($T_\ast$,$\log\,g_\ast$), the stellar mass $M$ and luminosity $L$  are chosen according to the evolutionary tracks calculated by \citet{Brott2011}. While the chemical composition is kept constant within the model grid, we also calculate specific models with adjusted C, N, O, Si abundances, when necessary. The SMC OB star grid is also available online\footnote{www.astro.physik.uni-potsdam.de/PoWR}. More details of the PoWR model grids for OB stars are given in Hainich et al.\,(2018).

 For all the 320 OB stellar spectra in the sample, we determined the best-fitting model by a careful iterative visual comparison of observed spectra with the model. We considered only single-star models for the fits. Our first step was to find the main parameters, i.e.,  the stellar temperature and surface gravity of the model that best represents the observed spectrum. The primary constraint we used is the helium and silicon ionization balance to determine whether the spectrum represents a very hot O star or cooler B star. For the hottest stars in the sample, we derived the stellar temperature by consistently fitting nitrogen emission lines and \ion{He}{ii} lines. In the case of cooler stars (10-20\,kK), we determined the temperature by fitting \ion{Si}{ii}, \ion{Mg}{ii}, and \ion{He}{i} lines. For intermediate temperature stars, we used the \ion{Si}{iii} to \ion{Si}{iv}  and \ion{He}{i}\,/\,\ion{He}{ii} line ratios. \changed{The precision in the temperature is limited by the grid resolution of $ \pm $1\,kK}. After getting a constraint on the temperature, we measured the surface gravity using the pressure-broadened wings of the Balmer lines. The main diagnostic lines are H$\gamma$ and H$\delta$, since they are less affected by wind as well as disk emission. The typical uncertainty for $\log\,g_\ast$ is $\pm$0.2\,dex. Since the ionization balance also reacts on $\log\,g_\ast$, the temperature might need to readjust accordingly. The uncertainty in $\log\,g_\ast$ therefore propagates to the temperature and leads to a total uncertainty of about $\pm $2\,kK. As example, the spectra of an O8 star (blue lines) is shown in Fig.\,\ref{fig:optfit} overplotted with a model (red lines) with $T _\ast =$\,35\,kK and $\log\,g_\ast = 4.2\,[\rm{cm\,s^{-2}}$]. It should be noted that  the above method of spectral fitting is not successful in the case of Be stars, since their Balmer absorption lines may be filled by disk emission. Therefore, the uncertainty in the stellar temperature and surface gravity of these stars is relatively high. 
  
After fixing $T_\ast$ and log\,$g_\ast$, we determined the wind parameters whenever possible. The mass-loss rate and terminal velocities ($\varv_\infty$) can be estimated from the UV P-Cygni profiles. For nine of our sample stars, UV spectra are available. We used the main diagnostic lines  \ion{C}{iv\,$\lambda\lambda$1548--1551} and \ion{Si}{iv\,$\lambda\lambda$1393--1403} in the HST\,/ IUE range and \ion{P}{v\,$\lambda\lambda$1118--1128}, \ion{C}{iv\,$\lambda$1169}, and \ion{C}{iii\,$\lambda$1176} in the FUSE range. The typical uncertainties are $\pm$0.2\,dex in $\log \dot{M}$ and $\pm$\,100 km\,s$^{-1}$ $\varv_\infty$. Since UV spectra are not available for most of the sample stars, we need to constrain $\dot{M}$ and $\varv_\infty$ based on optical lines such as H$\alpha$ and \ion{He}{ii\,$\lambda4686$}. In fact none of the stars in the sample shows wind emission in  H$\alpha$ or \ion{He}{ii\,$\lambda4686$} (do not confuse with disk emission in Be stars).  In these cases, the adopted mass-loss rate is only an upper limit. For stars with no available UV spectra,  we estimated the terminal velocities  theoretically from the escape velocity $\varv_{\mathrm{esc}}$. For Galactic massive stars with $T_\ast\geq$\,21\,kK, the ratio $\varv_\infty$/$\varv_{\mathrm{esc}}$ is 2.6 and for stars with $T_\ast <$\,21\,kK the ratio is $\approx$1.3 \citep{Lamers1995,Kudritzki2000}. We adopt a scaling for SMC metallicity using the relation, $\varv_\infty \, \propto (Z/Z_{\odot})^{q}$, where $q=0.13$ \citep{Leitherer1992}. The UV spectrum of the sample star SMCSGS-FS\,310 is shown in  Fig.\,\ref{fig:310uv}. The derived mass-loss rate and $\varv_\infty$ from the UV P-Cygni profiles are $\log \dot{M}$ = -6.6 [$ M_{\odot}\, \mathrm{yr}^{-1}$] and 550\,km\,s$^{-1}$, respectively.

We determined the luminosity $L$ and  color excess $E _{\rm B-V} $ of the individual OB stars  by fitting the model SED to the photometry (see first panel of Fig.\,\ref{fig:310uv}). In this case the model flux is diluted with the SMC distance modulus of 18.7\,mag. We consistently adjusted the color excess and luminosity of the model's SED to reproduce the observed data. The typical uncertainty in $\log\,L/L _{\odot}$ is about 0.2 dex. For stars with available flux-calibrated UV spectra (HST, IUE, or FUSE), the uncertainty in the luminosity is only $\approx$0.1 dex in these cases. The color excess of OB stars in the Wing of the SMC is very small, typically 0.05\,mag. 

Subsequently, we chose a best-fit model for each individual OB star in the sample and overplotted the selected model with the observed spectra (eg. Fig.\,\ref{fig:optfit} and Fig.\,\ref{fig:310uv}). There is a systematic shift between synthetic models and observed spectra owing to the radial velocities of sample stars. We measured the radial velocity of individual stars by fitting a number of absorption line centers of the synthetic spectra to the observation. The primary lines used for these measurements are absorption lines of \ion{He}{i}, \ion{He}{ii}, and \ion{Si }{iii}. The typical uncertainty of $\varv_{\rm rad}$ varies from $\pm$10 to 20 km\,s$^{-1}$. 

Finally, we estimated the projected rotation velocity $\varv\,\sin i$ of all OB stars from their line profile shapes. The measurements are based on the Fourier transform (FT) method using the \texttt{iacob-broad} tool \citep{Simon-diaz2014}. The primary lines selected for applying these methods are \ion{He}{i}, \ion{Si }{iv}  and \ion{Si }{iii}. The typical uncertainty in $\varv\,\sin i$ is $ \sim$\,10\%. We convolved our model spectra with measured $\varv\,\sin i$ to account for rotational broadening, which results in a consistent fit with the observations. For example, the model spectra shown in Fig.\,\ref{fig:optfit} is convolved with a $\varv\,\sin i$ of 300\,km\,s$^{-1}$.

We applied these spectral fitting methods  in an iterative manner for individual OB star spectrum. More detailed explanations of the fitting procedure for each parameter is given in \cite{Ramachandran2018b}. We also calculated individual models with refined stellar parameters and abundances for each of these stars, when necessary. The fitting procedure continued until no further improvement of the fit was possible. The final best-fit models yield the stellar and wind parameters of all OB stars in our sample.

%%---------------------------------------------------------------
\begin{figure*}
\centering
\includegraphics[scale=0.7]{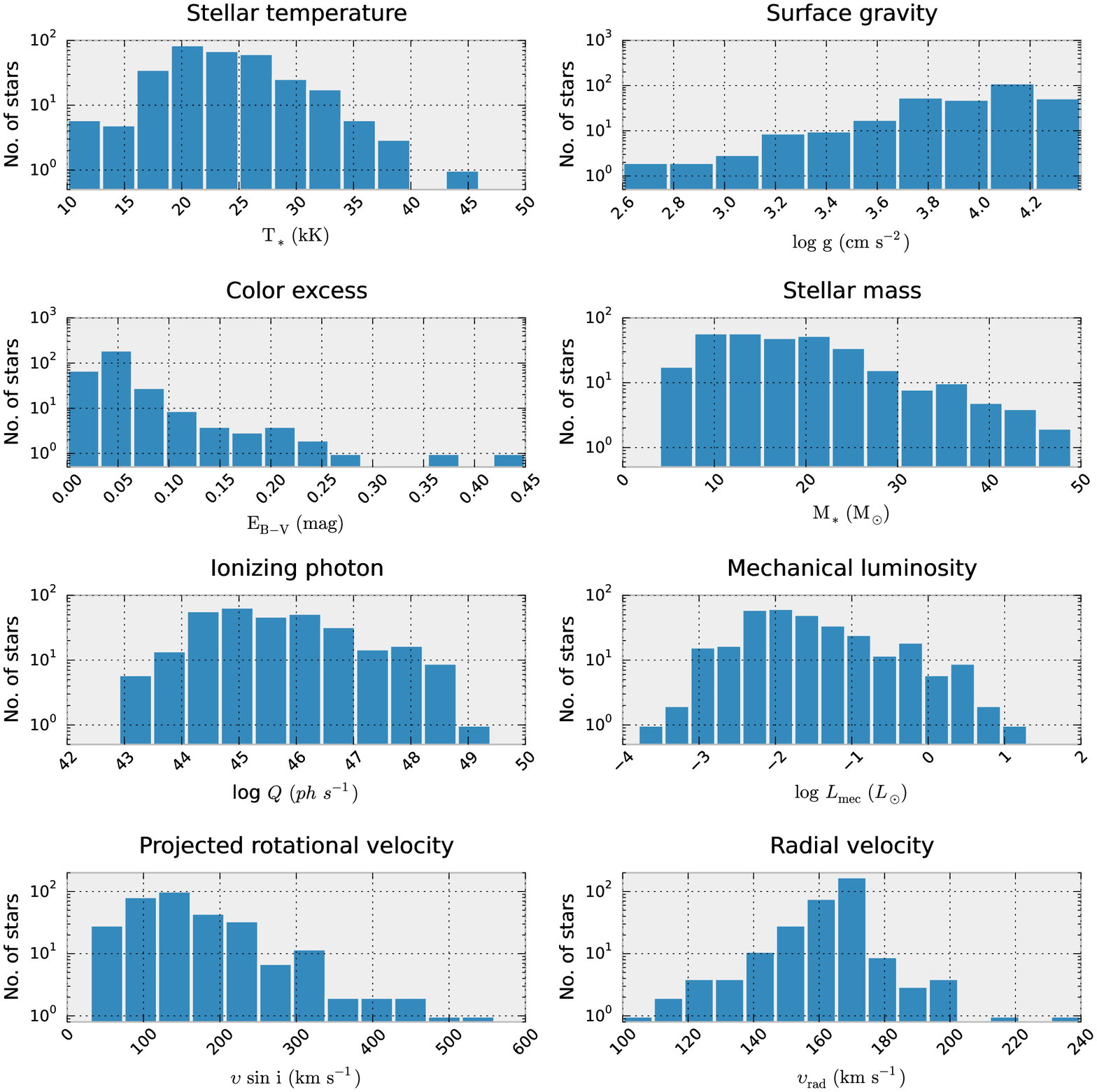}
\caption{Histograms of stellar temperature, surface gravity, color excess, projected rotational velocity, radial velocity, stellar mass, mass-loss rate, and mechanical luminosity of OB stars in SMC-SGS\,1. The bin widths used are 3\,kK, 0.18\,dex, 0.03\,mag, 4\,$M _{\odot}$, 0.6\,dex, 0.4\,dex, 45\,km\,s$^{-1}$, and 10\,km\,s$^{-1}$, respectively.  The number of stars in the y-axis are shown in logarithmic scale.}
\label{fig:hist}
\end{figure*}

%---------------------------------------------------------------

\section{Stellar parameters}
\label{sect:sparameters}

The stellar and wind parameters of individual OB stars derived from spectral analysis are  given in Table\,\ref{table:App_stellarparameters}. The PoWR model also gives the rate of hydrogen ionizing photons $Q_{0}$. We calculated the values of mass-loss rate, terminal velocity, and the mechanical luminosity $L_{\mathrm{mec}}\,=\,0.5\,\dot{M}\,\varv_\infty^{2}$ produced by the stellar winds only for nine stars with available UV spectra. For the rest of the stars we adopted values from the derived relation of these nine stars. The distribution of some of these derived parameters in the total sample is illustrated in Fig.\,\ref{fig:hist}.  Since we obtained only single-epoch spectra, we are not able to detect binarity. 

The stellar temperature statistics of the sample is shown in the top left panel of Fig.\,\ref{fig:hist}. The temperature of our sample stars ranges from 10\,kK to 46\,kK, with a peak around 20\,kK. It has to be emphasized that stars in the low temperature regime are incomplete owing to the observation limit. The earliest type O3 star has the highest stellar temperature of 46\,kK. The sample also contains late B supergiants of temperature $\approx 10$\,kK. The surface gravities of sample stars are in the range of $10^{2.6}\,\rm{to}\, 10^{4.4}\,\rm{cm\,s^{-2}}$, while most of the stars are found at $\log\,g_\ast$ of 4.2. This indicates that majority of the sample are young main sequence stars, while only a few percent of the stars are in evolved stages.

The mean color excess of our sample stars is only about $ 0.05$\,mag, which can be mainly attributed to the Galactic foreground. Hence the dust content in the Wing of the SMC is significantly low. Only six percent of the sample shows a color excess higher than 0.1\,mag. The star, SMCSGS-FS\,45, with highest $E_{\rm B-V} \approx 0.45$\,mag  is a B supergiant located close to the highly extincted \ion{H}{ii} region N\,88 \citep{Heydari-Malayeri1999}.

 The histogram of the stellar masses given in Fig.\,\ref{fig:hist} ranges from 4\,$M _{\odot} $ to 50\,$M _{\odot} $. In this figure the stellar mass refers to the spectroscopic mass calculated from $\log\,g_\ast$ and $R_\ast$ ($g_\ast=G\,M_\ast\,R_\ast^{-2}$).  We note that the lowest mass bins are incomplete owing to the observational constraint. Most of the stars in the sample have masses in the range of 10--20\,\,$M _{\odot} $. 
 
 The typical ionizing photon flux provided by an OB star in our sample is $10^{45}\,\rm{s^{-1}}$. One O3 star dominates the ionization of the region by releasing photons at a rate of $10^{49}\,\rm{s^{-1}}$. 
 
 We also plotted the statistics of the mechanical luminosity $L_{\rm mec}$ released by the stellar winds of our sample stars in Fig.\,\ref{fig:hist}. The OB stellar winds in SMC-SGS\,1 provide only  $L_{\rm mec} <10\,L_{\odot}$ to the surrounding ISM. Most of the stars produce mechanical luminosities a hundred times less than one solar luminosity. This is much lower than the mechanical luminosities of LMC OB stars \citep{Ramachandran2018}. 

\subsection{Observed $\varv\sin i$ distributions} 
\label{sec:vsini}

We derive the projected rotational velocities ($\varv\sin i$) of our sample stars from their optical line profiles.
 The $\varv\sin i$ distribution  shows a main peak at $\sim$120\,km\,s$^{-1}$, and the  tail of the distribution extends to high rotational velocities up to 550\,km\,s$^{-1}$.  The fastest rotator in the sample is a Be star SMCSGS-FS\,283, which has $\varv\sin i \approx 550$\,km\,s$^{-1}$. We identified 20 very rapidly rotating  stars with $\varv\sin i >300$\,km\,s$^{-1}$, which constitute 6\% of the whole sample. This is double compared to that of our LMC  \citep{Ramachandran2018b}. Among these rapid rotators, 12 stars have Oe/Be characteristics. Moreover, we can see a low but noteworthy peak around 300\,km\,s$^{-1}$ , which might be related to the effects of mergers and mass transfer in binary evolution \citep{deMink2013,deMink2014}.

%%---------------------------------------------------------------
\begin{figure}
\centering
\includegraphics[scale=0.5]{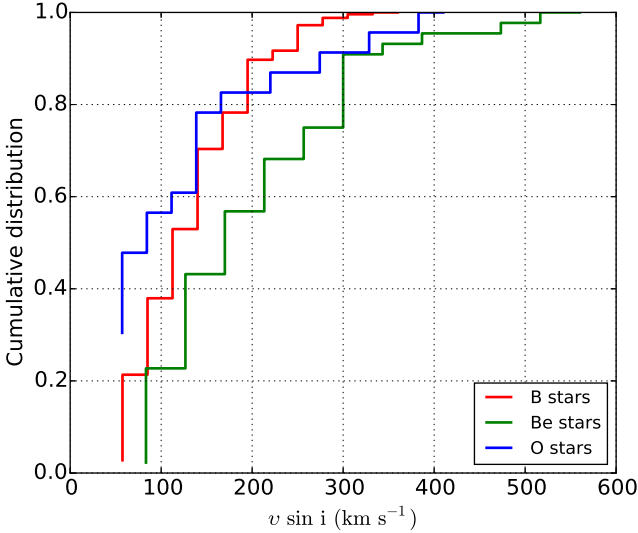}
\caption{Comparison of CDFs of  $\varv\sin i$ between B, Be, and O stars }
\label{fig:cdf_OB}
\end{figure}

%---------------------------------------------------------------

%%---------------------------------------------------------------
\begin{figure}
\centering
\includegraphics[scale=0.5]{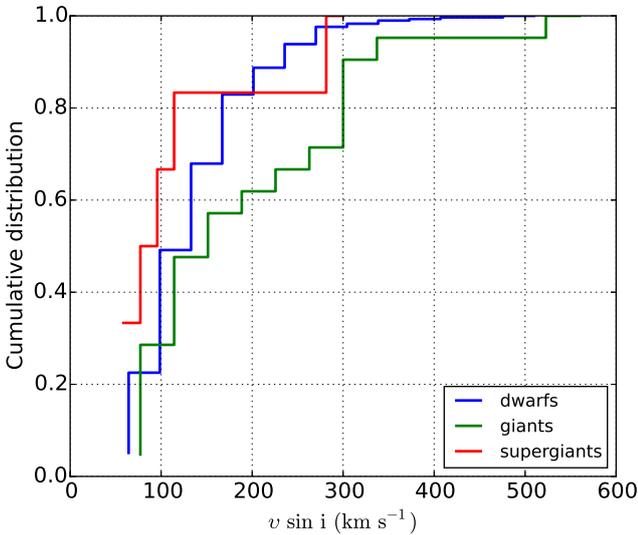}
\caption{Comparison of $\varv\sin i$ CDFs of  dwarfs, giants, and supergiants}
\label{fig:cdf_lumclass}
\end{figure}

%---------------------------------------------------------------

For further understanding, we constructed the cumulative distribution functions (CDFs) of the projected rotational velocities of our sample stars. The $\varv\sin i$ CDFs of O, B, and Be stars are compared in Fig.\,\ref{fig:cdf_OB}. As expected, Be stars exhibit a faster rotation compared to the normal B and O stars in the sample. The mean projected rotational velocities of Be stars in our SMC-SGS\,1 sample is $\approx$230\,km\,s$^{-1}$, which is significantly higher than the mean $\varv\sin i \approx$140\,km\,s$^{-1}$ of normal B stars. Most of the Be stars in this sample  rotate close to their critical velocity.

We constructed $\varv\sin i$ CDFs of stars with different luminosity classes as shown in Fig.\,\ref{fig:cdf_lumclass}. About 40\% of the giants have a $\varv\sin i >200$\,km\,s$^{-1}$. On average, they show a higher rotational velocity than unevolved dwarfs or evolved supergiants. Interestingly,  70\% of these giants are found to be emission line Be stars. In a study of early-type stars in the SMC, \citet{Mokiem2006} found that very fast and very slow rotators are unevolved with respect to the group of evolved objects.  However, in our sample, dwarfs are found to be less fast rotating than giants. 

\begin{figure*}
\sidecaption
\centering
\includegraphics[width=12cm]{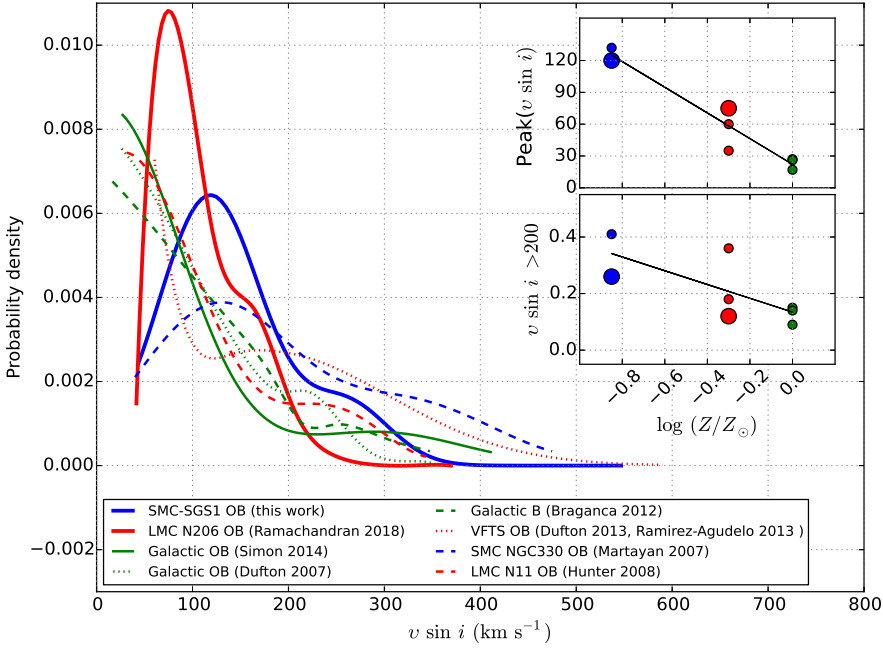}
\caption{Probability density distribution of projected rotational velocities
 for  metallicities  $Z =Z_{\odot}$ (Galactic, green curve), $Z =0.5Z_{\odot}$ (LMC, red curve), and  $Z =0.2Z_{\odot}$ (SMC, blue curve). The subplots in the top and bottom show the effect of metallicity with the peak of $\varv\sin i$ distribution\,/ most probable velocity and the fraction of stars with $\varv\sin i$ higher than 200\,km\,s$^{-1}$, respectively. The Galactic, LMC and, SMC samples are represented using green, red, and blue dots, respectively. Our SMC (this work) and LMC  \citep{Ramachandran2018b} samples are shown as enlarged dots. See Sect.\,\ref{sec:vsini} for further information. }
\label{fig:vrot_met_distri}
\end{figure*}

 The simulated distribution of projected rotational velocities  has a bimodal structure \citet{deMink2013}. Theory predicts that the majority of stars have low to intermediate rotational velocities, whereas nearly one-fourth have rotational velocities in excess of 200\,km\,s$^{-1}$. 
The VLT-FLAMES Tarantula Survey (VFTS) of massive stars in the LMC by \citet{Ramirez-Agudelo2013} and \citet{Dufton2013} also found a bimodal $\varv\sin i$ distribution. In order to check this trend in our sample, we plotted the probability density distribution of OB stars in Fig.\,\ref{fig:vrot_met_distri} (blue solid curve). This reveals a two-component structure with about 26\% of  stars have $\varv\sin i >200$\,km\,s$^{-1}$.

 For a comparison, the probability density distribution of $\varv\sin i$ of OB stars for various metallicities (Galaxy, LMC, and SMC) taken from our results and from previous papers are shown in  Fig.\,\ref{fig:vrot_met_distri}. All these distributions are bimodal. However, there are noticeable differences in the main peak of the distribution (most probable velocity) as well as in the tail of the distribution. At SMC metallicities ($0.14\,Z_{\odot}$), the distribution consists of two broad peaks and the tail of the distribution extends to velocities as high as 500\,km\,s$^{-1}$. The  $\varv\sin i$ distributions of LMC samples ($0.5\,Z_{\odot}$) have peaks at lower velocities than the the SMC sample. The Galactic samples show main peaks at relatively low velocities ($<50$\,km\,s$^{-1}$), the secondary peak is not very pronounced compared to the lower metallicity samples.   Obviously, OB stars in the SMC sample have higher rotational velocities. This is a possible consequence of lower SMC metallicity, where the stars are more compact, stellar winds are weaker compared to LMC stars, hence suffer less angular momentum loss and therefore rotate faster \citep{Meynet2002,Massey2003,Ekstrom2008}. 
 
 The subplot in Fig.\,\ref{fig:vrot_met_distri} (top) shows a linear relationship between the peak of the distribution with the metallicity as given below,
\begin{equation}
\rm{Peak}\,(\varv\sin i) / (\rm{km\,s^{-1}}) \simeq -121 \times \log\,(Z/Z_{\odot})+22.
\end{equation}

The fraction of fast rotating stars above 200\,km\,s$^{-1}$  also decreases with increasing metallicity, and can be written in the form, 
\begin{equation}
\rm{Fraction}\, (\varv\,\sin\,i >200) \simeq -0.2 \times \log\,(Z/Z_{\odot})+0.1.
\end{equation}
The OB star sample in the VFTS survey shows a very high fraction of rapidly rotating stars compared to other LMC samples. It should be emphasized that these  OB star samples from various papers have different selection criteria, observational biases, and different age distributions, etcetera.

%%---------------------------------------------------------------

\subsection{Radial velocity and runaway candidates}
\label{sec:radial}

%%---------------------------------------------------------------
\begin{figure}
\centering
\includegraphics[scale=0.52]{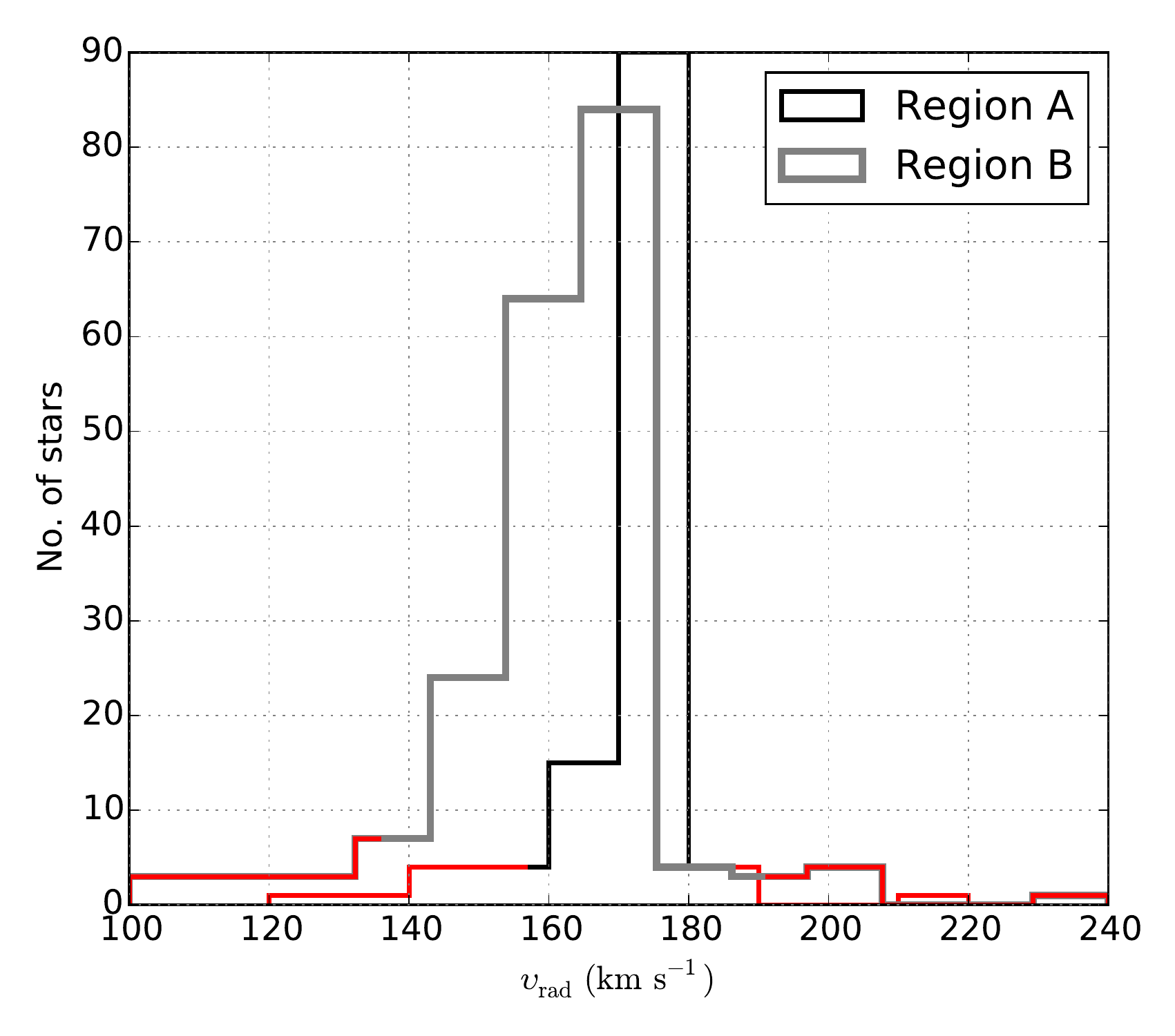}
\caption{Radial velocity distribution of stars located in region A and B. We used velocity bins of 10\,km\,s$^{-1}$ to construct the histogram.  The value $|\varv_{\rm rad}| > 3 \sigma$ is represented by red lines in both histograms.}
\label{fig:histcomp}
\end{figure}

%---------------------------------------------------------------

%%---------------------------------------------------------------
\begin{figure}
\centering
\includegraphics[scale=0.29]{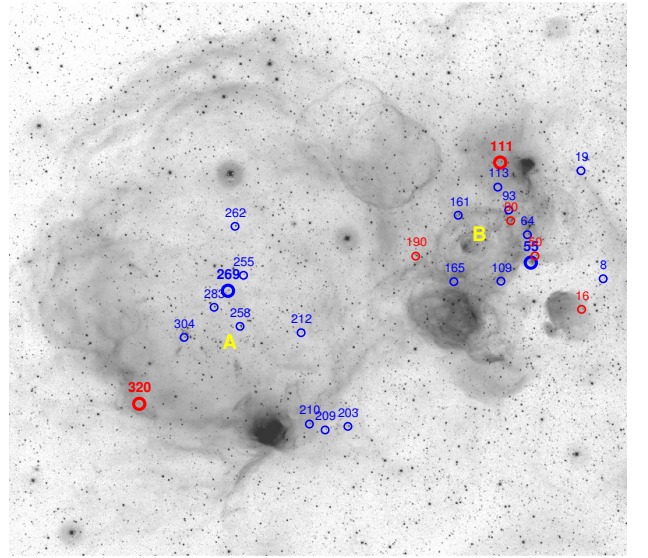}
\caption{Candidate runaway stars in the sample. The underlying H$\alpha$ image is from the Magellanic Cloud Emission-Line Survey \citep[MCELS, ][]{Smith2005}. stars with redshifted and blueshifted velocities are indicated with red and blue circles, respectively. Small circles represent candidate runaway stars with  $|\delta \varv_{\rm rad}| > 3 \sigma$, and large circles for stars with  $|\delta \varv_{\rm rad}| > 4 \sigma$.}
\label{fig:run}
\end{figure}

%---------------------------------------------------------------

 The radial velocities in our SMC-SGS\,1 sample range from 100 to 240\,km\,s$^{-1}$ (see Fig.\,\ref{fig:hist}). The peak of the distribution is about $\sim$170\,km\,s$^{-1}$, which is close to the central velocity of the associated \ion{H}{i} supershell \citep{Staveley-Smith1997}. The ionized gas shell  associated with SMC-SGS\,1 also has a heliocentric velocity of 173 km\,s$^{-1}$ \citep{Fulmer2019}. 
 The OB stars in the SMC show a radial velocity gradient, where the velocity distribution of stars in the Wing is more redshifted than in the Bar of the SMC \citep{Evans2008,Lamb2016}. In order to check for the velocity gradient within the SGS, we plotted the radial velocity distribution of stars located near the center of the SGS or close to NGC\,602 (region A) and stars located in the western part of the SGS close to N\,88 (region B).  We can see a small velocity gradient between stars in these two regions (see Fig.\,\ref{fig:histcomp}). Stars in region B exhibit lower radial velocities and higher dispersion than in region A. The region A is within the SGS and has a more sharply peaked velocity distribution. The velocity spread between the stars is similar to the expansion velocity of the present day shell  \citep[see][for details about velocity of the shell]{Fulmer2019}.

The radial velocities can be used to reveal runaway stars in a region. 
These are isolated stars or binaries that have escaped from clusters or OB associations as a result of a binary SN \citep{Blaauw1961} or  dynamical ejection \citep{Leonard1990}. In order to identify possible runaway candidates, we compared the radial velocity of each star to the mean and standard deviation of all sample stars. Since region A and B have different radial velocity distributions, we decided to check for runaway stars separately in these regions. We defined a velocity threshold of $|\delta \varv_{\rm rad}| > 3 \sigma$ so that stars with outlying radial velocities are identified as runaway candidates. By excluding these stars, we continued to recalculate the mean velocity and standard deviation, until no more stars with $|\delta \varv_{\rm rad}| > 3 \sigma$ remain. Using this method, we found 11 and 14 candidate runaway stars in region A and B, respectively. This also includes the HMXB SMCSGS-FS\,203 (alias SXP\,1062). Their positions are denoted in Fig.\,\ref{fig:run}, while their radial velocities and the deviation from the mean velocity are listed in Table\,\ref{table:runstars} along with their spectral types. Stars with positive and negative radial velocity deviations from the mean are denoted separately in Fig.\,\ref{fig:run}. The mean radial velocity of the OB stars (excluding the runaway candidates) in region A and region B are $169\pm$4\,km\,s$^{-1}$ and $163\pm$9\,km\,s$^{-1}$, respectively.  The velocity dispersions in both regions are very small considering their size. Therefore, we also defined a higher velocity threshold of  $|\delta \varv_{\rm rad}| > 4 \sigma$, which limits the possible runaway stars to four (large circles in Fig.\,\ref{fig:run}). The estimated radial velocity gradient in the region is $\lesssim 3$\,km\,s$^{-1}\,$deg$^{-1}$ along RA and  $\lesssim13$\,km\,s$^{-1}$\,deg$^{-1}$ along Dec.

\subsection{Spectral calibrations for OB stars at SMC metallicity}
\label{spec_calib}
With a sample of 320 OB stars in the Wing of the SMC, our study offers a unique opportunity to calibrate the physical parameters of OB stars. This will be helpful for characterizing massive stars in extragalactic dwarf galaxies. It is not possible to obtain  parameters such as temperature and ionizing flux  by using optical photometry alone. Using quantitative spectroscopy, we measured these parameters and then calibrated with spectral types. Such studies have been pursued for Galactic \citep{Martins2005,Repolust2004},  LMC \citep{Ramirez-Agudelo2017, Mokiem2007B}, and SMC \citep{Mokiem2006,Bouret2003} massive stars. However, these studies mainly focused on O stars.  In this work  we extend this up to late B-type stars with a larger sample at SMC metallicity. 
 %---------------------------------------------------------------
\begin{figure}
\centering
\includegraphics[scale=0.5]{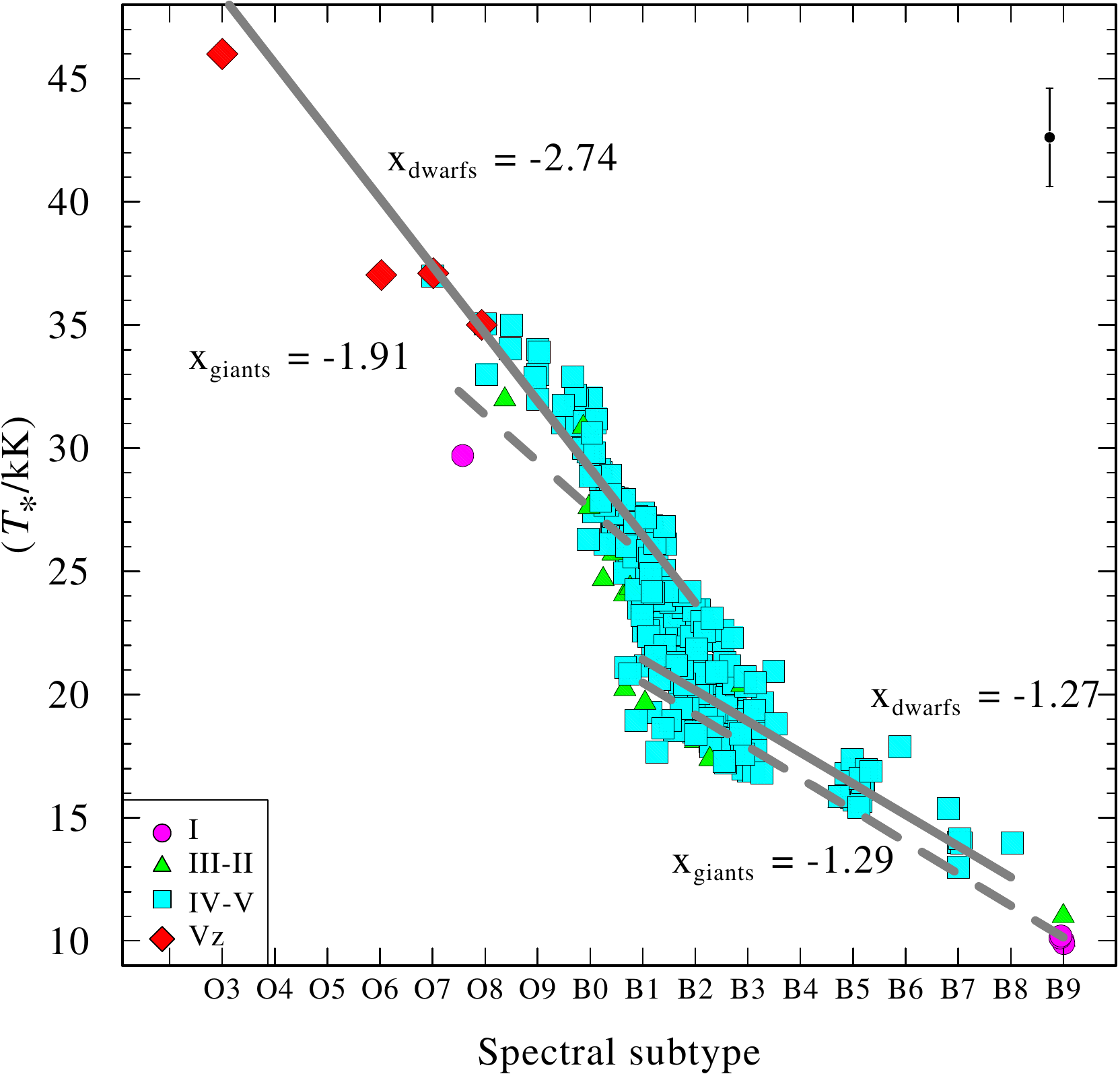}
\caption{Effective temperature vs. spectral type. The rhombi, squares, 
triangles, and, circles refer to luminosity classes Vz, V-IV, III-II, and, I, 
respectively. Typical uncertainties are indicated by the error bar in the top right corner.  Gray thick lines represent linear regressions for different subgroups of stars (see text). }
\label{fig:specT}
\end{figure}

%---------------------------------------------------------------

Figure\,\ref{fig:specT} shows how the effective temperature correlates with the
spectral subtypes. Different luminosity classes are denoted using different colors and symbols as given in the legends. As expected, the evolved stars (III-II and I) systematically possess lower stellar temperatures than dwarfs (Vz and V-IV) of corresponding spectral types. Moreover, we cannot define a single slope for temperature-spectral type relationship  from O3 to B9, rather it changes around  B1-B2. For stars with spectral type earlier than B1, the  relationship is much steeper, and stars with later spectral types than B2 show a shallower relation. For stars with spectral type B1-B2, the temperature determination is more uncertain. The linear regression  (see Fig.\,\ref{fig:specT}) yields the following relations:

For O3-B1 stars,

% \begin{equation}
% T _\ast=
% \begin{cases}
% 56600-2740 \times ST &\text{I-III} \\
% 46610 - 1910 \times \rm{ST} &\text{IV-Vz}
% \end{cases}
% \end{equation}
 \begin{equation}
     T _\ast [\mathrm{ kK}] =\left\{
                \begin{array}{ll}
                56.60 -2.74 \times \rm{ST} \hspace{1.5cm} (IV-Vz)\\
                46.61 - 1.91 \times \rm{ST}   \hspace{1.5cm} (I-III)\\
                \end{array}
              \right.
              .
  \end{equation}
  
For B2-B9 stars,

\begin{equation}
     T _\ast [\mathrm{ kK}] =\left\{
                \begin{array}{ll}
     35.40-1.27 \times \rm{ST}\hspace{1.5cm} (IV-Vz)\\
     34.67-1.29 \times \rm{ST} \hspace{1.5cm} (I-III)\\
     \end{array}
              \right.
              .
\end{equation}

In this case  the spectral subtype is represented by an integer; for
example, ST = 3 for an O3 star and ST = 13 for a B3 star. 
  Effective temperature scales of B-type stars in the SMC  presented by \citet{Trundle2007} agrees with our calibration within uncertainty limits of $\pm 2$\,kK.

%%---------------------------------------------------------------
\begin{figure}
\includegraphics[scale=0.57]{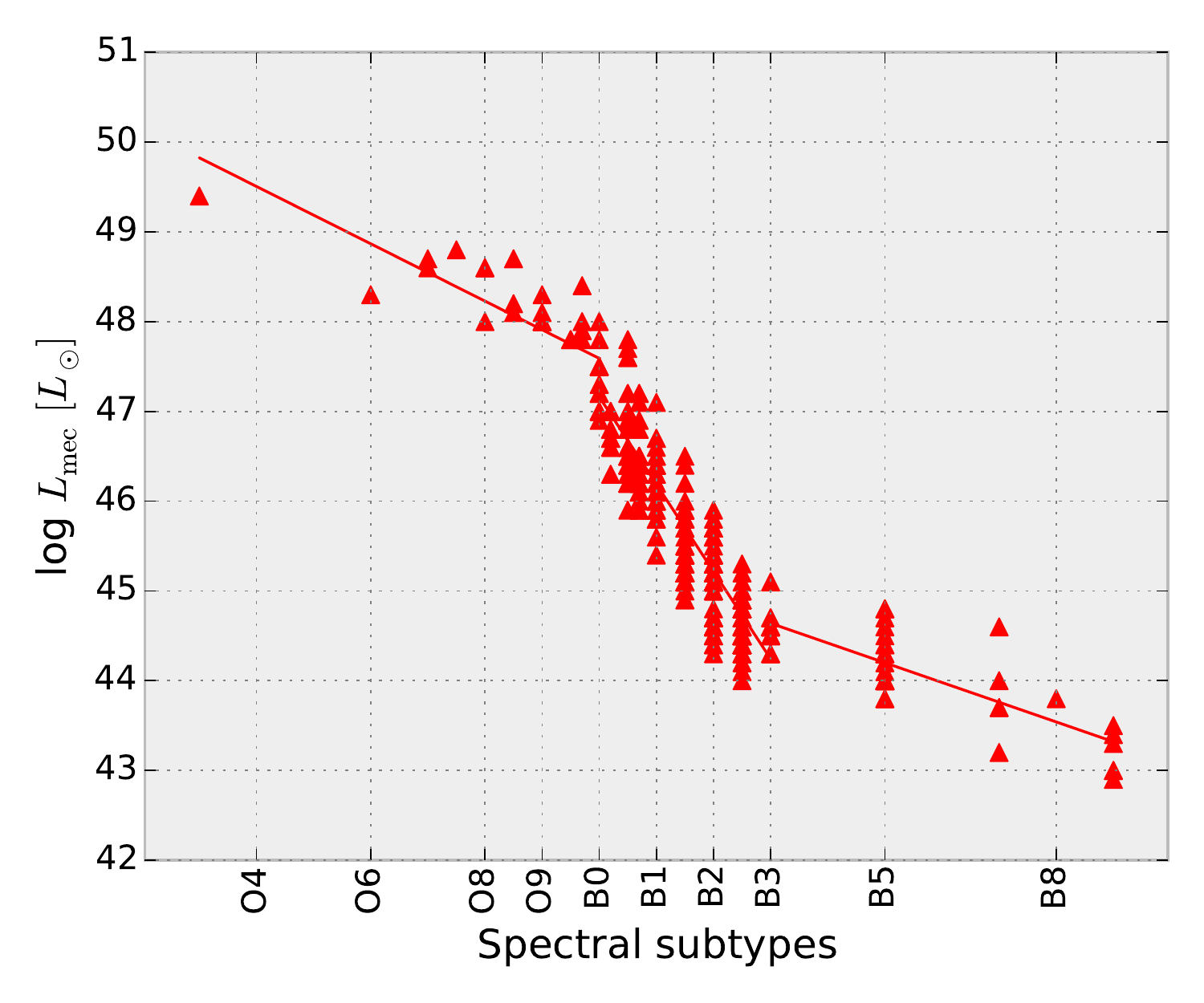}
\includegraphics[scale=0.57]{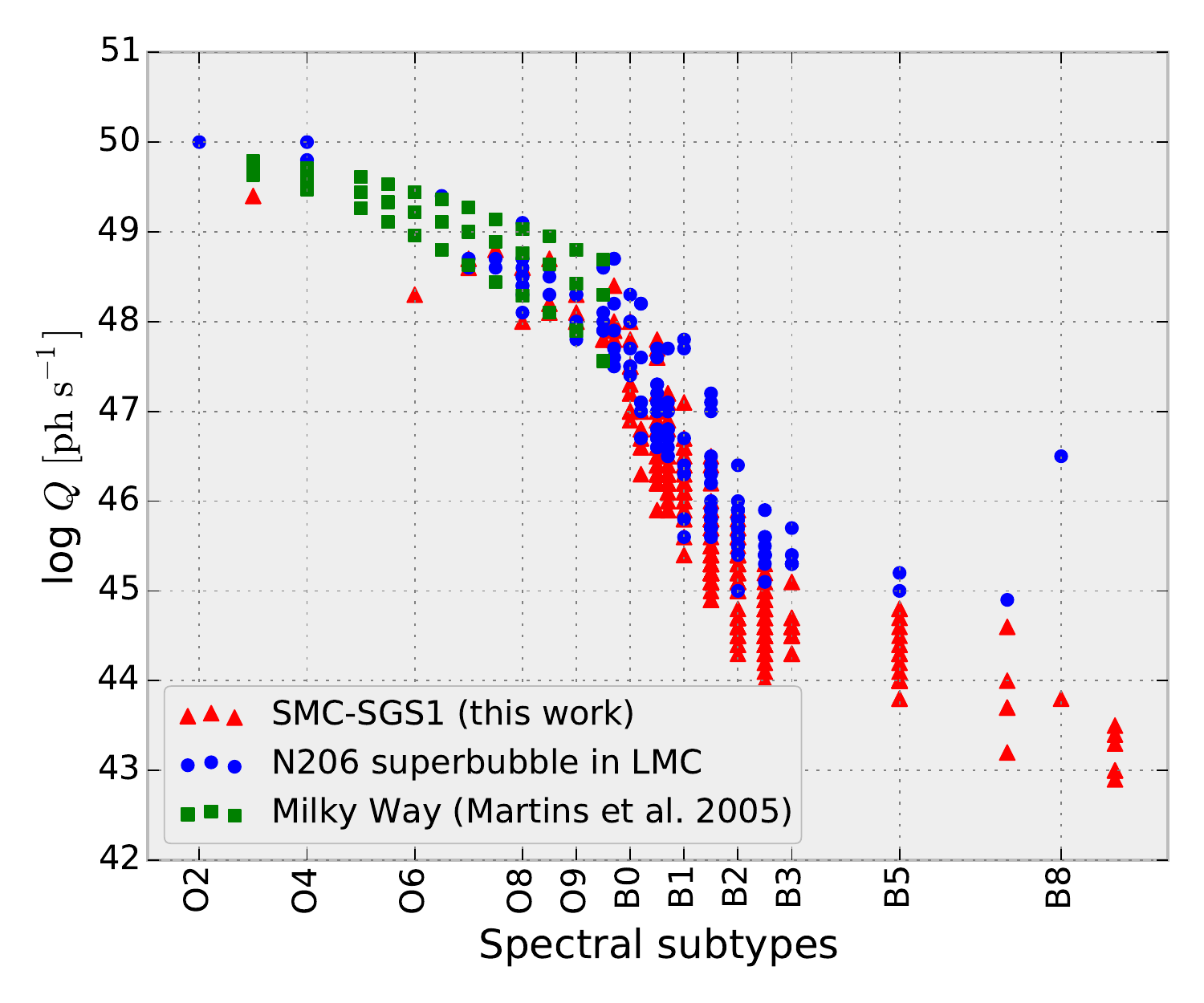}
\caption{Ionizing photon flux as a function of spectral subtypes (top panel) and the comparison with OB stars at LMC  \citep{Ramachandran2018b} and theoretically predicted values for O stars from \citet{Martins2005}.}
\label{fig:qlmec}
\end{figure}

%--------------------------------------------------------------- 

The PoWR models predict the rate of hydrogen ionizing photons ($Q$) for each individual OB star in our sample (see Table.\,\ref{table:App_stellarparameters}). We give   calibrations of $\log\,Q$ with spectral types in Fig.\,\ref{fig:qlmec} (top panel). Even though $\log\,Q$ decreases continuously to lower spectral subtypes, we can see breaks near B0 and B3. For O stars and late-type B stars ($\sim$ B3-B9), the slope of the relationship is much shallower than that for early-type B stars (B0-B3).  An approximate fit  to our data  (see top panel  in Fig.\,\ref{fig:qlmec}) is 

\begin{equation}
     \log\,Q \,[\rm{s}^{-1}]=\left\{
                \begin{array}{ll}           
                50.8 - 0.32\times \rm{ST}   \hspace{1.5cm} (O\,stars)\\
                56.9-0.97 \times \rm{ST} \hspace{1.5cm} (B0-B3)\\
                47.5 - 0.22\times \rm{ST}   \hspace{1.5cm} (B3-B9)\\
                \end{array}
              \right.
              .
\end{equation}

This $\log\,Q$ versus spectral subtype relation shows a similar trend as we found for the N\,206 complex in the LMC  \citep{Ramachandran2018b}. To check the effect of metallicity on the ionizing photon flux, we plotted the same diagram (see Fig.\,\ref{fig:qlmec}, bottom panel) but include values from  \cite{Ramachandran2018b} (LMC stars) and \citet{Martins2005} (Galactic stars). OB stars from the LMC sample (blue circles) also show similar breaks in the diagram.  For comparison, theoretically calculated $\log\,Q$ for O stars from \citet{Martins2005} (green squares) are also shown in the plot. Both LMC and Galactic data points match well. However, $\log\,Q$ values of SMC stars for a given spectral subtype are found to be systematically lower compared to the LMC and Galactic samples. Especially, $\log\,Q$ values of the late B stars in LMC and SMC, differ by more than 1\,dex.

%\section{Results and discussions}
%\label{sect:results}

\section{Indications for a dichotomy in massive star evolution independent of rotation}
\label{sect:hrd}

%---------------------------------------------------------------
 \begin{figure*}
\centering
\includegraphics[width=14cm]{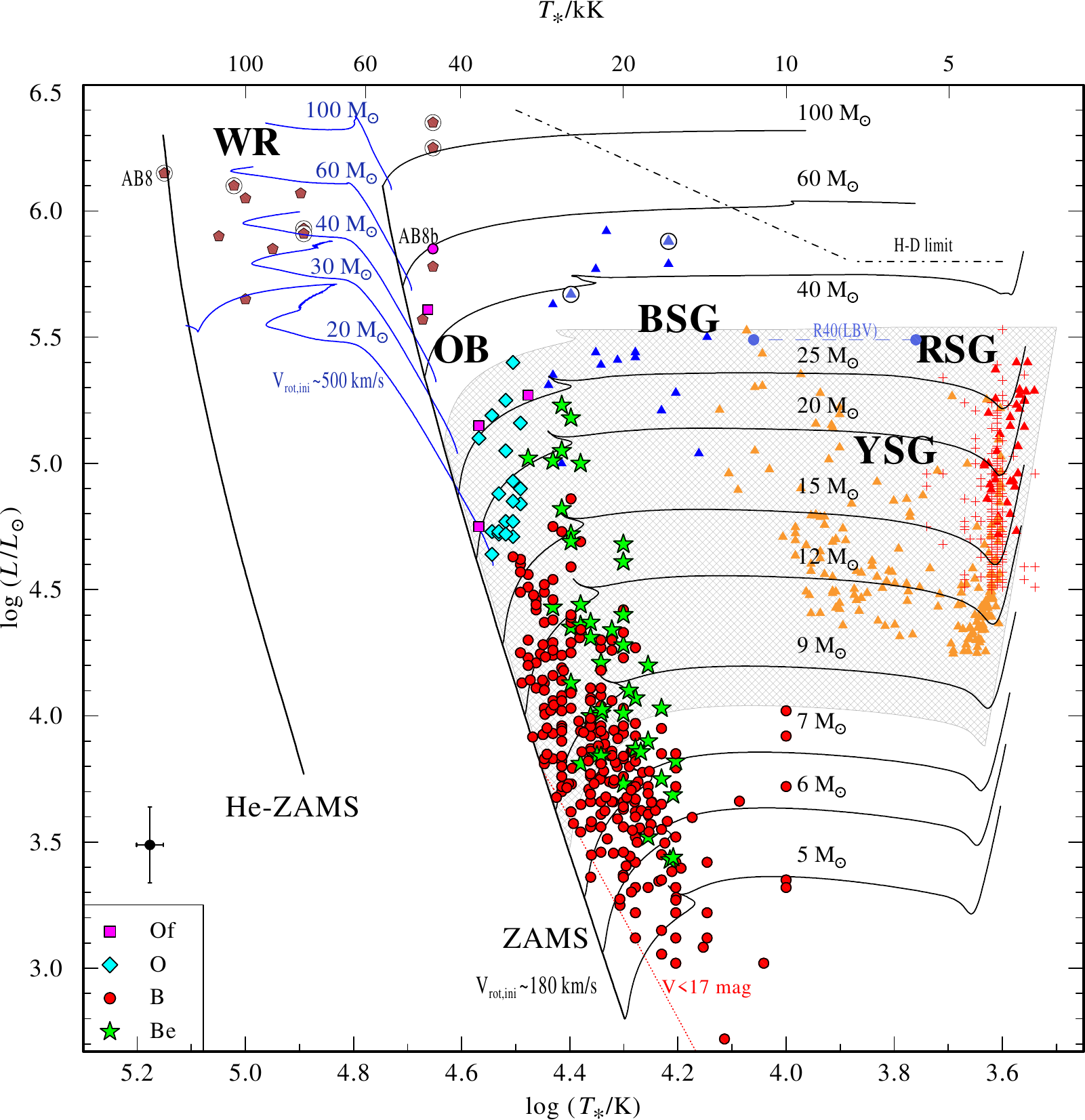} 
\caption{Hertzsprung-Russell diagram for massive stars in the SMC.  The OB stars analyzed in this work  are represented by different symbols as explained in the legend. The typical error bar is shown in the bottom left corner, above the legend. 
The brown pentagons represent WR stars (encircled if in binary systems) \citep{Hainich2015,Shenar2016}, while yellow symbols stand for YSGs \citep{Neugent2010}, blue triangles for BSG \citep{Trundle2004,Trundle2005} red crosses \citep{Davies2018} and red triangles \citep{Levesque2006} for RSGs. The HRD also includes the positions of the only confirmed 
LBV in the SMC, R40,  at different epochs \citep{Kalari2018}. Our OB sample is limited to the SGS in the Wing of the SMC, while as all other samples spread across the whole SMC. Theoretical stellar evolutionary tracks with $\varv_{\rm init}\approx180\,\rm{km\,s^{-1} }$ are shown by solid lines \citep{Brott2011,Kohler2015}. The initial masses are indicated above each track. The black 
tracks show standard evolutionary paths, while the blue tracks with $\varv_{\rm init}\approx550\,\rm{km\,s^{-1} }$show the tracks of quasi-chemically homogeneously evolving (QCHE) stars.
The ZAMS and He-ZAMS are also shown. 
The shaded gray area indicates the region where, according to standard stellar evolution tracks, stars become RSGs and explode as SNe\,II-P. However, the more massive star above the gray shaded area do not seem to follow the standard evolutionary tracks. Instead, their location on the HRD is in agreement with QCHE (blue tracks). These stars remain very massive and may undergo core collapse . The figure demonstrates the dichotomy between the SMC stars with initial 
masses above $\sim 30\,M_{\odot}$ and below, suggesting that the former experience QCHE (blue tracks), while the latter do not.}
\label{fig:hrd}
\end{figure*}
%---------------------------------------------------------------

\subsection{Observed Hertzsprung-Russell diagram}

\changed{Based on the results of our spectral analyses, we construct the 
Hertzsprung-Russell diagram (Fig.\,\ref{fig:hrd}) for our sample of 
320 OB stars in the SMC. The sample comprises 23 O-type stars, among them 
four classified as Of, and 297 B-type stars including 44 stars with
emission lines (type Be).
This gives a well populated upper HRD of hydrogen burning
stars with evolutionary masses  ($M_{\rm ev}$) in the range of
$\sim$5--50\,$M_{\odot}$ (see Table\,\ref{table:App_age}) and covering the parameter space from the ZAMS
to the terminal age main sequence (TAMS).
} 
The most massive star in our sample is an O3 star (Sk\,183), located
close to the ZAMS with $M_{\rm ev} \sim 47\,M_{\odot}$ and
$\log\,(L/L_{\odot}) \sim 5.6$. All other stars have $M_{\rm ev}
\lesssim 30\,M_{\odot}$ ($\log\,(L/L_{\odot}) \lesssim 5.4$). 
\changed{The sample is complete for $V\lesssim17$\,mag, which translates
into a steep limit in the HRD (red dotted line in Fig.\,\ref{fig:hrd})
due to the temperature dependence of the bolometric correction. 
}

To obtain a comprehensive picture of stellar evolution, we incorporate
into the HRD samples of red supergiants (RSG), yellow supergiants (YSG),
blue supergiants (BSG), luminous blue variables (LBV), and WR stars
from previous studies (references are given in the figure caption). 
\changed{While the OB stars of our 
sample all reside in the supergiant shell SGS1 in the Wing of the SMC
(cf.\ Sect.\,\ref{subsect:sgs}), the other samples are from the whole SMC.  
The WR (both single and binary) sample is
complete for the whole SMC.  The shown samples at  RSG, YSG, BSG and LBV
stages are also complete or nearly complete at high luminosities for
this entire galaxy.  In the Appendix (Fig.\,\ref{fig:hrd2})
we present a version of the HRD which additionally includes 
further samples of OB stars from various locations in the body of the
SMC, taken from the literature. 
}

All YSG, LBV and RSG  stars in the SMC were reported with
$\log\,(L/L_{\odot}) \lesssim 5.5$.  The sample of RSGs in the SMC
(complete for $\log\,(L/L_{\odot})> 5.0$)  has  maximum luminosity of
$\log\,(L/L_{\odot}) \sim 5.5$  \citep{Davies2018}. The region in the
HRD above $\log\,(L/L_{\odot}) \sim 5.6$ (or $M_{\rm init} \gtrsim
30\,M_{\odot}$) is populated mostly by stars close to the ZAMS or even
hotter.  Only six luminous BSGs  ($\log\,(L/L_{\odot}) \gtrsim 5.5$)
are seen to the right of ZAMS. These  supergiants  belong to the
brightest blue objects in the SMC, and therefore have $\sim$100\%
detection probability.  Two of  these supergiants are reported
\citep{Gvaramadze2011_run} as runaways (encircled blue triangles in
Fig.\,\ref{fig:hrd})  and, therefore, are  most likely products of 
binary evolution. Another three BSGs show radial velocities that differ
by more than $\pm 50\,\rm{km\,s^{-1} }$ from the systemic velocity of
the SMC.  Such large runaway velocity  likely also indicates a binary
past.  Clearly, further studies on the nature of these BSGs are
required. 
  
\changed{The empirical HRD (Fig.\,\ref{fig:hrd}) shows a remarkable
pattern: while a few very luminous O stars are found close to the main
sequence, all other OB stars as well as the BSGs, YSGs, and RSGs  obey 
strict luminosity limits ($\log L/L_\odot < 6.0$ for BSGs, and $< 5.5$
for the cooler classes). The comparison with single-star evolutionary
tracks suggest a dichotomy in the fate of massive stars in the SMC.
Only stars with an initial mass below $\sim 30\,M_\odot$ seem to evolve
from the main sequence to the cool side of the HRD to become a red
supergiant. In contrast, stars with initially more than $\sim
30\,M_\odot$ appear to stay always hot. 
}

\changed{
In comparison the tracks for single-star evolution, this would imply 
that only stars with initially $\lesssim30\,M_\odot$ evolve according to
the standard tracks to become red supergiants. For stars with
$\gtrsim30\,M_\odot$ such evolution towards the cool side of the HRD is
apparently inhibited. 
}

\changed{
A similar void region in the Galactic HRD is confined by the
Humphreys-Davidson (HD) limit, as was  empirically established by
\cite{HD1979} long ago. A couple of stars that are found in the Galaxy,
LMC, M31, and M33 which are close to or even slightly beyond the HD
limit are considered as unstable stars (LBVs)
\citep[e.g.][]{groh_fundamental_2013, martins_comparison_2013,
urbaneja_lmc_2017, kourniotis_evolutionary_2018,
humphreys_hrd_2017,humphreys_social_2016}. Our empirical HRD suggests
that for the SMC the void region is even more extended than for
galaxies with higher metallicity. 
}

\changed{Why the SMC stars with initially $\gtrsim30\,M_\odot$ stay at
the blue side of the HRD? A tentative explanation is offered by the
WR stars. The single WR stars, analyzed by \cite{Hainich2015}, all
reside between the ZAMS for hydrogen-rich stars and the theoretical
He-ZAMS for hypothetical pure-helium stars. Such HRD position can be
reached if a star evolves quasi chemically homogeneously, i.e.\
with efficient internal mixing, as demonstrated by the corresponding
tracks in Fig.\,\ref{fig:hrd}. 
}

\changed{
All single WR stars   in the SMC are hydrogen depleted, but not free of
hydrogen. In contrast, WN stars (i.e.\ WR stars of the nitrogen
sequence) in the Galaxy or LMC either reside at the cool side of the
ZAMS if showing some hydrogen, or on the hot side of the ZAMS if
hydrogen-free.  Remarkably, all WR stars in the SMC are very luminous
($\log L/L_\odot \gtrsim 5.6$).  There are no  WR stars  with lower
luminosities, in contrast  to the LMC and the Galaxy
\citep{Hainich2014,Hamann2006}. Considered as a whole, in the SMC the 
population of single WR stars  is  strikingly different from higher
metallicity environments. 
}

\changed{
Hence, the large void in the upper HRD, i.e.\ the  absence of
any cool stars at high luminosity, together with   the presence of
luminous, chemically homogeneous WR stars at such high luminosities, 
leads us to the working hypothesis that, at SMC metallicity, stellar
evolution is governed by internal mixing for initial mass
$\gtrsim30\,M_\odot$.   
}

\subsection{Predictions from stellar evolution models }

\begin{figure}
\centering
\includegraphics[width=9cm]{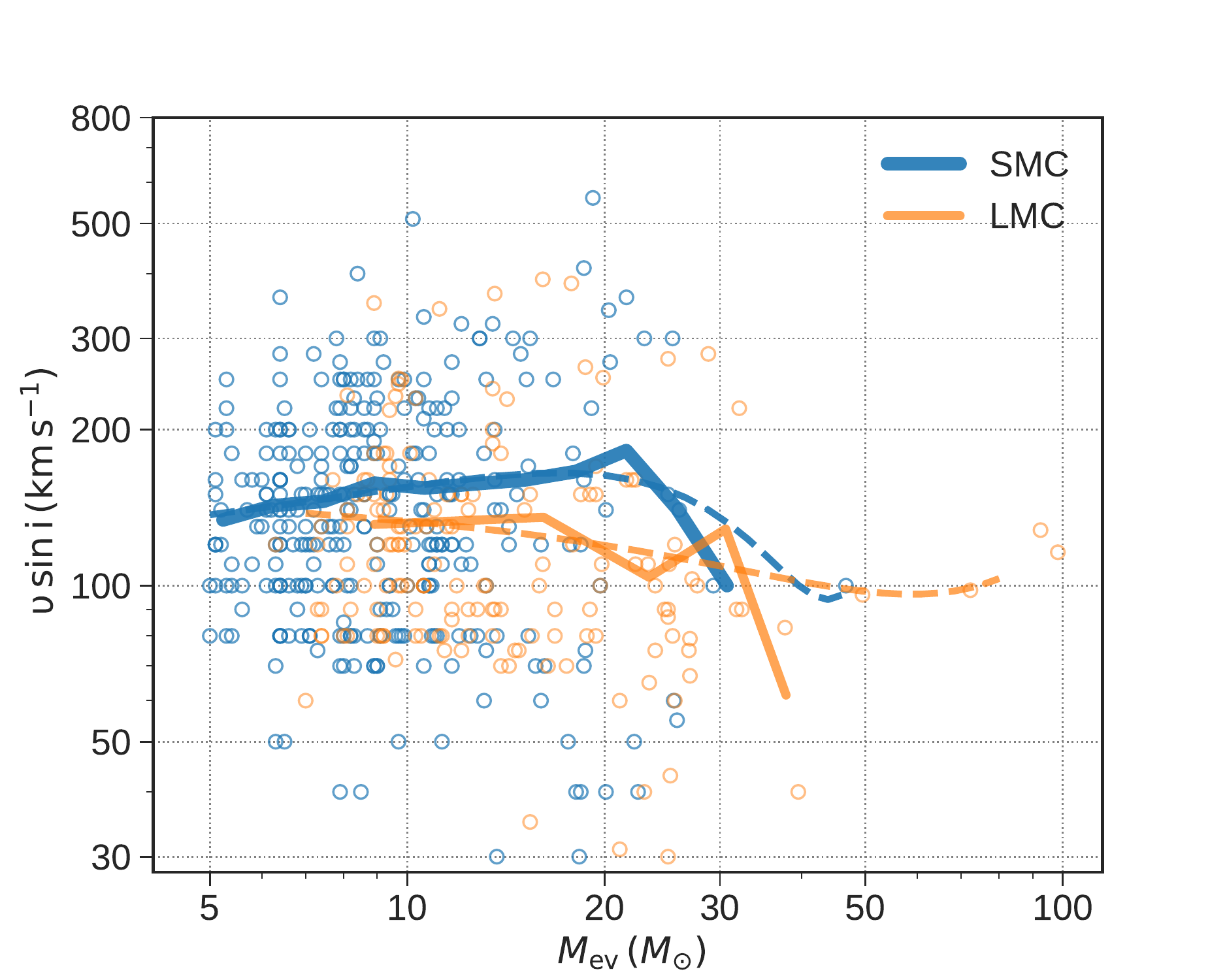}
\caption{Projected rotational velocities of OB stars as functions of stellar mass (blue circles) on logarithmic scales. The LMC OB star sample  \citep{Ramachandran2018b} is also shown for comparison. Solid lines represent mean velocities averaged over  logarithmic mass bins, while dashed lines give nonlinear fits  to the data.}
\label{fig:vrot}
\end{figure}

For comparison with the predictions of stellar evolution models, we consider recent stellar evolutionary tracks  \citep{Brott2011,Kohler2015}. The tracks  with an initial rotational velocity of  $\varv_{\rm init}\approx180\,\rm{km\,s^{-1} }$ which is consistent with the current average rotational velocity ($\approx 150\,\rm{km\,s^{-1} }$) of our sample (see Fig.\,\ref{fig:hrd}, \ref{fig:vrot})  show significant deviations from our empirical HRD for stars with initial masses $M_{\rm init}\gtrsim 30\,M_{\odot}$. 
The lack of luminous RSGs is supported by  detections of \mbox{SNe\,II-P}  progenitors that are consistent with an upper mass limit of $M_{\rm init}\lesssim25\,M_{\odot}$   \citep{Davies2018sn} for RSGs, even at low metallicity ($Z\lesssim0.1\,Z_\odot$) \citep{Anderson2018}.

While the standard evolutionary tracks cannot explain the location of putatively single  WR stars in the HRD \citep{Hainich2015,Martins2009}, these are better reproduced by QCHE tracks \citep[see however, discussions in][]{Schootemeijer2018}.  Even if a star is not fully mixed, a thin envelope could be removed by the  stellar wind. 
%\changed{Among them, only for one system, AB\,8 (WO+O4), QCHE might have played a
%role \citep{Shenar2016}. For all other systems, mass exchange must be invoked to explain
%the equal ages of the components. }
%It is also possible that the primary would initially have a thin hydrogen envelope on top  of the large stellar core. Such an envelope would start
%expanding while the star evolves towards the  terminal main sequence.
%In very close binaries, even a modest  expansion would lead to mass
%transfer. Having lost its envelope this way, the  primary continues to
%evolve quasi homogeneously.

%  However, idealized stars evolving strictly quasi-chemically homogeneously do not 
% expand, therefore no mass could be transferred in binaries.  It seems, 
% that the more realistic scenario is when the stars may have initially 
% large cores and relatively thin outer envelopes. When a thin outer layer is present, even stars more massive 
% than $\sim 30\,M_\odot$ start expanding while evolving towards the 
% terminal main sequence. However, in relatively close binaries, even modest 
% expansion results in mass transfer. Upon loosing its outer envelope, the 
% primary continues to evolve quasi-chemically homogeneous. 

We ask ourselves what could be the possible reasons for this apparent empiric dichotomy in stellar evolution.  Among the main factors that strongly affect massive star evolution, we focus on rotation, binarity, mass-loss rate (see Sect.\,\ref{sect:weakwind}) , and star formation history (see Sect.\,\ref{subsect:sgs}). Rapid rotation leads to efficient mixing of the stellar interior, leading to QCHE \citep{Maeder2000ARA} from the hydrogen ZAMS toward higher temperatures approaching the helium ZAMS toward the end of hydrogen burning \citep{Yoon2005}. Quasi-chemically homogeneous evolution is expected to be more common among massive stars at low $Z$ \citep{Brott2011}. The minimum rotation rate required for QCHE decreases with increasing mass \citep{Yoon2006}. 
At SMC metallicity, the theoretical minimum mass necessary for a star to experience QCHE is about $17 M_{\odot}$ with an initial $\varv_{\rm rot} \approx 550\,\rm{km\,s^{-1} }$ or higher (blue tracks in Fig.\,\ref{fig:hrd}). It should	be	noted	that	this	minimum	initial velocity	is	very	model	dependent,	in	particular,	it	depends	on	how	rotation	is	implemented	in	the	stellar	models \citep{Song2016}.

%%---------------------------------------------------------------

%---------------------------------------------------------------

\subsection{Impact of stellar rotation}

The distribution of stellar rotation rates of our sample stars  show a bimodal distribution (see Sect.\,\ref{sec:vsini}).  Approximately 30\%  of stars in the secondary peak are Be stars, which rotate close to their critical velocity. It is likely that some of the fast rotating stars, including members of the Be class, experienced binary interaction during their evolution \citep{deMink2013}.  According to single-star evolutionary tracks, their evolutionary masses range from  6 to 25\,$M_{\odot}$ and they reside close to the TAMS (indicated by the hooks in the evolutionary tracks in Figure\,\ref{fig:hrd}). These Be stars are also likely evolving toward a YSG/RSG phase.

Figure\,\ref{fig:vrot} shows $\varv\sin i$ for our sample of stars versus their evolutionary mass.  
 First, we consider stars with masses above $30\,M_\odot$. The $\varv\sin i$ values for these stars decline sharply with mass and become relatively low.  Such slow rotators are not expected  to experience QCHE. Yet,  we do see these stars close to the QCHE tracks in the HRD, suggesting that the mixing is not solely associated with rotation.
 Second, we consider stars with lower masses, $M_{\rm init}\sim 17-25\,M_\odot$.  Among these stars are a group of fast rotating OB stars with $\varv\sin i$ in the range 300--550\,km\,s$^{-1}$.  Despite their fast rotation, these stars do not follow  QCHE tracks, but are located on the cool side of the ZAMS. Therefore, we conclude that  SMC stars with $M_{\rm init}\lesssim 30\,M_\odot$ follow standard evolution, independent of their current rotation rates.
 	
The evolution of a star depends on  its initial rotation rather than  the current rotation. Unfortunately, it is hard to get a good handle on the	initial	rotation velocity of a star.  It  is different from the measured	surface rotation  because stellar mass loss (through line driven winds, mechanical mass loss by fast rotation, or induced by close binary evolution) can slow down the star throughout its evolution.	We can state that our sample stars are  still young and that  their present-day mass-loss rates are very small (see Sect.\,\ref{sect:weakwind}). Hence, in case their present mass-loss rate would be representative for their  whole previous lifetime, we would not expect a significant loss of angular momentum.  We conclude that  the evolutionary dichotomy  observed in the HRD is  independent of rotation.

 Significantly, the upper limits on the rotational velocities of WR stars in the SMC are lower than that predicted for QCHE \citep{Hainich2015}. We may speculate that these WN stars had higher rotational velocities while on the ZAMS, but had slowed down since then.  Our measurements of massive O stars do not support this (Fig.\,\ref{fig:vrot}); a similar trend is also observed at LMC metallicity \citep{Ramachandran2018b}. Therefore, rapid rotation alone cannot be the crucial factor responsible for the observed dichotomy. Our results confirm the spectropolarimetric study of WR stars in Magellanic clouds, where no evidence for rotationally induced QCHE was found  \citep{Vink2017}.

 \subsection{Impact of a companion}
 
Stellar evolution is also affected by the presence of a companion. So, the mass transfer in close binaries could also be another channel for the formation of WR stars. However, this channel cannot explain the apparently single massive WR stars in the SMC, which are all very massive and contain hydrogen, exactly as predicted by QCHE models. In case of WR stars in binary systems the situation is different.

While the spectra of our sample OB stars are satisfactorily fit by single-star models and do not clearly indicate binarity, it is likely that many of our targets are in fact binary systems \citep{Sana2012}. According to the Optical Gravitational Lensing Experiment (OGLE) survey \citep{Pawlak2016}, about 7\% of our sample are eclipsing binaries, setting a firm lower limit on the binary fraction. A significant binary fraction may indeed affect our results. Our derived luminosities would be overestimated (e.g., by 0.3\,dex for a binary with two identical stars)  and this may result  if  the continuum of a companion  dilutes the emission line spectrum. Lower luminosities would  imply lower masses, and the uncertainty in $T_\ast$ is irrelevant in the context of the observed dichotomy. We conclude that our main results are not affected by uncertain binary statistics.  However, close binary evolution  can  produce stars that occupy the same positions on the HRD as single stars, for example, via merging. In this case, single-star evolutionary models can mislead the interpretation. At present, the physics of merging is not well understood, but it is important to keep in mind that, in principle, merger products could  contaminate  our sample,  for example, as extraluminous BSGs \citep{Menon2017,Farrell2019}.

\medskip

Summarizing, our empirical HRD of the SMC and comparisons with 
evolutionary tracks indicate two different paths of massive star evolution 
at low metallicity.  Stars with initial mass below $30\,M_\odot $ evolve 
without full mixing toward a RSG phase, where they are likely to explode 
as SN\,II-P. In contrast, massive single stars ($M\gtrsim 30 M_\odot$)
expand only a little during their main sequence evolution, and then evolve 
toward WR stages remaining mixed, hot, and compact, in accordance with 
the predictions of the QCHE model. 
The lack of SNe with corresponding progenitors provides support to the idea that stars at the end of their 
QCHE may collapse to black holes directly \citep{Heger2003, Marchant2016, 
Hainich2018}. We suggest that the main factors for the evolutionary 
dichotomy are  low metallicity and the initial mass of the star, rather 
than rotation.

\section{Weak winds of massive stars at low metallicity}
\label{sect:weakwind}

\begin{figure}
\centering
\includegraphics[scale=0.58]{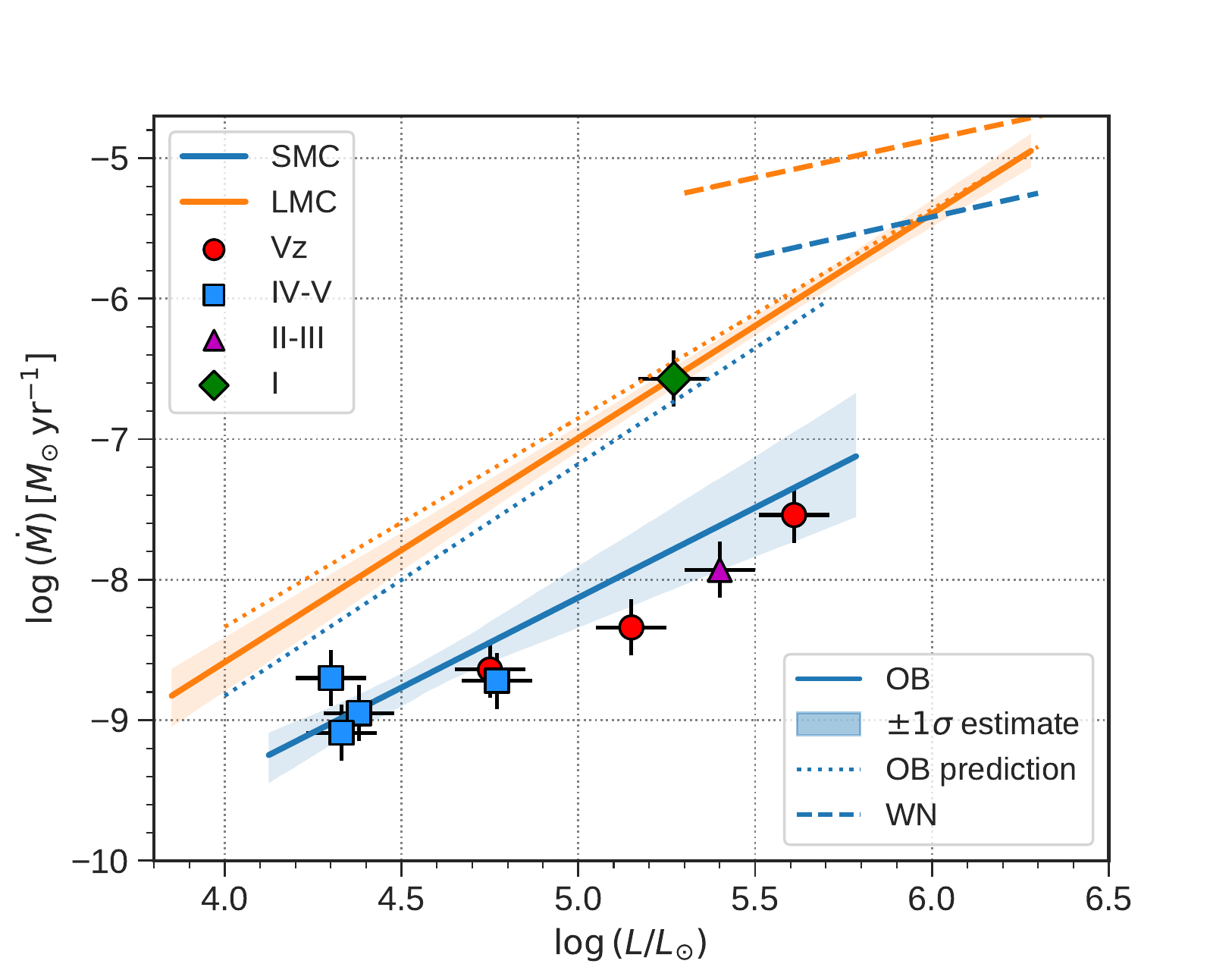}
\caption{Mass-loss rate as a function of stellar luminosity for the analyzed SMC OB stars with available UV data. Luminosity classes are distinguished by different symbols as given in the legend. A power-law fit to the empiric relation is represented by the blue solid line. For comparison, we include the empirical relation for LMC OB stars \citep{Ramachandran2018b} along with theoretical predictions \citep{Vink2001} (dotted lines) at SMC and LMC metallicities.  The empirical relations for WN stars in the SMC and LMC \citep{Hainich2017IAU} are also illustrated in the plot (dashed lines). }
\label{fig:wlr}
\end{figure}

The spectroscopic analysis allows us to quantitatively measure stellar 
mass-loss rates.  Observations in the UV are especially suitable for this. 
For nine stars with available UV spectra (see Sect.\,\ref{sect:spec}), we 
supplemented the optical spectra with the UV observations, and analyzed 
these spectra in a consistent manner using various  diagnostic lines. 
Figure\,\ref{fig:wlr} shows the mass-loss rate ($\dot{M}$) versus 
luminosity of these nine OB stars (B2 to O3). A linear regression to this 
$\log \dot{M} - \log L$ relation, which accounts for the individual error 
bars, shows an order of magnitude systematic offset compared to the 
theoretical predictions \citep{Vink2001} for SMC OB stars. The only 
exception is an Of supergiant (green diamond in Fig.\,\ref{fig:wlr}). 
Excluding this star would result in an even higher offset compared to the 
predictions. Our results confirm previous spectroscopic studies of massive 
stars in the SMC \citep{Bouret2003,Martins2004} and other low-metallicity 
dwarf galaxies like IC 1613 and WLM \citep{Bouret2015,Lucy2012}. This is 
also in agreement with  X-ray observations of NGC602 cluster 
\citep{Oskinova2013}, where they revealed that the emission is mainly 
coming from young low-mass stars, while the winds of the massive stars are 
not sufficient to power the detected extended X-ray emission.

In many OB-type dwarfs, the mass-loss rates derived empirically are much 
lower than predicted by the standard mass-loss recipes. At solar 
metallicity (or even LMC), this problem is limited to the late O- and 
early B-type dwarfs, and is often dubbed as the weak wind problem
\citep{Martins2005,Marcolino2009,Oskinova2011}.  

\changed{Among our subsample of OB stars with UV-based mass-loss rates, eight out
of nine stars exhibit the weak wind phenomenon (cf.\ 
Fig.\,\ref{fig:wlr}), with the exception of the supergiant
SMCSGS-FS\,310. The objects with weak wind comprise not
only dwarfs, but also one star of luminosity class III
(SMCSGS-FS\,288). One of the stars (the most luminous one indicated in
Fig.\,\ref{fig:wlr}) has even the very early subtype O3 (SMCSGS-FS\,231). 
%Previous studies encountered the weak-wind phenomenon for dwarf stars
%with $\log\,(L_\ast/L_{\odot}) \leq 5.2$
%\citep[e.g.]{Martins2005,Marcolino2009}. 
%Our subsample, however, exhibits weak winds thoroughly (except for the supergiant). 
In support of our finding, we mention that \citet{Bouret2013} obtained
a low $\dot{M}$ even for the most luminous O-star in the SMC, MPG\,355. 
Thus, it seems that at SMC metallicity the weak-wind phenomenon is
ubiquitous, and concerns all OB subtypes and all luminosities
except supergiants.  
}

%---------------------------------------------------------------
%---------------------------------------------------------------

\changed{All hot star winds are clumpy \citep{Hamann2007_clump,
Puls2008}. Our model calculations account for such inhomogeneities in
the  {\em microclumping} approximation. We adopt depth dependent
clumping  with a density contrast $D = 10$ (i.e. volume filling factor
$f=0.1$).  Details were given in Sect.\,\ref{sect:models}.  Clumping
enhances emission lines which are fed by recombination cascades
\citep{HK98}. Compensating for this effect, one 
derives a mass-loss rate that is smaller by the factor $\sqrt{D}$ than
with a smooth-wind model. Such clumping dependence would apply if we
would have derived $\dot{M}$ from, e.g., an H$\alpha$ emission
line. 
}

\changed{Fortunately, our empirical mass-loss rates shown in 
Fig.\,\ref{fig:wlr} and discussed above have been derived from
resonance lines in the UV. These wind lines are mainly formed by line
scattering. Since the corresponding opacities scale linearly with
density, these lines do not react on microclumping at all. Hence, the
mass-loss rates we obtained are independent of the adopted clumping
parameters. 
}

Theoretical considerations predict a power-law dependence of the
mass-loss  rate on metallicity, $\dot{M} \propto Z^{\alpha}$, with
$\alpha$=0.69 for  O-type stars \citep{Vink2001}, whereas empirical
studies found $\alpha$=0.83 \citep{Mokiem2007B}. However, our analysis
suggests a much steeper relation  with $\alpha\sim 2$. The WN winds show a
metallicity dependence of $\dot{M} \propto Z^{1.2}$
\citep{Hainich2017IAU}. Since WR  winds are much stronger than the
winds of O stars, it is important to  consider by how much the mass
loss increases while a star evolves from the O to the  WR stage.
Figure\,\ref{fig:wlr} shows empirical and theoretical mass-loss  rates
for O and WR stars in the LMC and SMC. It is striking that the 
difference between mass-loss rates of O and WR stars is significantly 
larger at the lower SMC metallicity compared to the LMC.

\changed{Very weak winds  of  OB stars should  have only  low  impact on their fates. However, it may change our current understanding on stellar evolution, because theoretical evolutionary tracks, as a rule, are calculated using the standard mass-loss recipes, and hence are based on strongly overestimated mass-loss rates. 
 After the first detection of gravitational waves, it has been argued that black
hole pairs as massive as $60\,M_\odot$ can only form if the mass loss is
relatively low \citep{abbott_2016}. 
For instance, according to SMC metallicity tracks by \citet{Eldridge2008}, a star with initially
$60\,M_\odot$ ends with only $22\,M_\odot$ short before core collapse.
This prediction depends on the adopted mass-loss rate prescriptions for
the different evolutionary phases. Additionally, a star might loose a
significant amount of mass by eruptions in the LBV stage
\citep{smith_owocki_2006}, which is not taken into account by
standard tracks.  
However, if a star evolves quasi chemically
homogeneously, the phase of LBV instability can be avoided.
The weak winds in the OB phase reported here are not sufficient to remove the hydrogen envelope, and therefore not lead to a WR phase. This is in agreement with the suggested dichotomy in evolution, where apparently single WR stars are  formed by strong internal mixing.} Very weak OB  winds and QCHE above a certain mass limit suggest that massive black holes can form even at SMC-like metallicity.

 \section{Mode of star formation in the SMC Wing}
 \label{subsect:sgs}

\begin{figure*}[!ht]
 \sidecaption
 \centering
\includegraphics[width=12cm]{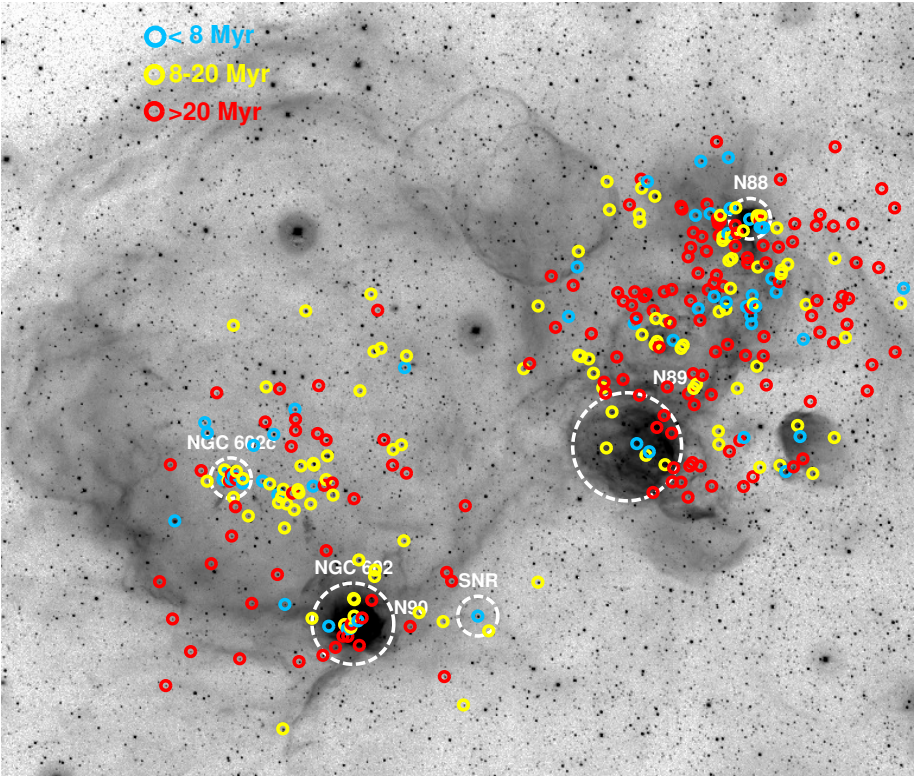}
\caption{Distribution of the OB stars in the SGS, color coded with their age}
\label{fig:agedis}
\end{figure*}

Spectroscopy of massive stars allows us to investigate the star formation modes in the SMC Wing and the SMC-SGS\,1. Using the evolutionary tracks and isochrones \citep{Brott2011}, we estimated the individual ages of all OB stars (see Fig.\,\ref{fig:agedis}, Table\,\ref{table:App_age}). The uncertainties in the age are mostly in the range of 20-40\%, which comes from the uncertainties in temperature, luminosity, and $\varv\sin i$. Five stars in this complex are found to be very young, and are less than 2\,Myr in age, including the O3-type star (Sk\,183) in NGC\,602. From ages and masses of the individual OB stars, and by extrapolation to lower masses ($0.5\,M_\odot$) using the Salpeter initial mass function (IMF), we estimate the present day star formation rate (SFR) of the complex to be $\approx 10^{-3}\,M_\odot\,\mathrm{yr}^{-1}$. With the area of the two observed fields in SMC-SGS\,1 ($\sim$0.2\,kpc$^{2}$) the SFR surface density becomes $\approx 4 \times 10^{-3}\,M_\odot\,\mathrm{yr}^{-1}\,\mathrm{kpc}^{-2}$. 

%%---------------------------------------------------------------

%---------------------------------------------------------------

%%---------------------------------------------------------------
\begin{figure}[!ht]
\centering
\includegraphics[width=8.7cm]{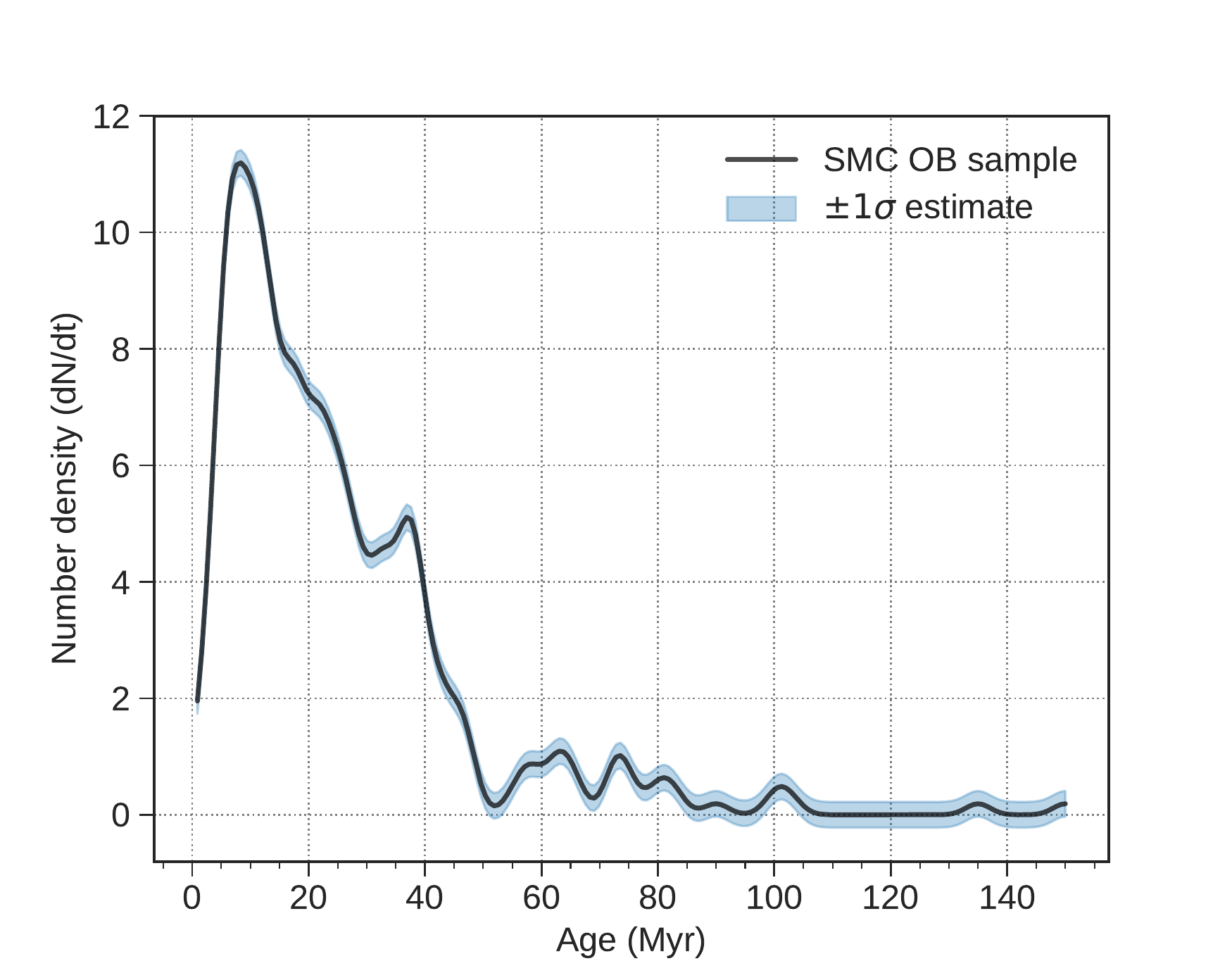}
\caption{Distribution of stellar ages of OB stars in our sample   }
\label{fig:histage}
\end{figure}

%---------------------------------------------------------------

A comparison of the empirical HRD with the evolutionary isochrones gives an age spread from  0 to 100\,Myr for the OB stars in the complex. The age histogram (Fig.\,\ref{fig:histage}) indicates persistent star formation in this large-scale low density region over a long time. The distribution shows active star formation ongoing since 30--40\,Myr, with a peak at about 6--7\,Myr ago. Of course, the present age distribution is biased by the expired lifetimes of the more massive stars. It is tempting to suggest that the feedback from massive stars that formed 30--40\,Myr ago, but already disappeared, have created this SGS, which expands and triggers the formation of present massive stars near its rim. However, as can be seen from Fig.\,\ref{fig:agedis}, there is no obvious age gradient across the SMC-SGS\,1 \citep[see also][]{Fulmer2019}. We found many young massive stars close to the central parts of the complex, especially in the NGC\,602c cluster, and also in the western regions near N\,88. On the other hand, there is a high concentration of the relatively old B-star population in the western nebular regions. Such spatial distribution in the complex nullifies SGS induced star formation at the rim. Instead, it supports a stochastic mode of star formation.  Altogether, we invalidate the central source/cluster scenario for the formation of the SGS. Moreover, we found that the H$\alpha$ emission in the complex does not fully trace the distribution of young massive stars. Only faint, filamentary H$\alpha$ emission is observed in the central part of SGS, which hosts a relatively high number of young massive stars.  

We ask ourselves whether the apparent dichotomy in the empirical HRDs  (Figs.\,\ref{fig:hrd} and \ref{fig:hrd2})  in fact reflects a specific star formation history instead of a bifurcation in stellar evolution.  We believe that this is unlikely because the stars in the empirical HRD are located in different parts of  the SMC;  while our  OB star sample is in the Wing of the SMC,  other stars are distributed throughout the SMC and include runaway and cluster members.  Therefore, there is no reason to suspect that 
all stars with masses above $\sim 30\,M_\odot$ are produced in the 
same coherent star burst episodes.

\section{Feedback from massive stars and the energy budget of the supergiant shell}
\label{sec:feedback}

The total ionizing photon flux produced by all the OB stars in our sample is $\Sigma Q \approx 8 \times 10^{49}\, \rm{s^{-1}}$,  and 30\% are contributed by the O3 star located in NGC\,602 alone. The SGS also encompasses AB\,8, the WO-type binary in the NGC\,602c cluster. By releasing a hydrogen ionizing photon flux of $Q \approx 1.5 \times 10^{50}\, \rm{s^{-1}}$ and \ion{He}{ii} ionizing photon flux of $\approx 5 \times 10^{47}\, \rm{s^{-1}}$, the WO+O4 system dominates the radiative feedback in SMC-SGS\,1 \citep{Shenar2016}. However, the \ion{H}{ii} region associated with NGC\,602c and its WO binary consists only of small, relatively faint emission regions scattered around the cluster. The integrated H$\alpha$ luminosity of the SGS \citep{Kennicutt1995} corresponds to an ionizing photon flux of $Q$ = $ 10^{50}$\,s$^{-1}$. This is only half of the total Lyman continuum flux emitted by the stars we  analyzed.

The winds of SMC OB stars are quite weak (Sect.\,\ref{sect:weakwind}). Therefore it is not surprising that the mechanical feedback by OB stars in this region is significantly lower than in star-forming regions in the LMC \citep{Ramachandran2018b} and the Galaxy. The total mechanical luminosity generated by all OB stars in our sample is estimated to be $ \Sigma L_{\rm mec} =\Sigma\,0.5\,\dot{M}\,\varv_\infty^{2}\sim \,3\times10^{35}\rm{erg\,s^{-1}}$. The accumulated mechanical energy from all sample OB stellar winds throughout their life is $\Sigma E_{\rm mec} =\Sigma\,0.5\,\dot{M}\,\varv_\infty^{2} t \,\approx 6 \times 10^{49}\,\rm{erg} $. We estimate the radial momentum contribution to be $\Sigma p =\dot{M}\,\varv_\infty t \,\approx 5 \times 10^{41}\,\rm{g\,cm\,s^{-1}} $. We emphasize that these feedback estimates only consider current OB stars in the complex and do not account for those massive stars that already disappeared. In addition, our sample does not include massive stars in the northern regions of the SGS.

The WO+O system, AB\,8, alone supplies a current mechanical luminosity of $\approx 7 \times 10^{37}\,\rm{erg\,s^{-1}}$. This is more than two orders of magnitude higher than the mechanical luminosity produced by all 320 OB stars in the sample together. In order to estimate the mechanical energy released by the WO star during its lifetime, we need information about its progenitor. We estimated this with the help of binary evolutionary tracks \citep{Eldridge2009}, by giving the current parameters of the primary and secondary as input. The mechanical energy contribution is $ \approx 1.6 \times 10^{51} \rm{erg} $ in the pre-WR phase and $\approx  4 \times 10^{50} \rm{erg}$ in the WR phase. 
Interestingly, the accumulated energy and momentum from this WO binary are  also two orders of magnitude higher than those generated  from all OB stars together (see Table\,\ref{table:LQ}). Hence, a single WR star (contribution of the secondary is negligible) is responsible for the vast majority of feedback in the complex, exceeding by far from the contribution from the whole OB population (see Table\,\ref{table:LQ}). Existing X-ray observations do not detect diffuse emission from a hot wind blown bubble around AB\,8 or in the SMC-SGS\,1 \citep{Oskinova2013}.  This is possibly  due to the low density in the SGS, or to leakage of hot gas.
%An upper limit on the X-ray luminosity of $5 \times 10^{32}$\,erg\,s$^{-1}$ could be established for its X-ray luminosity \citep{Oskinova2013}. This is $10^{5}$ times lower than the estimated value of the present mechanical luminosity from this star. Even if we assume low X-ray production efficiency $L_{\rm X} /L_{\rm mec}< $ 1\%, we expect a much higher X-ray luminosity than the detection limit.  

To estimate the full energy budget in the SGS, knowledge of SN rate in the complex is required. Two confirmed high mass X-ray binaries (HMXBs) in this complex serve as an evidence for past SNe. Interestingly, these HMXBs, SXP\,1062
 and RX\,J0123.4-7321, are located outside of the SGS. Following our empirical HRD (Sect.\,\ref{sect:hrd}), stars with masses in the range of 8 to $30\,M_{\odot}$ become SNe\,II-P. There are about 190 such stars in our sample as well as additional B stars without spectra in the photometric study \citep{Fulmer2019} which will become SNe\,II-P in the next 40\,Myr. Thus the current rate is $\gtrsim5$ SNe per Myr, assuming a constant SFR. If we only consider stars that are located inside the SGS, the rate becomes $\gtrsim 2.5$ per Myr. This is a lower limit since our spectroscopic survey did not cover the full extent of the SGS.  

 It is difficult to accurately estimate the age of SMC-SGS\,1. We do not detect SN remnants or HMXBs within the shell, and there is a high likelihood of off-center SNe along with the possibility of molecular clouds inside the shell that add extra cooling. It seems that an age of 20-40\,Myr is likely \citep{Fulmer2019}, which is in line with the dynamic age of the shell and the age of the stellar population (Fig.\,\ref{fig:histage}). We adopt an age of 30\,Myr for the feedback estimates, which is also the time-scale of active star formation in the complex. According to our estimated rate, at least 75 SNe must have contributed to the feedback in the SGS. A typical SNe\,II-P releases $10^{51}$\,erg. The accumulated mechanical energy by SNe is then $ E_{\rm{mec}}\approx  7.5 \times 10^{52} \,\rm{erg} $, which is huge compared to stellar wind contribution. However, the energy released by the SNe reduces over time owing to radiative cooling. In low density and low-metallicity environments, the radiative losses are small \citep{Geen2015}. Therefore, in 30\,Myr , the SN energy may decline only up to $ E_{\rm{SN}}\approx  10^{50} \,\rm{erg} $. Considering this effect, the total energy feedback from SNe to the SMC-SGS\,1 becomes $ \approx  10^{52}\,\rm{erg}$.  Supernovae also injects a huge amount of momentum into the ISM and drive turbulence, which cannot be radiated away \citep{Geen2015}. 
 %The typical value of momentum released from a supernova explosion is  $\sim 10^{43} \,\rm{g\, cm\, s^{-1 }}$  \citep{Geen2016,Martizzi2015}. However, the momentum ejection by the SNe also depends on the environment. The total momentum is high at low gas density, low metallicity regions \citep{Geen2015,Kimm2015}. Momentum per SN also increases with the total number of SNe \citep{Gentry2017,Kim2017}. The momentum by SNe at the end of the adiabatic phase significantly exceeds the input momentum at the free expansion phase, so that the total momentum boosted up to an order of magnitude 
 Following \citet{Kimm2014} and \citet{Kimm2015}, the total momentum released by SNe is $p \approx 5 \times 10^{45}\, \rm{g\,cm\,s^{-1}}$. This is again significantly higher than the momentum provided by massive star winds.

 \begin{table}
 
 \caption{Ionizing flux, mechanical luminosity, momentum, and accumulated mechanical energy contributions by  the OB stars,  WO binary, and previous SNe, compared to the values observed in the SGSl. }
 \label{table:LQ}
 \centering
 \setlength{\tabcolsep}{3.5pt}
\begin{tabular}{lllll}
\noalign{\vspace{1mm}}
\hline 
\noalign{\vspace{1mm}}
 & $Q$ & $L_{\rm mec}$  & $p$ & $E_{\rm mec}$ \\
 & [s$^{-1}$] & [erg\,s$^{-1}$] & $[\rm{g\,cm\,s^{-1}}$]& [erg] \\

 \noalign{\vspace{1mm}}
\hline 
\noalign{\vspace{1mm}}

 OB &$ 8 \times 10^{49}$  & $ 3\times10^{35}$ & $5\times 10^{41}$ & $ 6\times 10^{49}$\\ 

 WO & $ 1.5 \times 10^{50}$ & $ 7\times10^{37}$& $ 1.4\times 10^{43}$& $ 2\times 10^{51}$ \\ 
 SN & & &$ 5\times 10^{45}$& $   10^{52}$ \\
 Total &$  2 \times 10^{50}$ & $ 7\times10^{37}$& $ 5\times 10^{45}$& $ 1.2 \times 10^{52}$\\
 Observation &  $\sim 10^{50}$~\tablefootmark{(a)} &$<5 \times 10^{32}$~\tablefootmark{(b)}  &  $ 6\times 10^{45}$~\tablefootmark{(c)} & $ 6 \times 10^{51}$~\tablefootmark{(c)} \\
 
\hline 
\end{tabular} 
\tablefoot{
\tablefoottext{a}{ from H$\alpha$ luminosity}
\tablefoottext{b}{ from X-ray luminosity}
\tablefoottext{c}{from the expansion of the \ion{H}{i} shell }
}
 \end{table}
 
 From neutral hydrogen observations, the surface density of the \ion{H}{i} shell is $\sim 3\,M_{\odot}$\,pc$^{-2}$ \citep{Stanimirovic1999,Fulmer2019}. With a radius of about 300\,pc, the total mass of \ion{H}{i} gas in the SGS is $M_{\rm SGS} \sim 3 \times 10^{6}\,M_{\odot} $. At an expansion velocity of 10\,km\,s$^{-1}$, the total momentum and kinetic energy in the SGS is about $p_{\rm SGS} \approx 6\times 10^{45}\, \rm{g\,cm\,s^{-1}}$ 
 and $E_{\rm SGS} \approx 6 \times 10^{51} \rm{erg}$, i.e, comparable to the contributions from SNe explosions in the past 30\,Myr. The kinetic energy of the expanding SGS is a factor of two smaller compared to the estimated mechanical energy input. This might be because of radiative cooling or leakage of hot gas. These estimates show that the SN feedback plays a crucial role in the formation of the SMC-SGS\,1.

 Table \,\ref{table:LQ} summarizes the estimated energy feedback from OB stars, WR stars, and SNe and
includes a comparison with values derived from observations in the SGS. In our study of massive stars in the LMC superbubble N\,206 \citep{Ramachandran2018,Ramachandran2018b} we found that the contribution of young OB stars is very high in terms of both ionizing photon flux and mechanical luminosity. The contributions from massive stars and SNe were almost equal, while this SMC study reveals a different feedback situation. In SMC-SGS\,1, the WO star dominates all other OB stars in terms of ionization and mechanical feedback. The energy and momentum contribution from SNe over the last 30\,Myr (age of the SGS) is significantly higher than the stellar wind feedback in the complex and this contribution is similar to the observed energy and momentum stored in the SGS. We conclude that at low-metallicity dwarf galaxies, SN explosions seem to be the dominant source for most of the mechanical feedback. The low levels of feedback in metal-poor stellar populations may promote the 
growth and survival of molecular clouds, thereby allowing star formation episodes to persist over long timescales. Such extended star formation can result in a continuous supply of ionizing photons, which can leak out  into the circumgalactic medium via SN created holes or channels.

\section{Summary}
\label{sec:summary}

We present the results of a spectroscopic survey of massive stars in the SMC Wing. The spectra of 320 OB stars were analyzed using modern stellar atmosphere models to derive their fundamental stellar parameters as well as their contributions to feedback through ionizing radiation and stellar winds. We derive calibrations for the temperature and ionizing flux of OB stars as a functions of their spectral subtype.  Complementing the results of our study by previous analyses of evolved massive stars in the WR and RSG evolutionary stages, we obtain a well-populated empiric HRD of massive stars in the SMC. 

Comparison with evolutionary tracks reveals an apparent dichotomy in massive star evolution: stars initially less massive than $30\,M_\odot$ evolve to red supergiants. These stars are likely SNe\,II-P progenitors. 
In contrast, stars (single) initially more massive than $30\,M_\odot$ likely follow a QCHE. These stars might collapse and form black holes directly, without a SN explosion.

We find no evidence that the most-massive SMC stars evolving quasi-chemically  homogeneously are fast rotators.  Also, we do not find  evidence  that the fast rotating stars among the less massive objects evolve  quasi-chemically  homogeneously.   Hence, there is no empirical support for the idea that  QCHE is governed mainly by  rotation.  The bifurcation in evolutionary paths rather seems to  be unrelated to rotation.   We suggest that the main factors for the evolutionary dichotomy are  low metallicity and the initial mass of the star.
 
 The mass-loss rates of OB stars empirically estimated from the spectral analysis are significantly lower than theoretically predicted. This result calls for a revision of stellar evolutionary calculations that routinely rely on overestimated mass-loss rates at low metallicities.

The ages and spatial distribution of massive stars in the SMC Wing shows that star formation is proceeding in this quiescent low density region since more than 100\,Myr, and suggests that massive star formation is stochastic there and not spatially progressing. The weakness of OB-star winds at low metallicity make the SN explosions the dominant sources of mechanical energy and momentum input into the ISM. However, we find that the large population of OB stars produced during the star formation peak in the past $\sim$50\,Myr is fully capable of producing the SMC-SGS\,1 supershell.  Star-forming complexes with extended histories will leave their mark on the surrounding ISM even in low-metallicity systems.

%__________________________________________________________________
\begin{acknowledgements}
Based on observations collected at the European Organization for
Astronomical Research in the Southern Hemisphere
under ESO programme 086.D-0167(A) (P.I.:  L. Oskinova).
We would like to thank the anonymous referee for providing helpful comments. 
V.R. is grateful for financial support from Deutscher Akademischer Austauschdienst (DAAD), as a part of the Graduate School Scholarship Program. J.S.G appreciates support for this study from the University of Wisconsin-Madison College of Arts and Science and through his Rupple Bascom Professorship.  LMO acknowledges support by the DLR grant 50 OR 1508. A.A.C.S. is supported by the Deutsche Forschungsgemeinschaft (DFG) under grant HA 1455/26. TS acknowledges funding from the European Research Council (ERC) under the European Union's 
DLV\_772225\_MULTIPLES Horizon 2020 research and innovation programme. We thank C. J. Evans and V. H{\'e}nault-Brunet for taking the observations and helpful discussions.
This research made use of the VizieR catalog access tool, CDS,  Strasbourg, France. The original description of the VizieR service was published in A\&AS 143, 23. Some data presented in this paper were retrieved from the Mikulski Archive for Space Telescopes (MAST). STScI is operated by the 
Association of Universities for Research in Astronomy, Inc., under NASA contract 
NAS5-26555. Support for MAST for non-HST data is provided by the NASA Office of 
Space Science via grant NNX09AF08G and by other grants and contracts.
\end{acknowledgements}

\bibliographystyle{aa}
\bibliography{paper}

\begin{thebibliography}{136}
\expandafter\ifx\csname natexlab\endcsname\relax\def\natexlab#1{#1}\fi

\bibitem[{Abbott {et~al.}(2016)Abbott, Abbott, Abbott, Abernathy, {LIGO
  Scientific Collaboration}, \& {Virgo Collaboration}}]{abbott_2016}
Abbott, B.~P., Abbott, R., Abbott, T.~D., {et~al.} 2016, The Astrophysical
  Journal Letters, 818, L22

\bibitem[{{Anderson} {et~al.}(2018){Anderson}, {Dessart}, {Guti{\'e}rrez},
  {Kr{\"u}hler}, {Galbany}, {Jerkstrand}, {Smartt}, {Contreras}, {Morrell},
  {Phillips}, {Stritzinger}, {Hsiao}, {Gonz{\'a}lez-Gait{\'a}n}, {Agliozzo},
  {Castell{\'o}n}, {Chambers}, {Chen}, {Flewelling}, {Gonzalez},
  {Hosseinzadeh}, {Huber}, {Fraser}, {Inserra}, {Kankare}, {Mattila},
  {Magnier}, {Maguire}, {Lowe}, {Sollerman}, {Sullivan}, {Young}, \&
  {Valenti}}]{Anderson2018}
{Anderson}, J.~P., {Dessart}, L., {Guti{\'e}rrez}, C.~P., {et~al.} 2018, Nature
  Astronomy, 2, 574

\bibitem[{{Asplund} {et~al.}(2009){Asplund}, {Grevesse}, {Sauval}, \&
  {Scott}}]{Asplund2009}
{Asplund}, M., {Grevesse}, N., {Sauval}, A.~J., \& {Scott}, P. 2009, \araa, 47,
  481

\bibitem[{{Barkana}(2006)}]{Barkana2006Sci}
{Barkana}, R. 2006, Science, 313, 931

\bibitem[{{Blaauw}(1961)}]{Blaauw1961}
{Blaauw}, A. 1961, \bain, 15, 265

\bibitem[{{Bonanos} {et~al.}(2009){Bonanos}, {Massa}, {Sewilo}, {Lennon},
  {Panagia}, {Smith}, {Meixner}, {Babler}, {Bracker}, {Meade}, {Gordon},
  {Hora}, {Indebetouw}, \& {Whitney}}]{Bonanos2009}
{Bonanos}, A.~Z., {Massa}, D.~L., {Sewilo}, M., {et~al.} 2009, \aj, 138, 1003

\bibitem[{{Bouret} {et~al.}(2003){Bouret}, {Lanz}, {Hillier}, {Heap}, {Hubeny},
  {Lennon}, {Smith}, \& {Evans}}]{Bouret2003}
{Bouret}, J.-C., {Lanz}, T., {Hillier}, D.~J., {et~al.} 2003, \apj, 595, 1182

\bibitem[{{Bouret} {et~al.}(2015){Bouret}, {Lanz}, {Hillier}, {Martins},
  {Marcolino}, \& {Depagne}}]{Bouret2015}
{Bouret}, J.-C., {Lanz}, T., {Hillier}, D.~J., {et~al.} 2015, \mnras, 449, 1545

\bibitem[{{Bouret} {et~al.}(2013){Bouret}, {Lanz}, {Martins}, {Marcolino},
  {Hillier}, {Depagne}, \& {Hubeny}}]{Bouret2013}
{Bouret}, J.-C., {Lanz}, T., {Martins}, F., {et~al.} 2013, \aap, 555, A1

\bibitem[{{Brott} {et~al.}(2011){Brott}, {de Mink}, {Cantiello}, {Langer}, {de
  Koter}, {Evans}, {Hunter}, {Trundle}, \& {Vink}}]{Brott2011}
{Brott}, I., {de Mink}, S.~E., {Cantiello}, M., {et~al.} 2011, \aap, 530, A115

\bibitem[{{Castor} {et~al.}(1975){Castor}, {Abbott}, \& {Klein}}]{CAK1975}
{Castor}, J.~I., {Abbott}, D.~C., \& {Klein}, R.~I. 1975, \apj, 195, 157

\bibitem[{{Castro} {et~al.}(2018){Castro}, {Oey}, {Fossati}, \&
  {Langer}}]{Castro2018_smc}
{Castro}, N., {Oey}, M.~S., {Fossati}, L., \& {Langer}, N. 2018, \apj, 868, 57

\bibitem[{{Cignoni} {et~al.}(2009){Cignoni}, {Sabbi}, {Nota}, {Tosi},
  {Degl'Innocenti}, {Moroni}, {Angeretti}, {Carlson}, {Gallagher}, {Meixner},
  {Sirianni}, \& {Smith}}]{Cignoni2009}
{Cignoni}, M., {Sabbi}, E., {Nota}, A., {et~al.} 2009, \aj, 137, 3668

\bibitem[{{Cutri} {et~al.}(2012){Cutri}, {Skrutskie}, {van Dyk}, {Beichman},
  {Carpenter}, {Chester}, {Cambresy}, {Evans}, {Fowler}, {Gizis}, {Howard},
  {Huchra}, {Jarrett}, {Kopan}, {Kirkpatrick}, {Light}, {Marsh}, {McCallon},
  {Schneider}, {Stiening}, {Sykes}, {Weinberg}, {Wheaton}, {Wheelock}, \&
  {Zacharias}}]{Cutri2012}
{Cutri}, R.~M., {Skrutskie}, M.~F., {van Dyk}, S., {et~al.} 2012, VizieR Online
  Data Catalog, 2281, 0

\bibitem[{{Davies} \& {Beasor}(2018)}]{Davies2018sn}
{Davies}, B. \& {Beasor}, E.~R. 2018, \mnras, 474, 2116

\bibitem[{{Davies} {et~al.}(2018){Davies}, {Crowther}, \&
  {Beasor}}]{Davies2018}
{Davies}, B., {Crowther}, P.~A., \& {Beasor}, E.~R. 2018, \mnras, 478, 3138

\bibitem[{{Davies} {et~al.}(1976){Davies}, {Elliott}, \&
  {Meaburn}}]{Davies1976}
{Davies}, R.~D., {Elliott}, K.~H., \& {Meaburn}, J. 1976, \memras, 81, 89

\bibitem[{{de Mink} {et~al.}(2013){de Mink}, {Langer}, {Izzard}, {Sana}, \& {de
  Koter}}]{deMink2013}
{de Mink}, S.~E., {Langer}, N., {Izzard}, R.~G., {Sana}, H., \& {de Koter}, A.
  2013, \apj, 764, 166

\bibitem[{{de Mink} {et~al.}(2014){de Mink}, {Sana}, {Langer}, {Izzard}, \&
  {Schneider}}]{deMink2014}
{de Mink}, S.~E., {Sana}, H., {Langer}, N., {Izzard}, R.~G., \& {Schneider},
  F.~R.~N. 2014, \apj, 782, 7

\bibitem[{{Dufton} {et~al.}(2013){Dufton}, {Langer}, {Dunstall}, {Evans},
  {Brott}, {de Mink}, {Howarth}, {Kennedy}, {McEvoy}, {Potter},
  {Ram{\'{\i}}rez-Agudelo}, {Sana}, {Sim{\'o}n-D{\'{\i}}az}, {Taylor}, \&
  {Vink}}]{Dufton2013}
{Dufton}, P.~L., {Langer}, N., {Dunstall}, P.~R., {et~al.} 2013, \aap, 550,
  A109

\bibitem[{{Dufton} {et~al.}(2006){Dufton}, {Smartt}, {Lee}, {Ryans}, {Hunter},
  {Evans}, {Herrero}, {Trundle}, {Lennon}, {Irwin}, \& {Kaufer}}]{Dufton2006}
{Dufton}, P.~L., {Smartt}, S.~J., {Lee}, J.~K., {et~al.} 2006, \aap, 457, 265

\bibitem[{{Efremov} {et~al.}(1999){Efremov}, {Ehlerov{\'a}}, \& {Palou{\v s}
  }}]{Efremov1999}
{Efremov}, Y.~N., {Ehlerov{\'a}}, S., \& {Palou{\v s} }, J. 1999, \aap, 350,
  457

\bibitem[{{Ekstr{\"o}m} {et~al.}(2008){Ekstr{\"o}m}, {Meynet}, {Maeder}, \&
  {Barblan}}]{Ekstrom2008}
{Ekstr{\"o}m}, S., {Meynet}, G., {Maeder}, A., \& {Barblan}, F. 2008, \aap,
  478, 467

\bibitem[{{Eldridge} {et~al.}(2008){Eldridge}, {Izzard}, \&
  {Tout}}]{Eldridge2008}
{Eldridge}, J.~J., {Izzard}, R.~G., \& {Tout}, C.~A. 2008, \mnras, 384, 1109

\bibitem[{{Eldridge} \& {Stanway}(2009)}]{Eldridge2009}
{Eldridge}, J.~J. \& {Stanway}, E.~R. 2009, \mnras, 400, 1019

\bibitem[{{Elmegreen}(1997)}]{Elmegreen1997}
{Elmegreen}, B.~G. 1997, \apj, 477, 196

\bibitem[{{Evans} {et~al.}(2012){Evans}, {Hainich}, {Oskinova}, {Gallagher},
  {Chu}, {Gruendl}, {Hamann}, {H{\'e}nault-Brunet}, \& {Todt}}]{Evans2012}
{Evans}, C.~J., {Hainich}, R., {Oskinova}, L.~M., {et~al.} 2012, \apj, 753, 173

\bibitem[{{Evans} \& {Howarth}(2008)}]{Evans2008}
{Evans}, C.~J. \& {Howarth}, I.~D. 2008, \mnras, 386, 826

\bibitem[{{Evans} {et~al.}(2015){Evans}, {Kennedy}, {Dufton}, {Howarth},
  {Walborn}, {Markova}, {Clark}, {de Mink}, {de Koter}, {Dunstall},
  {H{\'e}nault-Brunet}, {Ma{\'{\i}}z Apell{\'a}niz}, {McEvoy}, {Sana},
  {Sim{\'o}n-D{\'{\i}}az}, {Taylor}, \& {Vink}}]{Evans2015}
{Evans}, C.~J., {Kennedy}, M.~B., {Dufton}, P.~L., {et~al.} 2015, \aap, 574,
  A13

\bibitem[{{Farrell} {et~al.}(2019){Farrell}, {Groh}, {Meynet}, {Kudritzki},
  {Eldridge}, {Georgy}, {Ekstr{\"o}m}, \& {Yoon}}]{Farrell2019}
{Farrell}, E.~J., {Groh}, J.~H., {Meynet}, G., {et~al.} 2019, \aap, 621, A22

\bibitem[{{Fulmer} {et~al.}(2019 submitted){Fulmer}, {Gallagher}, {Hamann}, ,
  {Oskinova}, \& {Ramachandran}}]{Fulmer2019}
{Fulmer}, L.~M., {Gallagher}, J.~S., {Hamann}, W.~R., {et~al.} 2019 submitted,
  \aap

\bibitem[{{Gallagher} {et~al.}(1984){Gallagher}, {Hunter}, \&
  {Tutukov}}]{Gallagher1984}
{Gallagher}, III, J.~S., {Hunter}, D.~A., \& {Tutukov}, A.~V. 1984, \apj, 284,
  544

\bibitem[{{Geen} {et~al.}(2015){Geen}, {Rosdahl}, {Blaizot}, {Devriendt}, \&
  {Slyz}}]{Geen2015}
{Geen}, S., {Rosdahl}, J., {Blaizot}, J., {Devriendt}, J., \& {Slyz}, A. 2015,
  \mnras, 448, 3248

\bibitem[{{Girard} {et~al.}(2011){Girard}, {van Altena}, {Zacharias}, {Vieira},
  {Casetti-Dinescu}, {Castillo}, {Herrera}, {Lee}, {Beers}, {Monet}, \&
  {L{\'o}pez}}]{Girard2011}
{Girard}, T.~M., {van Altena}, W.~F., {Zacharias}, N., {et~al.} 2011, \aj, 142,
  15

\bibitem[{{Gr{\"a}fener} {et~al.}(2002){Gr{\"a}fener}, {Koesterke}, \&
  {Hamann}}]{Graefener2002}
{Gr{\"a}fener}, G., {Koesterke}, L., \& {Hamann}, W.-R. 2002, \aap, 387, 244

\bibitem[{Groh {et~al.}(2013)Groh, Meynet, Georgy, \&
  Ekström}]{groh_fundamental_2013}
Groh, J.~H., Meynet, G., Georgy, C., \& Ekström, S. 2013, Astronomy \&
  Astrophysics, 558, A131

\bibitem[{{Gvaramadze} {et~al.}(2011){Gvaramadze}, {Pflamm-Altenburg}, \&
  {Kroupa}}]{Gvaramadze2011_run}
{Gvaramadze}, V.~V., {Pflamm-Altenburg}, J., \& {Kroupa}, P. 2011, \aap, 525,
  A17

\bibitem[{{Hainich} {et~al.}(2018){Hainich}, {Oskinova}, {Shenar}, {Marchant},
  {Eldridge}, {Sander}, {Hamann}, {Langer}, \& {Todt}}]{Hainich2018}
{Hainich}, R., {Oskinova}, L.~M., {Shenar}, T., {et~al.} 2018, \aap, 609, A94

\bibitem[{{Hainich} {et~al.}(2015){Hainich}, {Pasemann}, {Todt}, {Shenar},
  {Sander}, \& {Hamann}}]{Hainich2015}
{Hainich}, R., {Pasemann}, D., {Todt}, H., {et~al.} 2015, \aap, 581, A21

\bibitem[{{Hainich} {et~al.}(2014){Hainich}, {R{\"u}hling}, {Todt}, {Oskinova},
  {Liermann}, {Gr{\"a}fener}, {Foellmi}, {Schnurr}, \& {Hamann}}]{Hainich2014}
{Hainich}, R., {R{\"u}hling}, U., {Todt}, H., {et~al.} 2014, \aap, 565, A27

\bibitem[{{Hainich} {et~al.}(2017){Hainich}, {Shenar}, {Sander}, {Hamann}, \&
  {Todt}}]{Hainich2017IAU}
{Hainich}, R., {Shenar}, T., {Sander}, A., {Hamann}, W.-R., \& {Todt}, H. 2017,
  in IAU Symposium, Vol. 329, The Lives and Death-Throes of Massive Stars, ed.
  J.~J. {Eldridge}, J.~C. {Bray}, L.~A.~S. {McClelland}, \& L.~{Xiao}, 171--175

\bibitem[{{Hamann} {et~al.}(2008){Hamann}, {Feldmeier}, \&
  {Oskinova}}]{Hamann2007_clump}
{Hamann}, W.-R., {Feldmeier}, A., \& {Oskinova}, L.~M. 2008

\bibitem[{{Hamann} \& {Gr{\"a}fener}(2004)}]{Hamann2004}
{Hamann}, W.-R. \& {Gr{\"a}fener}, G. 2004, \aap, 427, 697

\bibitem[{{Hamann} {et~al.}(2006){Hamann}, {Gr{\"a}fener}, \&
  {Liermann}}]{Hamann2006}
{Hamann}, W.-R., {Gr{\"a}fener}, G., \& {Liermann}, A. 2006, \aap, 457, 1015

\bibitem[{{Hamann} \& {Koesterke}(1998)}]{HK98}
{Hamann}, W.-R. \& {Koesterke}, L. 1998, \aap, 335, 1003

\bibitem[{{Harris} \& {Zaritsky}(2008)}]{Harris2008}
{Harris}, J. \& {Zaritsky}, D. 2008, Publications of the Astronomical Society
  of Australia, 25, 116

\bibitem[{{Heger} {et~al.}(2003){Heger}, {Fryer}, {Woosley}, {Langer}, \&
  {Hartmann}}]{Heger2003}
{Heger}, A., {Fryer}, C.~L., {Woosley}, S.~E., {Langer}, N., \& {Hartmann},
  D.~H. 2003, \apj, 591, 288

\bibitem[{{Heydari-Malayeri} {et~al.}(1999){Heydari-Malayeri}, {Charmandaris},
  {Deharveng}, {Rosa}, \& {Zinnecker}}]{Heydari-Malayeri1999}
{Heydari-Malayeri}, M., {Charmandaris}, V., {Deharveng}, L., {Rosa}, M.~R., \&
  {Zinnecker}, H. 1999, \aap, 347, 841

\bibitem[{{Humphreys} \& {Davidson}(1979)}]{HD1979}
{Humphreys}, R.~M. \& {Davidson}, K. 1979, \apj, 232, 409

\bibitem[{Humphreys {et~al.}(2017)Humphreys, Davidson, Hahn, Martin, \&
  Weis}]{humphreys_hrd_2017}
Humphreys, R.~M., Davidson, K., Hahn, D., Martin, J.~C., \& Weis, K. 2017, The
  Astrophysical Journal, 844, 40

\bibitem[{Humphreys {et~al.}(2016)Humphreys, Weis, Davidson, \&
  Gordon}]{humphreys_social_2016}
Humphreys, R.~M., Weis, K., Davidson, K., \& Gordon, M.~S. 2016, The
  Astrophysical Journal, 825, 64

\bibitem[{{Hunter} {et~al.}(2008{\natexlab{a}}){Hunter}, {Brott}, {Lennon},
  {Langer}, {Dufton}, {Trundle}, {Smartt}, {de Koter}, {Evans}, \&
  {Ryans}}]{Hunter2008}
{Hunter}, I., {Brott}, I., {Lennon}, D.~J., {et~al.} 2008{\natexlab{a}}, \apjl,
  676, L29

\bibitem[{{Hunter} {et~al.}(2007){Hunter}, {Dufton}, {Smartt}, {Ryans},
  {Evans}, {Lennon}, {Trundle}, {Hubeny}, \& {Lanz}}]{Hunter2007}
{Hunter}, I., {Dufton}, P.~L., {Smartt}, S.~J., {et~al.} 2007, \aap, 466, 277

\bibitem[{{Hunter} {et~al.}(2008{\natexlab{b}}){Hunter}, {Lennon}, {Dufton},
  {Trundle}, {Sim{\'o}n-D{\'{\i}}az}, {Smartt}, {Ryans}, \&
  {Evans}}]{Hunter2008_B}
{Hunter}, I., {Lennon}, D.~J., {Dufton}, P.~L., {et~al.} 2008{\natexlab{b}},
  \aap, 479, 541

\bibitem[{{Kalari} {et~al.}(2018){Kalari}, {Vink, J. S.}, {Dufton, P. L.}, \&
  {Fraser, M.}}]{Kalari2018}
{Kalari}, V.~M., {Vink, J. S.}, {Dufton, P. L.}, \& {Fraser, M.} 2018, A\&A,
  618, A17

\bibitem[{{Kato} {et~al.}(2007){Kato}, {Nagashima}, {Nagayama}, {Kurita},
  {Koerwer}, {Kawai}, {Yamamuro}, {Zenno}, {Nishiyama}, {Baba}, {Kadowaki},
  {Haba}, {Hatano}, {Shimizu}, {Nishimura}, {Nagata}, {Sato}, {Murai},
  {Kawazu}, {Nakajima}, {Nakaya}, {Kandori}, {Kusakabe}, {Ishihara},
  {Kaneyasu}, {Hashimoto}, {Tamura}, {Tanab{\'e}}, {Ita}, {Matsunaga},
  {Nakada}, {Sugitani}, {Wakamatsu}, {Glass}, {Feast}, {Menzies}, {Whitelock},
  {Fourie}, {Stoffels}, {Evans}, \& {Hasegawa}}]{Kato2007}
{Kato}, D., {Nagashima}, C., {Nagayama}, T., {et~al.} 2007, \pasj, 59, 615

\bibitem[{{Kennicutt} {et~al.}(1995){Kennicutt}, {Bresolin}, {Bomans},
  {Bothun}, \& {Thompson}}]{Kennicutt1995}
{Kennicutt}, Jr., R.~C., {Bresolin}, F., {Bomans}, D.~J., {Bothun}, G.~D., \&
  {Thompson}, I.~B. 1995, \aj, 109, 594

\bibitem[{{Kimm} \& {Cen}(2014)}]{Kimm2014}
{Kimm}, T. \& {Cen}, R. 2014, \apj, 788, 121

\bibitem[{{Kimm} {et~al.}(2015){Kimm}, {Cen}, {Devriendt}, {Dubois}, \&
  {Slyz}}]{Kimm2015}
{Kimm}, T., {Cen}, R., {Devriendt}, J., {Dubois}, Y., \& {Slyz}, A. 2015,
  \mnras, 451, 2900

\bibitem[{{K{\"o}hler} {et~al.}(2015){K{\"o}hler}, {Langer}, {de Koter}, {de
  Mink}, {Crowther}, {Evans}, {Gr{\"a}fener}, {Sana}, {Sanyal}, {Schneider}, \&
  {Vink}}]{Kohler2015}
{K{\"o}hler}, K., {Langer}, N., {de Koter}, A., {et~al.} 2015, \aap, 573, A71

\bibitem[{Kourniotis {et~al.}(2018)Kourniotis, Kraus, Arias, Cidale, \&
  Torres}]{kourniotis_evolutionary_2018}
Kourniotis, M., Kraus, M., Arias, M.~L., Cidale, L., \& Torres, A.~F. 2018,
  Monthly Notices of the Royal Astronomical Society, 480, 3706

\bibitem[{{Kudritzki} {et~al.}(1989){Kudritzki}, {Pauldrach}, {Puls}, \&
  {Abbott}}]{Kudritzki1989}
{Kudritzki}, R.~P., {Pauldrach}, A., {Puls}, J., \& {Abbott}, D.~C. 1989, \aap,
  219, 205

\bibitem[{{Kudritzki} \& {Puls}(2000)}]{Kudritzki2000}
{Kudritzki}, R.-P. \& {Puls}, J. 2000, \araa, 38, 613

\bibitem[{{Lamb} {et~al.}(2016){Lamb}, {Oey}, {Segura-Cox}, {Graus}, {Kiminki},
  {Golden-Marx}, \& {Parker}}]{Lamb2016}
{Lamb}, J.~B., {Oey}, M.~S., {Segura-Cox}, D.~M., {et~al.} 2016, \apj, 817, 113

\bibitem[{{Lamers} {et~al.}(1995){Lamers}, {Snow}, \& {Lindholm}}]{Lamers1995}
{Lamers}, H.~J.~G.~L.~M., {Snow}, T.~P., \& {Lindholm}, D.~M. 1995, \apj, 455,
  269

\bibitem[{{Leitherer} {et~al.}(1992){Leitherer}, {Robert}, \&
  {Drissen}}]{Leitherer1992}
{Leitherer}, C., {Robert}, C., \& {Drissen}, L. 1992, \apj, 401, 596

\bibitem[{{Leonard} \& {Duncan}(1990)}]{Leonard1990}
{Leonard}, P.~J.~T. \& {Duncan}, M.~J. 1990, \aj, 99, 608

\bibitem[{{Levesque} {et~al.}(2006){Levesque}, {Massey}, {Olsen}, {Plez},
  {Meynet}, \& {Maeder}}]{Levesque2006}
{Levesque}, E.~M., {Massey}, P., {Olsen}, K.~A.~G., {et~al.} 2006, \apj, 645,
  1102

\bibitem[{{Lucy}(2012)}]{Lucy2012}
{Lucy}, L.~B. 2012, \aap, 543, A18

\bibitem[{{Maeder} \& {Meynet}(2000)}]{Maeder2000ARA}
{Maeder}, A. \& {Meynet}, G. 2000, \araa, 38, 143

\bibitem[{{Marchant} {et~al.}(2016){Marchant}, {Langer}, {Podsiadlowski},
  {Tauris}, \& {Moriya}}]{Marchant2016}
{Marchant}, P., {Langer}, N., {Podsiadlowski}, P., {Tauris}, T.~M., \&
  {Moriya}, T.~J. 2016, \aap, 588, A50

\bibitem[{{Marcolino} {et~al.}(2009){Marcolino}, {Bouret}, {Martins},
  {Hillier}, {Lanz}, \& {Escolano}}]{Marcolino2009}
{Marcolino}, W.~L.~F., {Bouret}, J.-C., {Martins}, F., {et~al.} 2009, \aap,
  498, 837

\bibitem[{{Martins} {et~al.}(2009){Martins}, {Hillier}, {Bouret}, {Depagne},
  {Foellmi}, {Marchenko}, \& {Moffat}}]{Martins2009}
{Martins}, F., {Hillier}, D.~J., {Bouret}, J.~C., {et~al.} 2009, \aap, 495, 257

\bibitem[{Martins \& Palacios(2013)}]{martins_comparison_2013}
Martins, F. \& Palacios, A. 2013, Astronomy \& Astrophysics, 560, A16

\bibitem[{{Martins} {et~al.}(2004){Martins}, {Schaerer}, {Hillier}, \&
  {Heydari-Malayeri}}]{Martins2004}
{Martins}, F., {Schaerer}, D., {Hillier}, D.~J., \& {Heydari-Malayeri}, M.
  2004, \aap, 420, 1087

\bibitem[{{Martins} {et~al.}(2005){Martins}, {Schaerer}, {Hillier},
  {Meynadier}, {Heydari-Malayeri}, \& {Walborn}}]{Martins2005}
{Martins}, F., {Schaerer}, D., {Hillier}, D.~J., {et~al.} 2005, \aap, 441, 735

\bibitem[{{Massey}(2002)}]{Massey2002}
{Massey}, P. 2002, \apjs, 141, 81

\bibitem[{{Massey}(2003)}]{Massey2003}
{Massey}, P. 2003, \araa, 41, 15

\bibitem[{{Matteucci} \& {Chiosi}(1983)}]{Matteucci1983}
{Matteucci}, F. \& {Chiosi}, C. 1983, \aap, 123, 121

\bibitem[{{McCumber} {et~al.}(2005){McCumber}, {Garnett}, \&
  {Dufour}}]{McCumber2005}
{McCumber}, M.~P., {Garnett}, D.~R., \& {Dufour}, R.~J. 2005, \aj, 130, 1083

\bibitem[{{McEvoy} {et~al.}(2015){McEvoy}, {Dufton}, {Evans}, {Kalari},
  {Markova}, {Sim{\'o}n-D{\'{\i}}az}, {Vink}, {Walborn}, {Crowther}, {de
  Koter}, {de Mink}, {Dunstall}, {H{\'e}nault-Brunet}, {Herrero}, {Langer},
  {Lennon}, {Ma{\'{\i}}z Apell{\'a}niz}, {Najarro}, {Puls}, {Sana},
  {Schneider}, \& {Taylor}}]{McEvoy2015}
{McEvoy}, C.~M., {Dufton}, P.~L., {Evans}, C.~J., {et~al.} 2015, \aap, 575, A70

\bibitem[{{McSwain} {et~al.}(2008){McSwain}, {Huang}, {Gies}, {Grundstrom}, \&
  {Townsend}}]{McSwain2008}
{McSwain}, M.~V., {Huang}, W., {Gies}, D.~R., {Grundstrom}, E.~D., \&
  {Townsend}, R.~H.~D. 2008, \apj, 672, 590

\bibitem[{{Meaburn}(1980)}]{Meaburn1980}
{Meaburn}, J. 1980, \mnras, 192, 365

\bibitem[{{Menon} \& {Heger}(2017)}]{Menon2017}
{Menon}, A. \& {Heger}, A. 2017, \mnras, 469, 4649

\bibitem[{{Meynet} \& {Maeder}(2002)}]{Meynet2002}
{Meynet}, G. \& {Maeder}, A. 2002, \aap, 390, 561

\bibitem[{{Mihalas} {et~al.}(1972){Mihalas}, {Hummer}, \&
  {Conti}}]{Walborn1972}
{Mihalas}, D., {Hummer}, D.~G., \& {Conti}, P.~S. 1972, \apjl, 175, L99

\bibitem[{{Mokiem} {et~al.}(2006){Mokiem}, {de Koter}, {Evans}, {Puls},
  {Smartt}, {Crowther}, {Herrero}, {Langer}, {Lennon}, {Najarro}, {Villamariz},
  \& {Yoon}}]{Mokiem2006}
{Mokiem}, M.~R., {de Koter}, A., {Evans}, C.~J., {et~al.} 2006, \aap, 456, 1131

\bibitem[{{Mokiem} {et~al.}(2007){Mokiem}, {de Koter}, {Vink}, {Puls}, {Evans},
  {Smartt}, {Crowther}, {Herrero}, {Langer}, {Lennon}, {Najarro}, \&
  {Villamariz}}]{Mokiem2007B}
{Mokiem}, M.~R., {de Koter}, A., {Vink}, J.~S., {et~al.} 2007, \aap, 473, 603

\bibitem[{{Neugent} {et~al.}(2010){Neugent}, {Massey}, {Skiff}, {Drout},
  {Meynet}, \& {Olsen}}]{Neugent2010}
{Neugent}, K.~F., {Massey}, P., {Skiff}, B., {et~al.} 2010, \apj, 719, 1784

\bibitem[{{Oskinova} {et~al.}(2013){Oskinova}, {Sun}, {Evans},
  {H{\'e}nault-Brunet}, {Chu}, {Gallagher}, {Guerrero}, {Gruendl}, {G{\"u}del},
  {Silich}, {Chen}, {Naz{\'e}}, {Hainich}, \& {Reyes-Iturbide}}]{Oskinova2013}
{Oskinova}, L.~M., {Sun}, W., {Evans}, C.~J., {et~al.} 2013, \apj, 765, 73

\bibitem[{{Oskinova} {et~al.}(2011){Oskinova}, {Todt}, {Ignace}, {Brown},
  {Cassinelli}, \& {Hamann}}]{Oskinova2011}
{Oskinova}, L.~M., {Todt}, H., {Ignace}, R., {et~al.} 2011, \mnras, 416, 1456

\bibitem[{{Pasquini} {et~al.}(2002){Pasquini}, {Avila}, {Blecha}, {Cacciari},
  {Cayatte}, {Colless}, {Damiani}, {de Propris}, {Dekker}, {di Marcantonio},
  {Farrell}, {Gillingham}, {Guinouard}, {Hammer}, {Kaufer}, {Hill}, {Marteaud},
  {Modigliani}, {Mulas}, {North}, {Popovic}, {Rossetti}, {Royer}, {Santin},
  {Schmutzer}, {Simond}, {Vola}, {Waller}, \& {Zoccali}}]{Pasquini2002}
{Pasquini}, L., {Avila}, G., {Blecha}, A., {et~al.} 2002, The Messenger, 110, 1

\bibitem[{{Pawlak} {et~al.}(2016){Pawlak}, {Soszy{\'n}ski}, {Udalski},
  {Szyma{\'n}ski}, {Wyrzykowski}, {Ulaczyk}, {Poleski}, {Pietrukowicz},
  {Koz{\l}owski}, {Skowron}, {Skowron}, {Mr{\'o}z}, \&
  {Hamanowicz}}]{Pawlak2016}
{Pawlak}, M., {Soszy{\'n}ski}, I., {Udalski}, A., {et~al.} 2016, \actaa, 66,
  421

\bibitem[{{Puls} {et~al.}(2008){Puls}, {Vink}, \& {Najarro}}]{Puls2008}
{Puls}, J., {Vink}, J.~S., \& {Najarro}, F. 2008, \aapr, 16, 209

\bibitem[{{Ramachandran} {et~al.}(2018{\natexlab{a}}){Ramachandran}, {Hainich},
  {Hamann}, {Oskinova}, {Shenar}, {Sander}, {Todt}, \&
  {Gallagher}}]{Ramachandran2018}
{Ramachandran}, V., {Hainich}, R., {Hamann}, W.-R., {et~al.}
  2018{\natexlab{a}}, \aap, 609, A7

\bibitem[{{Ramachandran} {et~al.}(2018{\natexlab{b}}){Ramachandran}, {Hamann},
  {Hainich}, {Oskinova}, {Shenar}, {Sander}, {Todt}, \&
  {Gallagher}}]{Ramachandran2018b}
{Ramachandran}, V., {Hamann}, W.~R., {Hainich}, R., {et~al.}
  2018{\natexlab{b}}, \aap, 615, A40

\bibitem[{{Ram{\'{\i}}rez-Agudelo} {et~al.}(2017){Ram{\'{\i}}rez-Agudelo},
  {Sana}, {de Koter}, {Tramper}, {Grin}, {Schneider}, {Langer}, {Puls},
  {Markova}, {Bestenlehner}, {Castro}, {Crowther}, {Evans}, {Garc{\'{\i}}a},
  {Gr{\"a}fener}, {Herrero}, {van Kempen}, {Lennon}, {Ma{\'{\i}}z
  Apell{\'a}niz}, {Najarro}, {Sab{\'{\i}}n-Sanjuli{\'a}n},
  {Sim{\'o}n-D{\'{\i}}az}, {Taylor}, \& {Vink}}]{Ramirez-Agudelo2017}
{Ram{\'{\i}}rez-Agudelo}, O.~H., {Sana}, H., {de Koter}, A., {et~al.} 2017,
  \aap, 600, A81

\bibitem[{{Ram{\'{\i}}rez-Agudelo} {et~al.}(2013){Ram{\'{\i}}rez-Agudelo},
  {Sim{\'o}n-D{\'{\i}}az}, {Sana}, {de Koter}, {Sab{\'{\i}}n-Sanjul{\'{\i}}an},
  {de Mink}, {Dufton}, {Gr{\"a}fener}, {Evans}, {Herrero}, {Langer}, {Lennon},
  {Ma{\'{\i}}z Apell{\'a}niz}, {Markova}, {Najarro}, {Puls}, {Taylor}, \&
  {Vink}}]{Ramirez-Agudelo2013}
{Ram{\'{\i}}rez-Agudelo}, O.~H., {Sim{\'o}n-D{\'{\i}}az}, S., {Sana}, H.,
  {et~al.} 2013, \aap, 560, A29

\bibitem[{{Repolust} {et~al.}(2004){Repolust}, {Puls}, \&
  {Herrero}}]{Repolust2004}
{Repolust}, T., {Puls}, J., \& {Herrero}, A. 2004, \aap, 415, 349

\bibitem[{{Runacres} \& {Owocki}(2002)}]{Runacres2002}
{Runacres}, M.~C. \& {Owocki}, S.~P. 2002, \aap, 381, 1015

\bibitem[{{Sab{\'{\i}}n-Sanjuli{\'a}n}
  {et~al.}(2014){Sab{\'{\i}}n-Sanjuli{\'a}n}, {Sim{\'o}n-D{\'{\i}}az},
  {Herrero}, {Walborn}, {Puls}, {Ma{\'{\i}}z Apell{\'a}niz}, {Evans}, {Brott},
  {de Koter}, {Garcia}, {Markova}, {Najarro}, {Ram{\'{\i}}rez-Agudelo}, {Sana},
  {Taylor}, \& {Vink}}]{Sabin-Sanjulian2014}
{Sab{\'{\i}}n-Sanjuli{\'a}n}, C., {Sim{\'o}n-D{\'{\i}}az}, S., {Herrero}, A.,
  {et~al.} 2014, \aap, 564, A39

\bibitem[{{Sana} {et~al.}(2012){Sana}, {de Mink}, {de Koter}, {Langer},
  {Evans}, {Gieles}, {Gosset}, {Izzard}, {Le Bouquin}, \&
  {Schneider}}]{Sana2012}
{Sana}, H., {de Mink}, S.~E., {de Koter}, A., {et~al.} 2012, Science, 337, 444

\bibitem[{{Sander} {et~al.}(2015){Sander}, {Shenar}, {Hainich},
  {G{\'{\i}}menez-Garc{\'{\i}}a}, {Todt}, \& {Hamann}}]{Sander2015}
{Sander}, A., {Shenar}, T., {Hainich}, R., {et~al.} 2015, \aap, 577, A13

\bibitem[{{Schneider} {et~al.}(2018){Schneider}, {Sana}, {Evans},
  {Bestenlehner}, {Castro}, {Fossati}, {Gr{\"a}fener}, {Langer},
  {Ram{\'{\i}}rez-Agudelo}, {Sab{\'{\i}}n-Sanjuli{\'a}n},
  {Sim{\'o}n-D{\'{\i}}az}, {Tramper}, {Crowther}, {de Koter}, {de Mink},
  {Dufton}, {Garcia}, {Gieles}, {H{\'e}nault-Brunet}, {Herrero}, {Izzard},
  {Kalari}, {Lennon}, {Ma{\'{\i}}z Apell{\'a}niz}, {Markova}, {Najarro},
  {Podsiadlowski}, {Puls}, {Taylor}, {van Loon}, {Vink}, \&
  {Norman}}]{Schneider2018}
{Schneider}, F.~R.~N., {Sana}, H., {Evans}, C.~J., {et~al.} 2018, Science, 359,
  69

\bibitem[{{Schootemeijer} \& {Langer}(2018)}]{Schootemeijer2018}
{Schootemeijer}, A. \& {Langer}, N. 2018, \aap, 611, A75

\bibitem[{{Seiden} {et~al.}(1979){Seiden}, {Schulman}, \&
  {Gerola}}]{Seiden1979}
{Seiden}, P.~E., {Schulman}, L.~S., \& {Gerola}, H. 1979, \apj, 232, 702

\bibitem[{{Shenar} {et~al.}(2016){Shenar}, {Hainich}, {Todt}, {Sander},
  {Hamann}, {Moffat}, {Eldridge}, {Pablo}, {Oskinova}, \&
  {Richardson}}]{Shenar2016}
{Shenar}, T., {Hainich}, R., {Todt}, H., {et~al.} 2016, \aap, 591, A22

\bibitem[{{Sim{\'o}n-D{\'{\i}}az} \& {Herrero}(2014)}]{Simon-diaz2014}
{Sim{\'o}n-D{\'{\i}}az}, S. \& {Herrero}, A. 2014, \aap, 562, A135

\bibitem[{Smith \& Owocki(2006)}]{smith_owocki_2006}
Smith, N. \& Owocki, S.~P. 2006, The Astrophysical Journal Letters, 645, L45

\bibitem[{{Smith} {et~al.}(2005){Smith}, {Points}, {Chu}, {Winkler}, {Leiton},
  \& {MCELS Team}}]{Smith2005}
{Smith}, R.~C., {Points}, S., {Chu}, Y.-H., {et~al.} 2005, in Bulletin of the
  American Astronomical Society, Vol.~37, American Astronomical Society Meeting
  Abstracts, \#145.01

\bibitem[{{Song} {et~al.}(2016){Song}, {Meynet}, {Maeder}, {Ekstr{\"o}m}, \&
  {Eggenberger}}]{Song2016}
{Song}, H.~F., {Meynet}, G., {Maeder}, A., {Ekstr{\"o}m}, S., \& {Eggenberger},
  P. 2016, \aap, 585, A120

\bibitem[{{Sota} {et~al.}(2014){Sota}, {Ma{\'{\i}}z Apell{\'a}niz}, {Morrell},
  {Barb{\'a}}, {Walborn}, {Gamen}, {Arias}, \& {Alfaro}}]{Sota2014}
{Sota}, A., {Ma{\'{\i}}z Apell{\'a}niz}, J., {Morrell}, N.~I., {et~al.} 2014,
  \apjs, 211, 10

\bibitem[{{Sota} {et~al.}(2011){Sota}, {Ma{\'{\i}}z Apell{\'a}niz}, {Walborn},
  {Alfaro}, {Barb{\'a}}, {Morrell}, {Gamen}, \& {Arias}}]{Sota2011}
{Sota}, A., {Ma{\'{\i}}z Apell{\'a}niz}, J., {Walborn}, N.~R., {et~al.} 2011,
  \apjs, 193, 24

\bibitem[{{Stanimirovic} {et~al.}(1999){Stanimirovic}, {Staveley-Smith},
  {Dickey}, {Sault}, \& {Snowden}}]{Stanimirovic1999}
{Stanimirovic}, S., {Staveley-Smith}, L., {Dickey}, J.~M., {Sault}, R.~J., \&
  {Snowden}, S.~L. 1999, \mnras, 302, 417

\bibitem[{{Stanimirovic} {et~al.}(2000){Stanimirovic}, {Staveley-Smith}, {van
  der Hulst}, {Bontekoe}, {Kester}, \& {Jones}}]{Stanimirovic2000}
{Stanimirovic}, S., {Staveley-Smith}, L., {van der Hulst}, J.~M., {et~al.}
  2000, \mnras, 315, 791

\bibitem[{{Staveley-Smith} {et~al.}(1997){Staveley-Smith}, {Sault},
  {Hatzidimitriou}, {Kesteven}, \& {McConnell}}]{Staveley-Smith1997}
{Staveley-Smith}, L., {Sault}, R.~J., {Hatzidimitriou}, D., {Kesteven}, M.~J.,
  \& {McConnell}, D. 1997, \mnras, 289, 225

\bibitem[{{Tenorio-Tagle}(1981)}]{Tenorio-Tagle1981}
{Tenorio-Tagle}, G. 1981, \aap, 94, 338

\bibitem[{{Tenorio-Tagle} {et~al.}(1986){Tenorio-Tagle}, {Bodenheimer},
  {Rozyczka}, \& {Franco}}]{Tenorio-Tagle1986}
{Tenorio-Tagle}, G., {Bodenheimer}, P., {Rozyczka}, M., \& {Franco}, J. 1986,
  \aap, 170, 107

\bibitem[{{Todt} {et~al.}(2015){Todt}, {Sander}, {Hainich}, {Hamann}, {Quade},
  \& {Shenar}}]{Todt2015}
{Todt}, H., {Sander}, A., {Hainich}, R., {et~al.} 2015, \aap, 579, A75

\bibitem[{{Trundle} {et~al.}(2007){Trundle}, {Dufton}, {Hunter}, {Evans},
  {Lennon}, {Smartt}, \& {Ryans}}]{Trundle2007}
{Trundle}, C., {Dufton}, P.~L., {Hunter}, I., {et~al.} 2007, \aap, 471, 625

\bibitem[{{Trundle} \& {Lennon}(2005)}]{Trundle2005}
{Trundle}, C. \& {Lennon}, D.~J. 2005, \aap, 434, 677

\bibitem[{{Trundle} {et~al.}(2004){Trundle}, {Lennon}, {Puls}, \&
  {Dufton}}]{Trundle2004}
{Trundle}, C., {Lennon}, D.~J., {Puls}, J., \& {Dufton}, P.~L. 2004, \aap, 417,
  217

\bibitem[{Urbaneja {et~al.}(2017)Urbaneja, Kudritzki, Gieren, Pietrzyński,
  Bresolin, \& Przybilla}]{urbaneja_lmc_2017}
Urbaneja, M.~A., Kudritzki, R.-P., Gieren, W., {et~al.} 2017, The Astronomical
  Journal, 154, 102

\bibitem[{{Vink} {et~al.}(2001){Vink}, {de Koter}, \& {Lamers}}]{Vink2001}
{Vink}, J.~S., {de Koter}, A., \& {Lamers}, H.~J.~G.~L.~M. 2001, \aap, 369, 574

\bibitem[{{Vink} \& {Harries}(2017)}]{Vink2017}
{Vink}, J.~S. \& {Harries}, T.~J. 2017, \aap, 603, A120

\bibitem[{{Walborn}(1973)}]{Walborn1973}
{Walborn}, N.~R. 1973, \aj, 78, 1067

\bibitem[{Walborn(2009)}]{walborn_2009}
Walborn, N.~R. 2009, Optically observable zero-age main-sequence O stars, ed.
  M.~Livio \& E.~Villaver, Space Telescope Science Institute Symposium Series
  (Cambridge University Press), 167–177

\bibitem[{{Walborn} {et~al.}(2010){Walborn}, {Howarth}, {Evans}, {Crowther},
  {Moffat}, {St-Louis}, {Fari{\~n}a}, {Bosch}, {Morrell}, {Barb{\'a}}, \& {van
  Loon}}]{Walborn2010}
{Walborn}, N.~R., {Howarth}, I.~D., {Evans}, C.~J., {et~al.} 2010, \aj, 139,
  1283

\bibitem[{{Walborn} {et~al.}(2002){Walborn}, {Howarth}, {Lennon}, {Massey},
  {Oey}, {Moffat}, {Skalkowski}, {Morrell}, {Drissen}, \&
  {Parker}}]{Walborn2002}
{Walborn}, N.~R., {Howarth}, I.~D., {Lennon}, D.~J., {et~al.} 2002, \aj, 123,
  2754

\bibitem[{{Walborn} {et~al.}(2014){Walborn}, {Sana}, {Sim{\'o}n-D{\'{\i}}az},
  {Ma{\'{\i}}z Apell{\'a}niz}, {Taylor}, {Evans}, {Markova}, {Lennon}, \& {de
  Koter}}]{Walborn2014}
{Walborn}, N.~R., {Sana}, H., {Sim{\'o}n-D{\'{\i}}az}, S., {et~al.} 2014, \aap,
  564, A40

\bibitem[{{Westerlund}(1964)}]{Westerlund1964}
{Westerlund}, B.~E. 1964, \mnras, 127, 429

\bibitem[{{Winkler} {et~al.}(2005){Winkler}, {Young}, {Braziunas}, {Condon},
  {Galle}, {Reaser}, {Leiton}, {Smith}, \& {MCELS Team}}]{Winkler2005}
{Winkler}, P.~F., {Young}, A.~L., {Braziunas}, D., {et~al.} 2005, in Bulletin
  of the American Astronomical Society, Vol.~37, American Astronomical Society
  Meeting Abstracts, 1380

\bibitem[{{Yoon} \& {Langer}(2005)}]{Yoon2005}
{Yoon}, S.-C. \& {Langer}, N. 2005, \aap, 443, 643

\bibitem[{{Yoon} {et~al.}(2006){Yoon}, {Langer}, \& {Norman}}]{Yoon2006}
{Yoon}, S.-C., {Langer}, N., \& {Norman}, C. 2006, \aap, 460, 199

\bibitem[{{Zacharias} {et~al.}(2012){Zacharias}, {Finch}, {Girard}, {Henden},
  {Bartlett}, {Monet}, \& {Zacharias}}]{Zacharias2012}
{Zacharias}, N., {Finch}, C.~T., {Girard}, T.~M., {et~al.} 2012, VizieR Online
  Data Catalog, 1322

\bibitem[{{Zaritsky} {et~al.}(2004){Zaritsky}, {Harris}, {Thompson}, \&
  {Grebel}}]{Zaritsky2004}
{Zaritsky}, D., {Harris}, J., {Thompson}, I.~B., \& {Grebel}, E.~K. 2004, \aj,
  128, 1606

\end{thebibliography}

\appendix

\section{Additional figures}

%---------------------------------------------------------------
 \begin{figure*}[htpb]
\centering
\includegraphics[width=13cm]{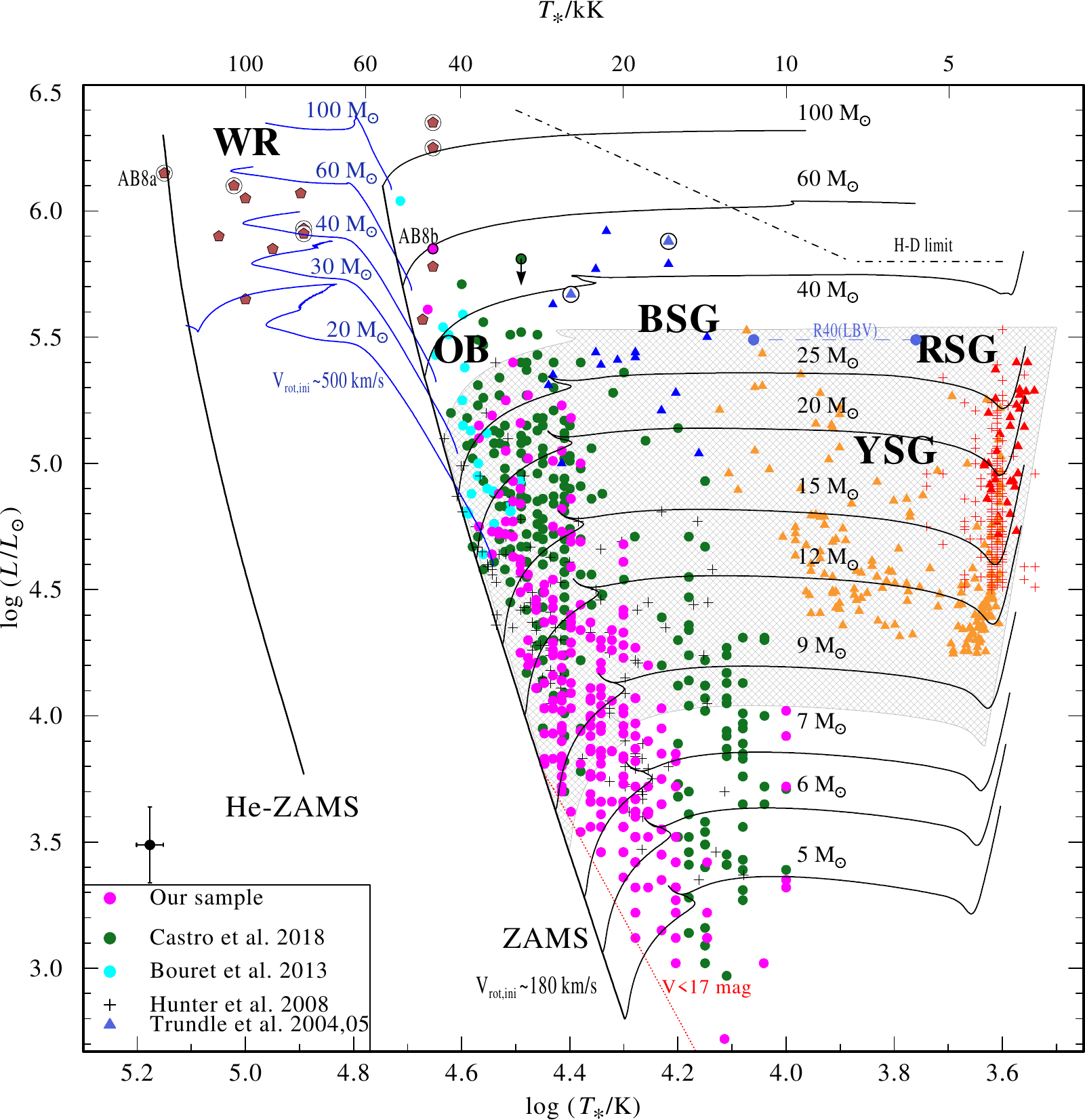} 
\caption{Same as Fig.\,\ref{fig:hrd}, but including additional OB stars
from \citet{Castro2018_smc},  \citet{Bouret2013}, and \citet{Hunter2008_B}, bringing the total number of OB stars to about $\sim 800$. While our sample is
restricted to the SGS in the Wing of the SMC (see
Sect.\,\ref{subsect:sgs}), the other samples spread all over the SMC.
The only outstanding star that seems to reside in the void region
was obviously confused in the catalog by \citep{Castro2018_smc}; we
added a downward arrow  because the observed SED definitely indicates
$\log L/L_\odot < 5.2$. The additional samples also support our
conclusions  on stellar evolution.}
\label{fig:hrd2}
\end{figure*}
%---------------------------------------------------------------

%
\section{Additional tables}
\label{sect:appendixa}

\longtab{2}{     
\setlength{\tabcolsep}{2pt}
\begin{longtable}{llcccS[table-format = <-1.3]cS[table-format = <-1.3]ccccccS[table-format = <-1.3]}
\caption{Stellar parameters of all OB stars in the SMC-SGS\,1}
 \label{table:App_stellarparameters} \\

%This is the header for the first page of the table...

\hline %-------------------------------------------------------------------------------
\hline%-------------------------------------------------------------------------
\noalign{\vspace{1mm}}
SMCSGS-FS & Spectral type &	$T _\ast$ 	& $\log\,L$ &	$\log\,g_\ast$ &	\multicolumn{1}{c}{$\log\,\dot{M}$\tablefootmark{(1)}} &	$E_{\rm B-V} $&	$M_{\mathrm{V}}$&$R _\ast$  &	
 $\varv_\infty$\tablefootmark{(2)} 	& $\varv\,\sin\,i$& $\varv_{\rm rad}$&$M_\ast$ &	$\log\,Q_{0}$ &$\log\,L_{\mathrm{mec}}$\tablefootmark{(3)}\\ 

\#& & [kK] & [$L _{\odot}$]&[cm s$ ^{-2} $]& \multicolumn{1}{c}{[$M _{\odot}\,\mathrm{yr}^{-1} $]} & 
[mag] &[mag]& [$R _{\odot}$] & [km\,s$^{-1}$]&[km\,s$^{-1}$]&[km\,s$^{-1}$]&[$M _{\odot} $]&[s$ ^{-1} $] 
&[$L _{\odot}$]\\

\noalign{\vspace{1mm}}
\hline %-------------------------------------------------------------------------------
\noalign{\vspace{1mm}}
\endfirsthead

%This is the header for the remaining page(s) of the table...
\caption{continued.}\\
\hline %-------------------------------------------------------------------------------
\hline%-------------------------------------------------------------------------
\noalign{\vspace{1mm}}
SMCSGS-FS & Spectral type &	$T _\ast$ 	& $\log\,L$ &	$\log\,g_\ast$ &	\multicolumn{1}{c}{$\log\,\dot{M}$\tablefootmark{(1)}} &	$E_{\rm B-V} $&	$M_{\mathrm{V}}$&$R _\ast$  &	
 $\varv_\infty$\tablefootmark{(2)} 	& $\varv\,\sin\,i$& $\varv_{\rm rad}$&$M_\ast$ &	$\log\,Q_{0}$ &$\log\,L_{\mathrm{mec}}$\tablefootmark{(3)}\\ 

\#& & [kK] & [$L _{\odot}$]&[cm s$ ^{-2} $]& \multicolumn{1}{c}{[$M _{\odot}\,\mathrm{yr}^{-1} $]} & 
[mag] &[mag]& [$R _{\odot}$] & [km\,s$^{-1}$]&[km\,s$^{-1}$]&[km\,s$^{-1}$]&[$M _{\odot} $]&[s$ ^{-1} $] 
&[$L _{\odot}$]\\

\noalign{\vspace{1mm}} 

\hline %-------------------------------------------------------------------------------

\noalign{\vspace{1mm}}
\noalign{\vspace{1mm}}
\endhead
\noalign{\vspace{1mm}}
%This is the footer for all pages except the last page of the table...
\hline
\endfoot

1	&	B0.7	V	&	27	&	3.83	&	4.4	&	-9.5	&	0.03	&	-1.8	&	3.8	&	2300	&	120	&	170	&	13	&	46.1	&	-1.9	\\
2	&	B0.2	V	&	29	&	4.21	&	4.4	&	-9.1	&	0.02	&	-3.0	&	5.1	&	2600	&	80	&	160	&	23	&	46.7	&	-1.2	\\
3	&	B2.5	V	&	19	&	3.52	&	4.0	&	-9.8	&	0.03	&	-2.2	&	5.3	&	900	&	130	&	150	&	10	&	44.5	&	-2.4	\\
4	&	B2	III	&	19	&	4.27	&	3.4	&	-9.0	&	0.03	&	-3.8	&	12.6	&	600	&	130	&	160	&	15	&	45.4	&	-1.1	\\
5	&	B2	V	&	20	&	3.56	&	4.2	&	-9.7	&	0.05	&	-1.9	&	5.0	&	1100	&	100	&	170	&	15	&	44.6	&	-2.4	\\
6	&	B1.5	IV	&	23	&	3.87	&	4.0	&	-9.4	&	0.03	&	-3.0	&	5.4	&	1700	&	70	&	170	&	11	&	45.6	&	-1.8	\\
7	&	B2	V	&	22	&	3.76	&	4.2	&	-9.5	&	0.05	&	-2.2	&	5.2	&	2200	&	70	&	170	&	16	&	45.2	&	-2.0	\\
8	&	B2	V	&	20	&	3.66	&	4.2	&	-9.6	&	0.05	&	-2.0	&	5.6	&	1100	&	130	&	120	&	18	&	44.7	&	-2.2	\\
9	&	B2	(IV)e	&	22	&	4.01	&	3.8	&	-9.3	&	0.09	&	-2.8	&	7.0	&	1500	&	250	&	170	&	11	&	45.6	&	-1.6	\\
10	&	B7	V	&	14	&	3.12	&	3.8	&	-10.2	&	0.05	&	-2.0	&	6.2	&	800	&	80	&	170	&	9	&	43.7	&	-3.1	\\
11	&	B2	V	&	20	&	3.66	&	4.2	&	-9.6	&	0.05	&	-2.0	&	5.6	&	1100	&	100	&	170	&	18	&	44.7	&	-2.2	\\
12	&	B8	V	&	14	&	3.22	&	3.8	&	-10.1	&	0.05	&	-2.1	&	6.9	&	800	&	100	&	170	&	11	&	43.8	&	-3.0	\\
13	&	B1.5	IV	&	22	&	4.08	&	3.6	&	-9.2	&	0.03	&	-3.0	&	7.6	&	1200	&	140	&	155	&	8	&	45.7	&	-1.4	\\
14	&	B2.5	V	&	19	&	3.52	&	4.0	&	-9.8	&	0.05	&	-2.2	&	5.3	&	900	&	200	&	170	&	10	&	44.5	&	-2.4	\\
15	&	B2.5	IV	&	20	&	3.63	&	3.6	&	-9.7	&	0.10	&	-2.2	&	5.5	&	500	&	140	&	160	&	4	&	44.9	&	-2.2	\\
16	&	B1.5	V	&	22	&	4.11	&	3.8	&	-9.2	&	0.05	&	-3.1	&	7.8	&	1600	&	150	&	200	&	14	&	45.7	&	-1.4	\\
17	&	B5	V	&	16	&	3.52	&	4.0	&	-9.8	&	0.05	&	-2.3	&	7.5	&	1000	&	130	&	140	&	21	&	44.2	&	-2.4	\\
18	&	B1.5	(IV)e	&	24	&	4.44	&	3.8	&	-8.9	&	0.09	&	-3.5	&	9.6	&	1800	&	320	&	170	&	21	&	46.4	&	-0.8	\\
19	&	B5	V	&	16	&	3.32	&	4.0	&	-10.0	&	0.02	&	-2.4	&	6.0	&	900	&	150	&	120	&	13	&	44.0	&	-2.8	\\
20	&	B5	V	&	16	&	3.12	&	4.0	&	-10.2	&	0.05	&	-1.9	&	4.7	&	800	&	140	&	160	&	8	&	43.8	&	-3.1	\\
21	&	B5	V	&	16	&	3.32	&	4.0	&	-10.0	&	0.03	&	-1.8	&	6.0	&	900	&	200	&	190	&	13	&	44.0	&	-2.8	\\
22	&	B0.5	V	&	27	&	4.54	&	4.0	&	-8.8	&	0.04	&	-3.8	&	8.5	&	2100	&	80	&	170	&	27	&	46.9	&	-0.6	\\
23	&	B9	III	&	11	&	3.02	&	3.5	&	-10.3	&	0.03	&	-2.2	&	8.9	&	600	&	110	&	180	&	9	&	43.0	&	-3.3	\\
24	&	B7	V	&	14	&	3.42	&	3.8	&	-9.9	&	0.05	&	-1.9	&	8.7	&	900	&	100	&	160	&	18	&	44.0	&	-2.6	\\
25	&	B3	V	&	18	&	3.32	&	3.8	&	-10.0	&	0.05	&	-2.0	&	4.7	&	600	&	180	&	170	&	5	&	44.3	&	-2.8	\\
26	&	B5	V	&	17	&	3.45	&	3.8	&	-9.9	&	0.04	&	-1.9	&	6.1	&	700	&	160	&	190	&	9	&	44.4	&	-2.6	\\
27	&	B2.5	V	&	19	&	3.52	&	4.2	&	-9.8	&	0.05	&	-1.9	&	5.3	&	1100	&	140	&	170	&	16	&	44.4	&	-2.4	\\
28	&	B5	V	&	17	&	3.65	&	3.8	&	-9.7	&	0.05	&	-2.7	&	7.7	&	800	&	220	&	160	&	14	&	44.6	&	-2.2	\\
29	&	B0.7	V	&	26	&	4.02	&	4.4	&	-9.3	&	0.03	&	-2.6	&	5.1	&	2700	&	230	&	170	&	23	&	46.2	&	-1.5	\\
30	&	B1	V	&	26	&	3.7	&	4.4	&	-9.6	&	0.02	&	-2.0	&	3.5	&	2200	&	70	&	170	&	11	&	45.9	&	-2.1	\\
31	&	O9	V	&	32	&	4.93	&	4.0	&	-8.4	&	0.02	&	-4.3	&	9.5	&	2000	&	340	&	160	&	33	&	48.1	&	0.1	\\
32	&	B0.7	(III)e	&	25	&	4.72	&	3.6	&	-8.6	&	0.22	&	-4.2	&	12.2	&	1500	&	280	&	170	&	22	&	46.9	&	-0.3	\\
33	&	B3	V	&	17	&	3.65	&	3.8	&	-9.7	&	0.05	&	-2.1	&	7.7	&	800	&	50	&	160	&	14	&	44.6	&	-2.2	\\
34	&	B3	(V)e	&	18	&	3.52	&	3.8	&	-9.8	&	0.08	&	-2.0	&	5.9	&	700	&	200	&	170	&	8	&	44.5	&	-2.4	\\
35	&	B1.5	V	&	22	&	3.76	&	4.2	&	-9.5	&	0.05	&	-2.6	&	5.2	&	2200	&	180	&	170	&	16	&	45.2	&	-2.0	\\
36	&	B0.5	(III)e	&	26	&	5.23	&	3.2	&	-8.1	&	0.19	&	-5.1	&	20.4	&	1000	&	300	&	160	&	24	&	47.8	&	0.6	\\
37	&	B0.7	V	&	26	&	4.14	&	4.2	&	-9.2	&	0.05	&	-3.1	&	5.8	&	2200	&	330	&	170	&	20	&	46.3	&	-1.3	\\
38	&	B0.2	V	&	28	&	4.37	&	4.2	&	-8.9	&	0.02	&	-3.0	&	6.5	&	2300	&	80	&	150	&	25	&	46.8	&	-0.9	\\
39	&	B0.5	V	&	27	&	4.29	&	4.2	&	-9.0	&	0.02	&	-3.0	&	6.4	&	2300	&	230	&	170	&	24	&	46.6	&	-1.1	\\
40	&	B0.7	V	&	25	&	4.29	&	4.0	&	-9.0	&	0.05	&	-3.6	&	7.5	&	2000	&	130	&	170	&	20	&	46.4	&	-1.1	\\
41	&	B2.5	V	&	19	&	3.22	&	4.2	&	-10.1	&	0.03	&	-1.5	&	3.8	&	900	&	160	&	170	&	8	&	44.1	&	-3.0	\\
42	&	B2	IV	&	20	&	4.03	&	3.8	&	-9.3	&	0.05	&	-3.1	&	8.6	&	900	&	80	&	170	&	17	&	45.1	&	-1.5	\\
43	&	B2	(V)e	&	20	&	4.01	&	3.8	&	-9.3	&	0.11	&	-3.3	&	8.4	&	900	&	180	&	170	&	16	&	45.7	&	-1.6	\\
44	&	B0	V	&	30	&	4.51	&	4.2	&	-8.8	&	0.03	&	-3.6	&	6.7	&	2300	&	130	&	160	&	26	&	47.2	&	-0.7	\\
45	&	B2	Ib	&	20	&	4.42	&	2.8	&	-8.9	&	0.45	&	-3.5	&	13.5	&	300	&	50	&	170	&	4	&	45.7	&	-0.8	\\
46	&	B0	V	&	30	&	4.3	&	4.4	&	-9.0	&	0.02	&	-3.2	&	5.2	&	2600	&	250	&	170	&	25	&	46.9	&	-1.0	\\
47	&	B2	V	&	22	&	3.86	&	4.2	&	-9.4	&	0.03	&	-2.7	&	5.9	&	2300	&	70	&	140	&	20	&	45.3	&	-1.8	\\
48	&	B2	V	&	22	&	3.96	&	4.2	&	-9.3	&	0.05	&	-3.0	&	6.6	&	2400	&	130	&	160	&	25	&	45.4	&	-1.7	\\
49	&	B3	(IV)e	&	17	&	3.75	&	3.4	&	-9.6	&	0.23	&	-2.6	&	8.7	&	500	&	150	&	170	&	7	&	44.7	&	-2.0	\\
50	&	B9	Ib	&	10	&	3.72	&	2.8	&	-9.6	&	0.03	&	-3.6	&	24.2	&	500	&	80	&	200	&	13	&	43.3	&	-2.1	\\
51	&	B1.5	(V)e	&	22	&	4.21	&	3.8	&	-9.1	&	0.08	&	-3.0	&	8.8	&	1700	&	250	&	170	&	18	&	45.8	&	-1.2	\\
52	&	B0.5	V	&	28	&	4.03	&	4.4	&	-9.3	&	0.05	&	-2.6	&	4.4	&	2500	&	150	&	170	&	18	&	46.4	&	-1.5	\\
53	&	B2	V	&	19	&	3.52	&	4.0	&	-9.8	&	0.02	&	-2.0	&	5.3	&	900	&	80	&	170	&	10	&	44.5	&	-2.4	\\
54	&	B1.5	V	&	23	&	3.66	&	4.4	&	-9.6	&	0.05	&	-2.0	&	4.3	&	2500	&	220	&	160	&	17	&	45.1	&	-2.2	\\
55	&	B1.5	IV	&	22	&	4.18	&	3.6	&	-9.1	&	0.02	&	-3.6	&	8.5	&	1300	&	270	&	100	&	10	&	45.8	&	-1.3	\\
56	&	B0.5	V	&	28	&	3.81	&	4.4	&	-9.5	&	0.09	&	-2.0	&	3.4	&	2200	&	100	&	150	&	11	&	46.2	&	-1.9	\\
57	&	B5	V	&	16	&	3.22	&	4.0	&	-10.1	&	0.03	&	-2.1	&	5.3	&	900	&	120	&	160	&	10	&	44.0	&	-3.0	\\
58	&	B2.5	V	&	19	&	3.32	&	4.2	&	-10.0	&	0.05	&	-2.2	&	4.2	&	1000	&	140	&	160	&	10	&	44.2	&	-2.8	\\
59	&	B0	V	&	29	&	4.42	&	4.2	&	-8.9	&	0.05	&	-3.3	&	6.4	&	2300	&	200	&	170	&	24	&	47.0	&	-0.8	\\
60	&	B0.2	V	&	28	&	3.93	&	4.4	&	-9.4	&	0.05	&	-2.6	&	3.9	&	2300	&	120	&	170	&	14	&	46.3	&	-1.7	\\
61	&	B1.5	V	&	23	&	4.07	&	4.0	&	-9.2	&	0.02	&	-4.1	&	6.8	&	1900	&	80	&	160	&	17	&	45.8	&	-1.5	\\
62	&	B0	V	&	29	&	4.46	&	4.2	&	-8.9	&	0.05	&	-3.7	&	6.7	&	2300	&	140	&	170	&	26	&	47.0	&	-0.8	\\
63	&	B0.5	V	&	27	&	4.03	&	4.4	&	-9.3	&	0.05	&	-2.6	&	4.7	&	2600	&	100	&	160	&	21	&	46.3	&	-1.5	\\
64	&	O9	V	&	33	&	5.05	&	4.0	&	-8.3	&	0.05	&	-4.6	&	10.3	&	2000	&	50	&	120	&	39	&	48.3	&	0.3	\\
65	&	B0.2	V	&	29	&	4.11	&	4.4	&	-9.2	&	0.03	&	-2.7	&	4.5	&	2500	&	120	&	170	&	19	&	46.6	&	-1.4	\\
66	&	B2	V	&	19	&	3.92	&	4.2	&	-9.4	&	0.09	&	-2.5	&	8.4	&	1400	&	250	&	170	&	41	&	44.8	&	-1.7	\\
67	&	B0	V	&	31	&	4.6	&	4.2	&	-8.7	&	0.05	&	-3.9	&	6.9	&	2300	&	70	&	150	&	28	&	47.5	&	-0.5	\\
68	&	B2	V	&	20	&	3.56	&	4.2	&	-9.7	&	0.05	&	-2.0	&	5.0	&	1100	&	130	&	160	&	15	&	44.6	&	-2.4	\\
69	&	B0.5	(II)e	&	26	&	5.05	&	3.2	&	-8.3	&	0.05	&	-5.0	&	16.6	&	900	&	100	&	160	&	16	&	47.6	&	0.3	\\
70	&	B0	V	&	30	&	4.56	&	4.2	&	-8.8	&	0.05	&	-3.4	&	7.1	&	2300	&	170	&	160	&	29	&	47.3	&	-0.6	\\
71	&	B1.5	IV	&	22	&	3.61	&	3.8	&	-9.7	&	0.05	&	-2.0	&	4.4	&	1200	&	250	&	160	&	4	&	45.2	&	-2.3	\\
72	&	B2	(II)e	&	18	&	4.2	&	3.0	&	-9.1	&	0.05	&	-3.9	&	13.0	&	400	&	150	&	170	&	6	&	45.2	&	-1.2	\\
73	&	B0	(IV)e	&	30	&	5.02	&	3.8	&	-8.3	&	0.08	&	-4.6	&	12.0	&	1700	&	270	&	160	&	33	&	48.0	&	0.2	\\
74	&	B2	V	&	19	&	3.52	&	4.2	&	-9.8	&	0.02	&	-2.2	&	5.3	&	1100	&	160	&	170	&	16	&	44.4	&	-2.4	\\
75	&	B1.5	V	&	22	&	3.86	&	4.2	&	-9.4	&	0.05	&	-2.2	&	5.9	&	2300	&	250	&	170	&	20	&	45.3	&	-1.8	\\
76	&	B0.7	V	&	26	&	3.96	&	4.2	&	-9.3	&	0.03	&	-2.7	&	4.7	&	2000	&	160	&	150	&	13	&	46.1	&	-1.7	\\
77	&	B1.5	IV	&	22	&	4.27	&	3.8	&	-9.0	&	0.02	&	-3.8	&	9.4	&	1800	&	160	&	150	&	20	&	45.9	&	-1.1	\\
78	&	B2	V	&	20	&	3.56	&	4.2	&	-9.7	&	0.03	&	-2.2	&	5.0	&	1100	&	140	&	160	&	15	&	44.6	&	-2.4	\\
79	&	B2.5	V	&	19	&	3.52	&	4.2	&	-9.8	&	0.05	&	-2.2	&	5.3	&	1100	&	120	&	160	&	16	&	44.4	&	-2.4	\\
80	&	B2.5	V	&	19	&	3.52	&	4.0	&	-9.8	&	0.07	&	-2.3	&	5.3	&	900	&	160	&	160	&	10	&	44.5	&	-2.4	\\
81	&	B1.5	V	&	23	&	3.96	&	4.4	&	-9.3	&	0.05	&	-1.9	&	6.0	&	3000	&	180	&	190	&	33	&	45.4	&	-1.7	\\
82	&	B2	IV	&	20	&	3.83	&	3.6	&	-9.5	&	0.08	&	-2.7	&	6.9	&	600	&	120	&	160	&	7	&	45.1	&	-1.9	\\
83	&	B1.5	V	&	23	&	3.76	&	4.4	&	-9.5	&	0.05	&	-2.0	&	4.8	&	2600	&	170	&	160	&	21	&	45.2	&	-2.0	\\
84	&	B1.5	V	&	23	&	3.76	&	4.4	&	-9.5	&	0.05	&	-2.2	&	4.8	&	2600	&	250	&	170	&	21	&	45.2	&	-2.0	\\
85	&	B0.7	V	&	26	&	4.19	&	4.2	&	-9.1	&	0.03	&	-3.3	&	6.2	&	2300	&	70	&	170	&	22	&	46.3	&	-1.2	\\
86	&	B9	III	&	10	&	3.35	&	3.4	&	-10.0	&	0.03	&	-2.9	&	15.8	&	800	&	80	&	170	&	23	&	43.0	&	-2.7	\\
87	&	B2	V	&	22	&	3.66	&	4.2	&	-9.6	&	0.03	&	-2.4	&	4.7	&	2000	&	120	&	180	&	13	&	45.1	&	-2.2	\\
88	&	B0.5	V	&	26	&	3.72	&	4.4	&	-9.6	&	0.05	&	-2.2	&	3.6	&	2200	&	220	&	170	&	12	&	45.9	&	-2.1	\\
89	&	O9.5	V	&	31	&	4.84	&	4.0	&	-8.5	&	0.02	&	-5.1	&	9.1	&	2000	&	120	&	160	&	31	&	47.8	&	-0.1	\\
90	&	O9.7	V	&	31	&	4.9	&	4.0	&	-8.4	&	0.02	&	-4.6	&	9.8	&	2100	&	70	&	200	&	35	&	47.9	&	0.0	\\
91	&	B0.7	(IV)e	&	25	&	4.35	&	3.8	&	-9.0	&	0.09	&	-3.2	&	8.0	&	1600	&	150	&	170	&	15	&	46.5	&	-1.0	\\
92	&	B0.5	V	&	28	&	3.83	&	4.4	&	-9.5	&	0.05	&	-2.4	&	3.5	&	2200	&	90	&	160	&	11	&	46.2	&	-1.9	\\
93	&	B2.5	V	&	19	&	3.42	&	4.2	&	-9.9	&	0.08	&	-2.0	&	4.7	&	1000	&	140	&	130	&	13	&	44.3	&	-2.6	\\
94	&	B1.5	V	&	23	&	4.06	&	4.4	&	-9.3	&	0.05	&	-3.1	&	6.8	&	3100	&	80	&	160	&	42	&	45.5	&	-1.5	\\
95	&	B1.5	V	&	23	&	3.76	&	4.4	&	-9.5	&	0.05	&	-2.2	&	4.8	&	2600	&	220	&	160	&	21	&	45.2	&	-2.0	\\
96	&	B1.5	V	&	22	&	4.3	&	4.0	&	-8.7$^\ast$	&	0.08	&	-3.7	&	9.8	&	900$^\ast$	&	250	&	170	&	35	&	46.5	&	-1.0	$^\ast$	\\
97	&	B1.5	V	&	23	&	3.56	&	4.4	&	-9.7	&	0.05	&	-1.6	&	3.8	&	2300	&	170	&	170	&	13	&	45.0	&	-2.4	\\
98	&	B0.5	V	&	26	&	4.08	&	4.2	&	-9.2	&	0.02	&	-2.9	&	5.4	&	2100	&	130	&	150	&	17	&	46.2	&	-1.4	\\
99	&	B1	IV	&	24	&	4.69	&	3.6	&	-8.6	&	0.05	&	-4.8	&	12.8	&	1600	&	120	&	170	&	24	&	46.7	&	-0.4	\\
100	&	B2.5	V	&	19	&	3.42	&	4.2	&	-9.9	&	0.02	&	-2.2	&	4.7	&	1000	&	180	&	160	&	13	&	44.3	&	-2.6	\\
101	&	B2.5	V	&	20	&	3.56	&	4.2	&	-9.7	&	0.05	&	-2.3	&	5.0	&	1100	&	200	&	170	&	15	&	44.6	&	-2.4	\\
102	&	B3	V	&	17	&	3.35	&	3.8	&	-10.0	&	0.02	&	-1.8	&	5.5	&	700	&	200	&	150	&	7	&	44.3	&	-2.7	\\
103	&	B2.5	V	&	19	&	3.42	&	4.2	&	-9.9	&	0.05	&	-2.0	&	4.7	&	1000	&	150	&	140	&	13	&	44.3	&	-2.6	\\
104	&	B5	V	&	16	&	3.02	&	4.0	&	-10.3	&	0.05	&	-1.8	&	4.2	&	800	&	250	&	160	&	7	&	43.8	&	-3.3	\\
105	&	B2.5	V	&	20	&	3.76	&	4.2	&	-9.5	&	0.02	&	-2.4	&	6.3	&	1200	&	130	&	160	&	23	&	44.8	&	-2.0	\\
106	&	B1.5	V	&	23	&	3.86	&	4.4	&	-9.4	&	0.02	&	-2.1	&	5.4	&	2800	&	180	&	160	&	27	&	45.3	&	-1.8	\\
107	&	B0	V	&	31	&	4.62	&	4.2	&	-8.7	&	0.05	&	-3.6	&	7.1	&	2300	&	60	&	160	&	29	&	47.5	&	-0.5	\\
108	&	B2.5	V	&	20	&	3.46	&	4.2	&	-9.8	&	0.04	&	-2.0	&	4.5	&	1000	&	140	&	170	&	12	&	44.5	&	-2.5	\\
109	&	B2.5	V	&	19	&	3.62	&	4.2	&	-9.7	&	0.05	&	-1.7	&	6.0	&	1200	&	120	&	110	&	21	&	44.5	&	-2.3	\\
110	&	B2.5	(III)e	&	18	&	3.9	&	3.2	&	-9.4	&	0.07	&	-3.4	&	9.2	&	400	&	220	&	160	&	5	&	45.0	&	-1.8	\\
111	&	O8.5	V	&	35	&	4.73	&	4.4	&	-8.6	&	0.05	&	-3.7	&	6.3	&	2700	&	220	&	240	&	37	&	48.1	&	-0.3	\\
112	&	B3	V	&	19	&	3.62	&	4.2	&	-9.7	&	0.05	&	-2.1	&	6.0	&	1200	&	100	&	160	&	21	&	44.5	&	-2.3	\\
113	&	B7	V	&	13	&	2.72	&	3.8	&	-10.6	&	0.03	&	-1.1	&	4.5	&	600	&	100	&	110	&	5	&	43.2	&	-3.8	\\
114	&	B2.5	V	&	19	&	3.52	&	4.0	&	-9.8	&	0.03	&	-2.2	&	5.3	&	900	&	200	&	170	&	10	&	44.5	&	-2.4	\\
115	&	B1.5	V	&	22	&	3.46	&	4.2	&	-9.8	&	0.05	&	-2.0	&	3.7	&	1800	&	120	&	160	&	8	&	44.9	&	-2.5	\\
116	&	B2.5	V	&	19	&	3.52	&	4.2	&	-9.8	&	0.03	&	-2.0	&	5.3	&	1100	&	80	&	160	&	16	&	44.4	&	-2.4	\\
117	&	B5	(IV)e	&	16	&	3.82	&	3.4	&	-9.5	&	0.08	&	-2.8	&	10.6	&	600	&	180	&	170	&	10	&	44.8	&	-1.9	\\
118	&	B2.5	V	&	19	&	3.67	&	4.0	&	-9.6	&	0.05	&	-2.5	&	6.3	&	1000	&	170	&	150	&	15	&	44.9	&	-2.2	\\
119	&	B0	V	&	32	&	4.63	&	4.0	&	-8.7	&	0.05	&	-3.3	&	6.7	&	1700	&	250	&	160	&	17	&	47.8	&	-0.5	\\
120	&	B0.2	V	&	29	&	4.44	&	4.2	&	-8.9	&	0.05	&	-2.8	&	6.6	&	2300	&	30	&	170	&	25	&	47.0	&	-0.8	\\
121	&	O9.7	V	&	32	&	4.71	&	4.2	&	-8.6	&	0.05	&	-3.9	&	7.4	&	2300	&	50	&	170	&	32	&	47.8	&	-0.3	\\
122	&	B2	V	&	20	&	3.46	&	4.2	&	-9.8	&	0.05	&	-1.9	&	4.5	&	1000	&	80	&	160	&	12	&	44.5	&	-2.5	\\
123	&	B2	IV	&	20	&	3.93	&	3.6	&	-9.4	&	0.05	&	-3.0	&	7.7	&	600	&	100	&	170	&	9	&	45.2	&	-1.7	\\
124	&	B2.5	(V)e	&	19	&	4.07	&	3.8	&	-9.2	&	0.08	&	-2.9	&	10.0	&	900	&	200	&	180	&	23	&	45.1	&	-1.5	\\
125	&	B0.5	V	&	27	&	4.24	&	4.2	&	-9.1	&	0.05	&	-2.7	&	6.0	&	2300	&	220	&	160	&	21	&	46.5	&	-1.2	\\
126	&	B5	V	&	16	&	3.42	&	3.8	&	-9.9	&	0.05	&	-1.7	&	6.7	&	800	&	100	&	160	&	10	&	44.3	&	-2.6	\\
127	&	B2.5	V	&	20	&	3.83	&	3.8	&	-9.5	&	0.05	&	-2.8	&	6.9	&	800	&	280	&	160	&	11	&	44.9	&	-1.9	\\
128	&	O9	V	&	34	&	4.72	&	4.2	&	-8.6	&	0.05	&	-3.2	&	6.6	&	2100	&	410	&	170	&	25	&	48.0	&	-0.3	\\
129	&	B3	V	&	18	&	3.62	&	3.8	&	-9.7	&	0.05	&	-2.6	&	6.7	&	800	&	70	&	180	&	10	&	44.6	&	-2.3	\\
130	&	B2.5	V	&	19	&	3.52	&	4.2	&	-9.8	&	0.02	&	-2.1	&	5.3	&	1100	&	160	&	160	&	16	&	44.4	&	-2.4	\\
131	&	B1.5	V	&	22	&	3.46	&	4.2	&	-9.8	&	0.05	&	-1.5	&	3.7	&	1800	&	100	&	170	&	8	&	44.9	&	-2.5	\\
132	&	B2	V	&	22	&	3.86	&	4.2	&	-9.4	&	0.05	&	-2.4	&	5.9	&	2300	&	150	&	170	&	20	&	45.3	&	-1.8	\\
133	&	B2	V	&	22	&	3.86	&	4.2	&	-9.4	&	0.05	&	-1.9	&	5.9	&	2300	&	120	&	160	&	20	&	45.3	&	-1.8	\\
134	&	B2.5	V	&	20	&	3.56	&	4.2	&	-9.7	&	0.05	&	-1.8	&	5.0	&	1100	&	200	&	160	&	15	&	44.6	&	-2.4	\\
135	&	B2	IV	&	20	&	3.93	&	3.8	&	-9.4	&	0.05	&	-3.1	&	7.7	&	800	&	170	&	160	&	14	&	45.0	&	-1.7	\\
136	&	B1	(IV)e	&	23	&	4.31	&	3.8	&	-9.0	&	0.05	&	-3.2	&	9.0	&	1700	&	120	&	170	&	19	&	46.2	&	-1.0	\\
137	&	B1.5	IV	&	21	&	4.26	&	3.6	&	-9.1	&	0.05	&	-3.6	&	10.2	&	1500	&	180	&	170	&	15	&	45.7	&	-1.1	\\
138	&	B1.5	V	&	22	&	3.76	&	4.2	&	-9.5	&	0.05	&	-2.1	&	5.2	&	2200	&	270	&	140	&	16	&	45.2	&	-2.0	\\
139	&	B2.5	(V)e	&	20	&	3.73	&	3.8	&	-9.6	&	0.08	&	-2.3	&	6.1	&	700	&	150	&	170	&	9	&	44.8	&	-2.1	\\
140	&	B2	V	&	20	&	3.64	&	4.2	&	-9.7	&	0.05	&	-2.0	&	5.5	&	1100	&	80	&	160	&	18	&	44.7	&	-2.2	\\
141	&	B1.5	V	&	22	&	3.96	&	4.2	&	-9.3	&	0.05	&	-3.0	&	6.6	&	2400	&	200	&	170	&	25	&	45.4	&	-1.7	\\
142	&	B2	V	&	21	&	3.86	&	4.0	&	-9.4	&	0.05	&	-2.5	&	6.4	&	1900	&	130	&	170	&	15	&	45.1	&	-1.8	\\
143	&	O8	V	&	35	&	5.19	&	4.0	&	-8.1	&	0.03	&	-4.8	&	10.7	&	2000	&	60	&	150	&	42	&	48.6	&	0.5	\\
144	&	B2.5	V	&	20	&	3.36	&	4.2	&	-9.9	&	0.05	&	-1.4	&	4.0	&	1000	&	200	&	140	&	9	&	44.4	&	-2.7	\\
145	&	B2	V	&	22	&	3.86	&	4.2	&	-9.4	&	0.05	&	-2.7	&	5.9	&	2300	&	250	&	150	&	20	&	45.3	&	-1.8	\\
146	&	B2.5	V	&	20	&	3.66	&	4.2	&	-9.6	&	0.05	&	-2.0	&	5.6	&	1100	&	120	&	170	&	18	&	44.7	&	-2.2	\\
147	&	B1	V	&	26	&	3.9	&	4.4	&	-9.4	&	0.05	&	-2.3	&	4.4	&	2500	&	100	&	170	&	18	&	46.1	&	-1.8	\\
148	&	B2	V	&	21	&	3.86	&	4.0	&	-9.4	&	0.10	&	-2.8	&	6.4	&	1900	&	200	&	170	&	15	&	45.1	&	-1.8	\\
149	&	B0.5	V	&	27	&	4.03	&	4.4	&	-9.3	&	0.05	&	-2.3	&	4.7	&	2600	&	100	&	160	&	21	&	46.3	&	-1.5	\\
150	&	B3	V	&	18	&	3.61	&	3.8	&	-9.7	&	0.05	&	-2.1	&	6.6	&	800	&	140	&	170	&	10	&	44.5	&	-2.3	\\
151	&	B1.5	V	&	22	&	3.76	&	4.2	&	-9.5	&	0.02	&	-2.3	&	5.2	&	2200	&	200	&	170	&	16	&	45.2	&	-2.0	\\
152	&	B2.5	V	&	19	&	3.52	&	4.2	&	-9.8	&	0.02	&	-2.4	&	5.3	&	1100	&	160	&	170	&	16	&	44.4	&	-2.4	\\
153	&	B2.5	IV	&	20	&	4.23	&	3.8	&	-9.1	&	0.02	&	-4.1	&	10.9	&	1000	&	250	&	150	&	27	&	45.3	&	-1.2	\\
154	&	B2	V	&	22	&	3.96	&	4.2	&	-9.3	&	0.05	&	-2.8	&	6.6	&	2400	&	150	&	160	&	25	&	45.4	&	-1.7	\\
155	&	B0.5	V	&	27	&	4.06	&	4.2	&	-9.3	&	0.05	&	-2.7	&	4.9	&	2000	&	100	&	150	&	14	&	46.3	&	-1.5	\\
156	&	B0.2	V	&	29	&	4.21	&	4.4	&	-9.1	&	0.03	&	-2.9	&	5.1	&	2600	&	160	&	170	&	23	&	46.7	&	-1.2	\\
157	&	B2.5	(V)e	&	17	&	4.03	&	3.8	&	-9.3	&	0.08	&	-3.8	&	12.0	&	1000	&	150	&	160	&	33	&	45.0	&	-1.5	\\
158	&	B0.7	(III)e	&	26	&	4.82	&	3.4	&	-8.5	&	0.11	&	-4.2	&	12.7	&	1100	&	250	&	150	&	15	&	47.2	&	-0.1	\\
159	&	B0.7	V	&	26	&	3.76	&	4.2	&	-9.5	&	0.05	&	-1.9	&	3.7	&	1800	&	110	&	140	&	8	&	45.9	&	-2.0	\\
160	&	B1	IV	&	25	&	4.29	&	4.0	&	-9.0	&	0.03	&	-3.5	&	7.5	&	2000	&	220	&	150	&	20	&	46.4	&	-1.1	\\
161	&	B5	IV	&	16	&	3.85	&	3.8	&	-9.5	&	0.02	&	-3.0	&	11.0	&	1000	&	140	&	130	&	28	&	44.7	&	-1.8	\\
162	&	B1	V	&	25	&	3.62	&	4.4	&	-9.7	&	0.05	&	-2.1	&	3.5	&	2200	&	250	&	150	&	11	&	45.6	&	-2.3	\\
163	&	B2.5	V	&	19	&	3.12	&	4.2	&	-10.2	&	0.05	&	-0.9	&	3.4	&	900	&	80	&	150	&	7	&	44.0	&	-3.1	\\
164	&	B1	V	&	25	&	4.09	&	4.0	&	-9.2	&	0.05	&	-2.9	&	5.9	&	1800	&	230	&	170	&	13	&	46.2	&	-1.4	\\
165	&	B5	V	&	16	&	3.32	&	4.0	&	-10.0	&	0.05	&	-2.3	&	6.0	&	900	&	120	&	130	&	13	&	44.0	&	-2.8	\\
166	&	O8	IV	&	33	&	5.25	&	3.8	&	-8.1	&	0.02	&	-5.1	&	12.9	&	1700	&	140	&	150	&	39	&	48.6	&	0.6	\\
167	&	B0.7	(II)e	&	24	&	5	&	3.2	&	-8.3	&	0.05	&	-5.1	&	18.3	&	1100	&	120	&	150	&	19	&	47.2	&	0.2	\\
168	&	B1.5	V	&	23	&	3.96	&	4.4	&	-9.3	&	0.05	&	-3.0	&	6.0	&	3000	&	190	&	160	&	33	&	45.4	&	-1.7	\\
169	&	B1	IV	&	24	&	4.34	&	3.8	&	-9.0	&	0.05	&	-3.4	&	8.6	&	1700	&	120	&	170	&	17	&	46.3	&	-1.0	\\
170	&	B0.7	V	&	27	&	4.16	&	4.2	&	-9.2	&	0.05	&	-3.0	&	5.5	&	2200	&	200	&	170	&	18	&	46.5	&	-1.3	\\
171	&	B1	(V)e	&	25	&	4.13	&	3.8	&	-9.2	&	0.22	&	-2.6	&	6.2	&	1400	&	510	&	160	&	9	&	46.3	&	-1.3	\\
172	&	B0	V	&	30	&	4.23	&	4.2	&	-9.1	&	0.09	&	-2.6	&	4.8	&	1900	&	80	&	170	&	14	&	47.0	&	-1.2	\\
173	&	B2.5	V	&	19	&	3.62	&	4.2	&	-9.7	&	0.05	&	-1.8	&	6.0	&	1200	&	250	&	150	&	21	&	44.5	&	-2.3	\\
174	&	B2.5	V	&	19	&	3.62	&	4.2	&	-9.7	&	0.03	&	-2.6	&	6.0	&	1200	&	280	&	150	&	21	&	44.5	&	-2.3	\\
175	&	B9	Ib	&	10	&	3.32	&	3.0	&	-10.0	&	0.03	&	-3.1	&	15.3	&	500	&	120	&	150	&	9	&	42.9	&	-2.8	\\
176	&	B7	V	&	14	&	3.12	&	3.8	&	-10.2	&	0.03	&	-2.1	&	6.2	&	800	&	220	&	170	&	9	&	43.7	&	-3.1	\\
177	&	B5	(V)e	&	17	&	3.45	&	3.8	&	-9.9	&	0.05	&	-1.8	&	6.1	&	700	&	110	&	150	&	9	&	44.4	&	-2.6	\\
178	&	B2	V	&	22	&	3.56	&	4.2	&	-9.7	&	0.05	&	-1.5	&	4.2	&	1900	&	120	&	160	&	10	&	45.0	&	-2.4	\\
179	&	B2.5	(IV)e	&	19	&	4.07	&	3.4	&	-9.2	&	0.10	&	-3.0	&	10.0	&	600	&	250	&	170	&	9	&	45.2	&	-1.5	\\
180	&	B2	V	&	21	&	4.3	&	4.0	&	-9.0	&	0.00	&	-3.9	&	10.7	&	2400	&	140	&	160	&	42	&	45.6	&	-1.0	\\
181	&	B2	V	&	22	&	3.66	&	4.2	&	-9.6	&	0.05	&	-1.8	&	4.7	&	2000	&	130	&	170	&	13	&	45.1	&	-2.2	\\
182	&	B1	V	&	26	&	3.72	&	4.4	&	-9.6	&	0.07	&	-2.1	&	3.6	&	2200	&	70	&	160	&	12	&	45.9	&	-2.1	\\
183	&	B5	(V)e	&	16	&	3.32	&	4.0	&	-10.0	&	0.05	&	-2.6	&	6.0	&	900	&	160	&	170	&	13	&	44.0	&	-2.8	\\
184	&	B2.5	V	&	20	&	3.73	&	3.8	&	-9.6	&	0.05	&	-2.6	&	6.1	&	700	&	180	&	170	&	9	&	44.8	&	-2.1	\\
185	&	B0.7	V	&	27	&	4.06	&	4.2	&	-9.3	&	0.05	&	-2.5	&	4.9	&	2000	&	80	&	170	&	14	&	46.3	&	-1.5	\\
186	&	B0.7	V	&	27	&	3.93	&	4.4	&	-9.4	&	0.05	&	-2.6	&	4.2	&	2400	&	220	&	170	&	16	&	46.2	&	-1.7	\\
187	&	B0.5	IV	&	28	&	4.49	&	4.0	&	-8.8	&	0.03	&	-3.8	&	7.5	&	1900	&	140	&	170	&	20	&	47.0	&	-0.7	\\
188	&	B5	V	&	17	&	3.55	&	3.8	&	-9.8	&	0.02	&	-2.2	&	6.9	&	800	&	160	&	160	&	11	&	44.5	&	-2.4	\\
189	&	B2	(III)e	&	20	&	4.4	&	3.4	&	-8.9	&	0.08	&	-4.0	&	13.2	&	600	&	120	&	160	&	16	&	45.7	&	-0.9	\\
190	&	B0.7	IV	&	25	&	4.86	&	3.6	&	-8.5	&	0.05	&	-4.8	&	14.4	&	1600	&	70	&	200	&	30	&	47.1	&	-0.1	\\
191	&	B0.5	V	&	28	&	4.13	&	4.4	&	-9.2	&	0.05	&	-2.7	&	4.9	&	2600	&	110	&	160	&	22	&	46.5	&	-1.3	\\
192	&	B1.5	IV	&	23	&	4.11	&	4.0	&	-9.2	&	0.05	&	-3.1	&	7.2	&	2000	&	170	&	160	&	19	&	45.9	&	-1.4	\\
193	&	B0.7	V	&	27	&	3.93	&	4.4	&	-9.4	&	0.05	&	-2.9	&	4.2	&	2400	&	80	&	160	&	16	&	46.2	&	-1.7	\\
194	&	B5	V	&	16	&	3.27	&	4.0	&	-10.0	&	0.02	&	-2.0	&	5.6	&	900	&	120	&	170	&	12	&	44.0	&	-2.9	\\
195	&	O9.7	IV	&	32	&	4.85	&	4.0	&	-8.5	&	0.05	&	-4.5	&	8.7	&	1900	&	30	&	170	&	28	&	48.0	&	-0.1	\\
196	&	B3	IV	&	18	&	3.62	&	3.8	&	-9.7	&	0.03	&	-2.7	&	6.7	&	800	&	110	&	170	&	10	&	44.6	&	-2.3	\\
197	&	B1.5	IV	&	21	&	3.66	&	3.6	&	-9.6	&	0.05	&	-2.0	&	5.1	&	1000	&	100	&	170	&	4	&	45.1	&	-2.2	\\
198	&	B1	V	&	25	&	4.37	&	4.0	&	-8.9	&	0.02	&	-3.5	&	8.2	&	2100	&	150	&	170	&	24	&	46.4	&	-0.9	\\
199	&	B1	V	&	25	&	4.03	&	4.2	&	-9.3	&	0.05	&	-2.9	&	5.5	&	2200	&	100	&	160	&	18	&	46.0	&	-1.5	\\
200	&	B2	V	&	22	&	3.76	&	4.2	&	-9.5	&	0.05	&	-2.5	&	5.2	&	2200	&	130	&	140	&	16	&	45.2	&	-2.0	\\
201	&	B0.7	V	&	27	&	4.03	&	4.4	&	-9.3	&	0.03	&	-2.8	&	4.7	&	2600	&	220	&	150	&	21	&	46.3	&	-1.5	\\
202	&	B1	V	&	26	&	4.24	&	4.2	&	-9.1	&	0.05	&	-3.2	&	6.5	&	2300	&	80	&	170	&	25	&	46.4	&	-1.2	\\
203	&	B0.5	(III)e	&	25	&	5.18	&	3.2	&	-8.1	&	0.25	&	-5.1	&	20.8	&	1100	&	450	&	130	&	25	&	47.6	&	0.5	\\
204	&	B1	V	&	26	&	3.95	&	4.4	&	-9.4	&	0.02	&	-2.5	&	4.7	&	2500	&	80	&	170	&	20	&	46.1	&	-1.7	\\
205	&	B2.5	V	&	19	&	3.62	&	4.0	&	-9.7	&	0.03	&	-2.1	&	6.0	&	900	&	360	&	170	&	13	&	44.6	&	-2.3	\\
206	&	B3	IV	&	18	&	3.72	&	3.8	&	-9.6	&	0.03	&	-2.9	&	7.5	&	800	&	80	&	170	&	13	&	44.7	&	-2.1	\\
207	&	B3	IV	&	19	&	3.97	&	3.4	&	-9.3	&	0.05	&	-3.1	&	8.9	&	500	&	140	&	170	&	7	&	45.1	&	-1.6	\\
208	&	B2.5	V	&	20	&	3.84	&	4.2	&	-9.5	&	0.05	&	-2.5	&	6.9	&	1300	&	75	&	170	&	28	&	44.9	&	-1.9	\\
209	&	B1.5	V	&	23	&	3.76	&	4.4	&	-9.5	&	0.05	&	-2.1	&	4.8	&	2600	&	80	&	150	&	21	&	45.2	&	-2.0	\\
210	&	B0.5	III	&	28	&	4.52	&	3.6	&	-8.8	&	0.18	&	-3.4	&	7.8	&	1100	&	320	&	145	&	9	&	47.2	&	-0.7	\\
211	&	B5	V	&	17	&	3.85	&	3.8	&	-9.5	&	0.18	&	-2.4	&	9.7	&	900	&	150	&	170	&	22	&	44.8	&	-1.8	\\
212	&	B2.5	V	&	20	&	3.66	&	4.2	&	-9.6	&	0.05	&	-2.1	&	5.6	&	1100	&	150	&	150	&	18	&	44.7	&	-2.2	\\
213	&	B1.5	(V)e	&	24	&	3.81	&	4.0	&	-9.5	&	0.05	&	-2.0	&	4.7	&	1600	&	400	&	170	&	8	&	45.8	&	-1.9	\\
214	&	B1	(V)e	&	23	&	4	&	4.0	&	-9.3	&	0.16	&	-2.5	&	6.3	&	1900	&	200	&	170	&	15	&	45.8	&	-1.6	\\
215	&	O9.7	IV	&	33	&	4.77	&	4.2	&	-8.6	&	0.05	&	-3.8	&	7.4	&	2300	&	40	&	170	&	32	&	47.9	&	-0.2	\\
216	&	B0.7	IV	&	27	&	4.38	&	4.0	&	-9.0$^\ast$	&	0.07	&	-3.4	&	7.1	&	1200$^\ast$	&	110	&	170	&	18	&	46.8	&	-0.9	$^\ast$	\\
217	&	B1	IV	&	25	&	4.39	&	4.0	&	-8.9	&	0.05	&	-3.5	&	8.4	&	2100	&	110	&	160	&	26	&	46.5	&	-0.9	\\
218	&	B3	V	&	17	&	3.55	&	3.8	&	-9.8	&	0.05	&	-2.4	&	6.9	&	800	&	130	&	160	&	11	&	44.5	&	-2.4	\\
219	&	B9	Ia	&	10	&	4.02	&	2.6	&	-9.3	&	0.05	&	-5.0	&	34.2	&	400	&	40	&	170	&	17	&	43.5	&	-1.5	\\
220	&	B1.5	V	&	24	&	3.94	&	4.2	&	-9.4	&	0.05	&	-2.4	&	5.4	&	2200	&	70	&	160	&	17	&	45.8	&	-1.7	\\
221	&	B2.5	V	&	19	&	3.6	&	4.2	&	-9.7	&	0.05	&	-2.1	&	5.8	&	1200	&	50	&	170	&	20	&	44.4	&	-2.3	\\
222	&	B0.5	V	&	28	&	4.03	&	4.4	&	-9.3	&	0.05	&	-2.8	&	4.4	&	2500	&	80	&	170	&	18	&	46.4	&	-1.5	\\
223	&	B0.5	(IV)e	&	27	&	4.43	&	3.6	&	-8.9	&	0.05	&	-3.2	&	7.5	&	1100	&	300	&	170	&	8	&	46.9	&	-0.8	\\
224	&	B1.5	(II)e	&	20	&	4.68	&	3.0	&	-8.6	&	0.05	&	-4.5	&	18.3	&	400	&	160	&	170	&	12	&	46.0	&	-0.4	\\
225	&	B2	(V)e	&	22	&	3.96	&	4.2	&	-9.3	&	0.05	&	-2.7	&	6.6	&	2400	&	150	&	160	&	25	&	45.4	&	-1.7	\\
226	&	B1	(IV)e	&	25	&	4.37	&	3.8	&	-8.9	&	0.27	&	-2.9	&	8.2	&	1600	&	270	&	170	&	15	&	46.5	&	-0.9	\\
227	&	B2	V	&	20	&	3.96	&	4.2	&	-9.3	&	0.05	&	-2.6	&	8.0	&	1400	&	200	&	170	&	37	&	45.0	&	-1.7	\\
228	&	B2.5	V	&	19	&	3.62	&	4.2	&	-9.7	&	0.05	&	-2.2	&	6.0	&	1200	&	100	&	170	&	21	&	44.5	&	-2.3	\\
229	&	B1.5	V	&	23	&	3.77	&	4.0	&	-9.5	&	0.02	&	-2.3	&	4.8	&	1600	&	140	&	160	&	9	&	45.5	&	-2.0	\\
230	&	B1	IV	&	25	&	4.59	&	4.0	&	-8.7	&	0.05	&	-4.6	&	10.5	&	2400	&	180	&	170	&	41	&	46.7	&	-0.5	\\
231	&	O3	V((f*))z	&	46	&	5.61	&	4.1	&	-7.5$^\ast$	&	0.08	&	-5.1	&	10.1	&	2900$^\ast$	&	100	&	170	&	47	&	49.4	&	1.3$^\ast$		\\
232	&	B0.5	V	&	27	&	4.03	&	4.4	&	-9.3	&	0.05	&	-2.5	&	4.7	&	2600	&	110	&	170	&	21	&	46.3	&	-1.5	\\
233	&	B1.5	(IV)e	&	23	&	4.37	&	3.8	&	-8.9	&	0.15	&	-3.8	&	9.7	&	1800	&	120	&	170	&	22	&	46.2	&	-0.9	\\
234	&	B3	IV	&	18	&	3.62	&	3.8	&	-9.7	&	0.05	&	-2.6	&	6.7	&	800	&	100	&	170	&	10	&	44.6	&	-2.3	\\
235	&	B2	(IV)e	&	21	&	4.34	&	3.6	&	-9.0	&	0.15	&	-3.7	&	11.2	&	1500	&	120	&	170	&	18	&	45.8	&	-1.0	\\
236	&	B2.5	V	&	19	&	3.72	&	4.2	&	-9.6	&	0.05	&	-2.4	&	6.7	&	1200	&	200	&	170	&	26	&	44.6	&	-2.1	\\
237	&	B0.7	V	&	27	&	4.26	&	4.2	&	-9.1	&	0.05	&	-3.3	&	6.2	&	2300	&	200	&	170	&	22	&	46.5	&	-1.1	\\
238	&	O8.5	V	&	34	&	4.88	&	4.2	&	-8.4	&	0.05	&	-4.6	&	8.0	&	2300	&	140	&	170	&	37	&	48.2	&	0.0	\\
239	&	O9	V	&	34	&	4.73	&	4.2	&	-8.6	&	0.05	&	-3.4	&	6.7	&	2100	&	75	&	170	&	26	&	48.0	&	-0.3	\\
240	&	B2	V	&	22	&	4.04	&	4.2	&	-9.3	&	0.05	&	-3.4	&	7.2	&	2500	&	90	&	170	&	30	&	45.5	&	-1.5	\\
241	&	O7	V	&	37	&	5.1	&	4.2	&	-8.2	&	0.09	&	-4.8	&	8.7	&	2300	&	150	&	170	&	43	&	48.6	&	0.4	\\
242	&	B1.5	V	&	23	&	3.69	&	4.4	&	-9.6	&	0.05	&	-1.8	&	4.4	&	2500	&	85	&	170	&	18	&	45.2	&	-2.1	\\
243	&	B2.5	V	&	19	&	3.82	&	4.0	&	-9.5	&	0.05	&	-3.4	&	7.5	&	1000	&	110	&	160	&	21	&	44.8	&	-1.9	\\
244	&	B2	V	&	21	&	3.86	&	4.0	&	-9.4	&	0.05	&	-2.4	&	6.4	&	1900	&	100	&	170	&	15	&	45.1	&	-1.8	\\
245	&	B1.5	(IV)e	&	22	&	3.84	&	3.8	&	-9.5	&	0.10	&	-2.3	&	5.7	&	1400	&	300	&	160	&	8	&	45.5	&	-1.9	\\
246	&	B1	V	&	25	&	4.25	&	4.2	&	-9.1	&	0.03	&	-3.4	&	7.1	&	2500	&	180	&	180	&	29	&	46.2	&	-1.1	\\
247	&	B5	V	&	16	&	3.42	&	3.8	&	-9.9	&	0.05	&	-2.2	&	6.7	&	800	&	90	&	160	&	10	&	44.3	&	-2.6	\\
248	&	B0.2	V	&	30	&	4.2	&	4.4	&	-9.1	&	0.05	&	-2.5	&	4.7	&	2500	&	120	&	170	&	20	&	46.8	&	-1.2	\\
249	&	B2	(III)e	&	20	&	4.61	&	3.2	&	-8.7	&	0.10	&	-4.4	&	16.9	&	600	&	300	&	170	&	16	&	45.9	&	-0.5	\\
250	&	B2	V	&	22	&	3.81	&	3.8	&	-9.5	&	0.05	&	-2.2	&	5.5	&	1400	&	120	&	170	&	7	&	45.4	&	-1.9	\\
251	&	B1.5	V	&	23	&	3.76	&	4.4	&	-9.5	&	0.05	&	-1.7	&	4.8	&	2600	&	80	&	170	&	21	&	45.2	&	-2.0	\\
252	&	B9	Ia	&	10	&	3.92	&	2.6	&	-9.4	&	0.05	&	-4.4	&	30.5	&	400	&	100	&	170	&	13	&	43.4	&	-1.7	\\
253	&	B1.5	V	&	23	&	3.92	&	4.4	&	-9.4	&	0.05	&	-1.9	&	5.8	&	2900	&	130	&	170	&	30	&	45.4	&	-1.7	\\
254	&	B1.5	V	&	21	&	4.06	&	4.0	&	-9.3	&	0.05	&	-3.4	&	8.1	&	2100	&	120	&	170	&	24	&	45.3	&	-1.5	\\
255	&	B2	V	&	20	&	3.56	&	4.2	&	-9.7	&	0.05	&	-2.2	&	5.0	&	1100	&	80	&	140	&	15	&	44.6	&	-2.4	\\
256	&	B1.5	V	&	23	&	3.86	&	4.4	&	-9.4	&	0.05	&	-2.4	&	5.4	&	2800	&	80	&	170	&	27	&	45.3	&	-1.8	\\
257	&	B0.5	V	&	27	&	4.46	&	4.2	&	-8.9	&	0.05	&	-4.0	&	7.8	&	2600	&	100	&	170	&	35	&	46.8	&	-0.8	\\
258	&	B0.5	(III)e	&	25	&	4.69	&	3.4	&	-8.6	&	0.12	&	-4.8	&	11.8	&	1100	&	300	&	150	&	13	&	47.0	&	-0.4	\\
259	&	B0.7	V	&	27	&	4.03	&	4.4	&	-9.3	&	0.05	&	-2.5	&	4.7	&	2600	&	110	&	175	&	21	&	46.3	&	-1.5	\\
260	&	B0.5	V	&	27	&	4.26	&	4.2	&	-9.1	&	0.05	&	-2.9	&	6.2	&	2300	&	160	&	160	&	22	&	46.5	&	-1.1	\\
261	&	B0.7	V	&	26	&	3.82	&	4.4	&	-9.5	&	0.05	&	-2.3	&	4.0	&	2400	&	80	&	170	&	15	&	46.0	&	-1.9	\\
262	&	B2	(III)e	&	20	&	4.28	&	3.2	&	-9.0	&	0.10	&	-3.6	&	11.5	&	500	&	180	&	140	&	8	&	45.6	&	-1.1	\\
263	&	B0.2	V	&	29	&	4.44	&	4.2	&	-8.9	&	0.05	&	-3.5	&	6.6	&	2300	&	80	&	170	&	25	&	47.0	&	-0.8	\\
264	&	B2.5	III	&	17	&	3.95	&	3.2	&	-9.4	&	0.05	&	-3.3	&	10.9	&	500	&	80	&	170	&	7	&	45.0	&	-1.7	\\
265	&	B0.5	IV	&	27	&	4.75	&	3.8	&	-8.6	&	0.08	&	-4.4	&	10.9	&	1800	&	120	&	180	&	27	&	47.2	&	-0.3	\\
266	&	B1	V	&	25	&	4.35	&	4.0	&	-9.0	&	0.02	&	-3.6	&	8.0	&	2100	&	150	&	160	&	23	&	46.4	&	-1.0	\\
267	&	B2	V	&	20	&	3.64	&	4.2	&	-9.7	&	0.05	&	-2.2	&	5.5	&	1100	&	80	&	170	&	18	&	44.7	&	-2.2	\\
268	&	B2	V	&	20	&	3.64	&	4.2	&	-9.7	&	0.05	&	-2.1	&	5.5	&	1100	&	80	&	170	&	18	&	44.7	&	-2.2	\\
269	&	O8	Vz	&	35	&	4.64	&	4.2	&	-8.7	&	0.05	&	-3.5	&	5.7	&	2000	&	160	&	120	&	19	&	48.0	&	-0.4	\\
270	&	B0.7	V	&	26	&	3.76	&	4.2	&	-9.5	&	0.05	&	-2.1	&	3.7	&	1800	&	70	&	170	&	8	&	45.9	&	-2.0	\\
271	&	B3	(IV)e	&	18	&	3.62	&	3.8	&	-9.7	&	0.07	&	-2.5	&	6.7	&	800	&	120	&	170	&	10	&	44.6	&	-2.3	\\
272	&	B2.5	IV	&	19	&	3.67	&	3.8	&	-9.6	&	0.05	&	-2.4	&	6.3	&	700	&	90	&	170	&	9	&	44.7	&	-2.2	\\
273	&	B1.5	V	&	23	&	3.66	&	4.4	&	-9.6	&	0.05	&	-1.9	&	4.3	&	2500	&	200	&	170	&	17	&	45.1	&	-2.2	\\
274	&	B0.5	V	&	28	&	4.03	&	4.4	&	-9.3	&	0.05	&	-2.6	&	4.4	&	2500	&	120	&	160	&	18	&	46.4	&	-1.5	\\
275	&	B0.7	V	&	26	&	4.06	&	4.2	&	-9.3	&	0.05	&	-2.8	&	5.3	&	2100	&	210	&	170	&	16	&	46.2	&	-1.5	\\
276	&	B0.5	V	&	26	&	4.14	&	4.2	&	-9.2	&	0.05	&	-2.8	&	5.8	&	2200	&	100	&	170	&	20	&	46.3	&	-1.3	\\
277	&	B1	IV	&	25	&	4.4	&	3.6	&	-8.9	&	0.05	&	-3.2	&	8.5	&	1200	&	200	&	170	&	10	&	46.6	&	-0.9	\\
278	&	B1	IV	&	26	&	4.73	&	3.6	&	-8.6	&	0.05	&	-4.2	&	11.5	&	1400	&	300	&	170	&	19	&	47.1	&	-0.3	\\
279	&	B2	V	&	22	&	3.96	&	4.2	&	-9.3	&	0.05	&	-2.6	&	6.6	&	2400	&	220	&	180	&	25	&	45.4	&	-1.7	\\
280	&	B1	V	&	26	&	3.94	&	4.2	&	-9.4	&	0.05	&	-2.1	&	4.6	&	2000	&	50	&	170	&	12	&	46.1	&	-1.7	\\
281	&	B0.2	V	&	29	&	4.11	&	4.4	&	-9.2	&	0.05	&	-2.8	&	4.5	&	2500	&	70	&	170	&	19	&	46.6	&	-1.4	\\
282	&	B2.5	(IV)e	&	19	&	3.87	&	3.8	&	-9.4	&	0.06	&	-2.7	&	8.0	&	800	&	150	&	170	&	15	&	44.9	&	-1.8	\\
283	&	B0.5	(III)e	&	27	&	5.01	&	3.4	&	-8.3	&	0.38	&	-4.6	&	14.7	&	1100	&	550	&	140	&	20	&	47.7	&	0.2	\\
284	&	B1.5	IV	&	20	&	4.33	&	3.8	&	-9.1$^\ast$	&	0.02	&	-3.8	&	12.2	&	752$^\ast$	&	140	&	170	&	34	&	45.9	&	-1.0$^\ast$		\\
285	&	B1.5	V	&	23	&	3.76	&	4.4	&	-9.5	&	0.05	&	-2.1	&	4.8	&	2600	&	100	&	170	&	21	&	45.2	&	-2.0	\\
286	&	B7	IV	&	16	&	3.72	&	3.8	&	-9.6	&	0.08	&	-3.1	&	9.5	&	900	&	120	&	170	&	21	&	44.6	&	-2.1	\\
287	&	O6	V((f))z	&	37	&	4.75	&	4.2	&	-8.6$^\ast$	&	0.05	&	-3.6	&	5.8	&	2500$^\ast$	&	40	&	160	&	19	&	48.3	&	-0.3$^\ast$		\\
288	&	O8.5	III	&	32	&	5.4	&	3.6	&	-7.9$^\ast$	&	0.06	&	-5.7	&	16.4	&	1900$^\ast$	&	100	&	170	&	39	&	48.7	&	0.9$^\ast$		\\
289	&	B1.5	(IV)e	&	23	&	3.96	&	3.8	&	-9.3	&	0.20	&	-2.6	&	6.0	&	1400	&	300	&	160	&	8	&	45.8	&	-1.7	\\
290	&	B1	V	&	24	&	3.54	&	4.2	&	-9.8	&	0.05	&	-2.0	&	3.4	&	1700	&	150	&	170	&	7	&	45.4	&	-2.4	\\
291	&	B1	V	&	26	&	3.84	&	4.2	&	-9.5	&	0.02	&	-2.6	&	4.1	&	1900	&	300	&	170	&	10	&	46.0	&	-1.9	\\
292	&	O7	V((f))z	&	37	&	5.15	&	4.2	&	-8.3$^\ast$	&	0.05	&	-4.8	&	9.2	&	2500$^\ast$	&	55	&	170	&	49	&	48.7	&	0.5$^\ast$		\\
293	&	B2	V	&	19	&	3.52	&	4.2	&	-9.8	&	0.05	&	-2.1	&	5.3	&	1100	&	150	&	170	&	16	&	44.4	&	-2.4	\\
294	&	B1	V	&	24	&	3.94	&	4.2	&	-9.4	&	0.05	&	-2.5	&	5.4	&	2200	&	180	&	170	&	17	&	45.8	&	-1.7	\\
295	&	B2	V	&	21	&	3.86	&	4.0	&	-9.4	&	0.05	&	-2.7	&	6.4	&	1900	&	100	&	170	&	15	&	45.1	&	-1.8	\\
296	&	B5	V	&	19	&	3.42	&	4.2	&	-9.9	&	0.05	&	-2.1	&	4.7	&	1000	&	100	&	170	&	13	&	44.3	&	-2.6	\\
297	&	B1.5	V	&	23	&	3.76	&	4.4	&	-9.5	&	0.05	&	-2.4	&	4.8	&	2600	&	170	&	170	&	21	&	45.2	&	-2.0	\\
298	&	O9.5	V	&	32	&	4.77	&	4.2	&	-8.7$^\ast$	&	0.08	&	-3.8	&	7.9	&	1700$^\ast$	&	180	&	170	&	36	&	47.8	&	-0.2$^\ast$		\\
299	&	B1.5	V	&	23	&	3.76	&	4.4	&	-9.5	&	0.09	&	-2.5	&	4.8	&	2600	&	200	&	170	&	21	&	45.2	&	-2.0	\\
300	&	B1.5	V	&	22	&	3.86	&	4.2	&	-9.4	&	0.05	&	-2.8	&	5.9	&	2300	&	80	&	160	&	20	&	45.3	&	-1.8	\\
301	&	B1.5	V	&	23	&	3.99	&	4.4	&	-9.3	&	0.01	&	-2.6	&	6.2	&	3000	&	230	&	170	&	36	&	45.5	&	-1.6	\\
302	&	B0	V	&	31	&	4.57	&	4.2	&	-8.7	&	0.02	&	-3.7	&	6.7	&	2200	&	80	&	170	&	26	&	47.5	&	-0.6	\\
303	&	B1.5	V	&	23	&	3.86	&	4.4	&	-9.4	&	0.05	&	-2.6	&	5.4	&	2800	&	230	&	180	&	27	&	45.3	&	-1.8	\\
304	&	B0.5	V	&	27	&	4.29	&	4.2	&	-9.0	&	0.05	&	-3.4	&	6.4	&	2300	&	120	&	150	&	24	&	46.6	&	-1.1	\\
305	&	B1.5	V	&	23	&	3.66	&	4.4	&	-9.6	&	0.05	&	-2.1	&	4.3	&	2500	&	150	&	170	&	17	&	45.1	&	-2.2	\\
306	&	B0	V	&	31	&	4.49	&	4.4	&	-8.8	&	0.05	&	-3.3	&	6.1	&	2800	&	150	&	170	&	34	&	47.3	&	-0.7	\\
307	&	B0.5	V	&	28	&	3.81	&	4.4	&	-9.5	&	0.05	&	-2.1	&	3.4	&	2200	&	150	&	170	&	11	&	46.2	&	-1.9	\\
308	&	B5	V	&	17	&	3.15	&	3.8	&	-10.1	&	0.06	&	-1.9	&	4.3	&	600	&	100	&	160	&	4	&	44.1	&	-3.1	\\
309	&	B2.5	V	&	20	&	3.56	&	4.2	&	-9.7	&	0.05	&	-2.1	&	5.0	&	1100	&	180	&	170	&	15	&	44.6	&	-2.4	\\
310	&	O7.5	In(f)p	&	30	&	5.27	&	3.2	&	-6.6$^\ast$	&	0.06	&	-5.3	&	16.0	&	600$^\ast$	&	300	&	170	&	15	&	48.8	&	0.7$^\ast$		\\
311	&	B1	V	&	26	&	3.8	&	4.4	&	-9.5	&	0.05	&	-2.2	&	3.9	&	2300	&	80	&	170	&	14	&	46.0	&	-1.9	\\
312	&	O9	IV	&	33	&	4.72	&	4.0	&	-8.6	&	0.05	&	-3.6	&	7.0	&	1700	&	40	&	170	&	18	&	48.0	&	-0.3	\\
313	&	B2.5	V	&	20	&	3.76	&	4.2	&	-9.5	&	0.05	&	-2.1	&	6.3	&	1200	&	160	&	170	&	23	&	44.8	&	-2.0	\\
314	&	B2	V	&	19	&	3.52	&	4.0	&	-9.8	&	0.05	&	-2.4	&	5.3	&	900	&	180	&	170	&	10	&	44.5	&	-2.4	\\
315	&	O9.7	(III)e	&	31	&	5.16	&	3.6	&	-8.2	&	0.22	&	-4.8	&	13.2	&	1300	&	440	&	170	&	25	&	48.4	&	0.5	\\
316	&	B2	(V)e	&	19	&	3.62	&	4.0	&	-9.7	&	0.09	&	-2.4	&	6.0	&	900	&	100	&	170	&	13	&	44.6	&	-2.3	\\
317	&	B0	V	&	31	&	4.25	&	4.2	&	-9.1	&	0.05	&	-2.5	&	4.6	&	1900	&	75	&	170	&	12	&	47.2	&	-1.1	\\
318	&	B2	V	&	19	&	3.42	&	4.2	&	-9.9	&	0.05	&	-1.8	&	4.7	&	1000	&	150	&	170	&	13	&	44.3	&	-2.6	\\
319	&	B1.5	V	&	22	&	3.94	&	4.2	&	-9.4	&	0.15	&	-2.2	&	6.4	&	2400	&	40	&	170	&	24	&	45.4	&	-1.7	\\
320	&	B2	V	&	20	&	3.66	&	4.2	&	-9.6	&	0.05	&	-2.0	&	5.6	&	1100	&	100	&	220	&	18	&	44.7	&	-2.2	\\

\end{longtable}
\tablefoot{
\tablefoottext{1}{$\log\,M _{\odot}$ values with $^\ast$ are determined from UV P-Cygni profiles. We derived a relation for these measured $\log\,M _{\odot}$ and $\log\,L/L _{\odot}$ and applied for rest of the stars.}
\tablefoottext{2}{$\varv_\infty$ values with $^\ast$ are determined from UV P-Cygni profiles. Other 
values are theoretically calculated from $\varv_{\mathrm{esc}}$.}
\tablefoottext{3}{$\log\,L_{\mathrm{mec}}$  values with $^\ast$ are calculated for nine stars with UV spectra.  For the rest
of the stars we adopted values from the derived relation of these nine stars.}
}
}

\begin{table}
\caption{Candidate runaway OB stars in the SMC supergiant shell.} 
\label{table:runstars}     
\centering
\begin{tabular}{ccS[table-format = <-1.3]l}
\hline%-------------------------------------------------------------------------
\hline%-------------------------------------------------------------------------
\noalign{\vspace{1mm}}
 SMCSGS-FS & $\varv_{\rm rad}$  &  \multicolumn{1}{c}{$\varv_{\rm rad} -  \bar{\varv}_{\rm rad}$}& spectral type\\
 \#  & (km\,s$^{-1}$) & \multicolumn{1}{c}{(km\,s$^{-1}$)} &\\
\noalign{\vspace{1mm}}
\hline %-------------------------------------------------------------------------------
\noalign{\vspace{1mm}}
203 & 130  & -39   &  B0.5	(III)e  \\
                                                
209 & 150  & -19   &  B1.5	V       \\
                                                
210 & 145  & -24   &  B0.5	III     \\
                                                
212 & 150  & -19   & B2.5	V       \\
                                                
255 & 140  & -29   &  B2	V       \\
                                                
258 & 150  & -19   &  B0.5	(III)e  \\
                                                
262 & 140  & -29   &  B2	(III)e  \\
                                                
269 & 120  & -49   &   O8	Vz      \\
                                                
283 & 140  & -29   &  B0.5	(III)e  \\
                                                
304 & 150  & -19   &  B0.5	V       \\
                                                
320 & 220  & 51   &  B2	V       \\
\hline%-------------------------------------------------------------------------                                  
8   & 120  &  -43  &  B2	V   \\
                                            
16  & 200  &  37  &   B1.5	V   \\
                                            
19  & 120  &  -43  &   B5	V   \\
                                            
50  & 200  &  37  &  B9	Ib  \\
                                            
55  & 100  &  -63  &   B1.5	IV  \\
                                            
64  & 120  &  -43  &   O9	V   \\
                                            
90  & 200  &  37  &   O9.7	V   \\
                                            
93  & 130  & -33  &   B2.5	V   \\
                                            
109 & 110  & -53  &   B2.5	V   \\
                                            
111 & 240  &  77  &  O8.5	V   \\
                                            
113 & 110  &  -53  &  B7	V   \\
                                            
161 & 130  &  -33  &  B5	IV  \\
                                            
165 & 130  &  -33  &  B5	V   \\
                                            
190 & 200  &  37  &   B0.7	IV  \\

\hline%-------------------------------------------------------------------------
\end{tabular}
\end{table}
%%%%%%%%%%%%%%%%%%%%%%%%%%%%%%%%%%%%%%%%%
%%%% table ofstars
%%%%%%%%%%%%%%%%%%%%%%%%%%%%%%%%%%%%%%%%%

\begin{center}

\tablehead{
\hline%-------------------------------------------------------------------------
\hline%-------------------------------------------------------------------------
\noalign{\vspace{1mm}}
 SMCSGS-FS& Age&$M_{\rm ev}$   \\
 \# &[Myr]& [$M_{\odot}$]  \\
\noalign{\vspace{1mm}}
\hline 
%-------------------------------------------------------------------------------
\noalign{\vspace{1mm}}}
%\tabletail{\hline
%\multicolumn{4}{r}{\small\sl Table 1 continued...}\\
%\hline}
\tabletail{\hline }
\topcaption{Ages and evolutionary masses of the OB stars determined from stellar evolutionary tracks and isochrones. }
\label{table:App_age}
\begin{supertabular}{cccl}
1	&     6.0	  &     9.3       \\
2	&     8.6	  &     12.0      \\
3	&     38.5	  &     6.4       \\
4	&     22.0	  &     10.1      \\
5	&     36.9	  &     6.6       \\
6	&     21.4	  &     8.3       \\
7	&     21.0	  &     7.9       \\
8	&     32.4	  &     7.0       \\
9	&     22.4	  &     8.9       \\
10	&     101.0	  &     5.3       \\
11	&     32.4	  &      7.0      \\
12	&     100.0	  &      5.1      \\
13	&     17.6	  &     9.4       \\
14	&     38.5	  &      6.4      \\
15	&     32.9	  &     6.8       \\
16	&     18.9	  &      9.5      \\
17	&     63.3	  &      5.9      \\
18	&     13.4	  &      12.1     \\
19	&     72.9	  &      5.1      \\
20	&     80.0	  &     5.2       \\
21	&     72.9	  &      5.1      \\
22	&     10.6	  &      13.1     \\
23	&     135.0	  &     5.4       \\
24	&     80.0	  &      5.4      \\
25	&     56.6	  &     5.4       \\
26	&     55.2	  &     5.8       \\
27	&     38.5	  &      6.4      \\
28	&     43.4	  &      6.5      \\
29	&     13.1	  &      10.3     \\
30	&     5.1	  &      8.9      \\
31	&     5.6	  &      20.3     \\
32	&     10.7	  &      14.9     \\
33	&     43.4	  &      6.5      \\
34	&     42.4	  &     6.3       \\
35	&     21.0	  &      7.9      \\
36	&     6.1	  &      23.0     \\
37	&     13.0	  &      10.6     \\
38	&     10.1	  &      12.8     \\
39	&     12.0	  &      11.7     \\
40	&     14.0	  &      11.1     \\
41	&     37.2	  &     5.6       \\
42	&     24.7	  &      8.7      \\
43	&     23.2	  &      8.6      \\
44	&     7.9	  &      14.3     \\
45	&     16.7	  &     11.3      \\
46	&     6.3	  &      13.2     \\
47	&     24.6	  &      8.0      \\
48	&     22.6	  &      8.6      \\
49	&     39.8	  &     6.9       \\
50	&     47.2	  &      6.4      \\
51	&     19.2	  &      9.7      \\
52	&     7.6	  &      11.1     \\
53	&     38.5	  &      6.4      \\
54	&     10.4	  &      7.9      \\
55	&     18.0	  &      9.2      \\
56	&     0.9	  &      9.4      \\
57	&     75.8	  &      5.1      \\
58	&     46.3	  &      5.7      \\
59	&     8.3	  &      13.6     \\
60	&     5.2	  &      10.2     \\
61	&     17.5	  &      9.7      \\
62	&     8.3	  &      13.9     \\
63	&     10.0	  &      10.8     \\
64	&     5.0	  &      22.2     \\
65	&     5.9	  &      11.7     \\
66	&     27.6	  &      7.9      \\
67	&     6.3	  &      15.7     \\
68	&     36.9	  &      6.6      \\
69	&     7.3	  &      19.7     \\
70	&     8.0	  &      15.3     \\
71	&     21.0	  &     7.4       \\
72	&     21.8	  &     9.4       \\
73	&     6.4	  &      20.4     \\
74	&     38.5	  &      6.4      \\
75	&     24.6	  &      8.0      \\
76	&     12.2	  &      9.9      \\
77	&     15.7	  &      10.4     \\
78	&     30.9	  &      6.6      \\
79	&     38.5	  &      6.4      \\
80	&     38.5	  &      6.4      \\
81	&     20.0	  &      8.9      \\
82	&     28.9	  &     7.2       \\
83	&     17.5	  &      8.2      \\
84	&     17.5	  &      8.2      \\
85	&     13.1	  &      10.6     \\
86	&     82.0	  &      5.0      \\
87	&     15.3	  &      7.6      \\
88	&     6.0	  &      8.9      \\
89	&     6.3	  &      17.7     \\
90	&     6.2	  &      18.6     \\
91	&     13.5	  &      11.6     \\
92	&     1.0	  &      9.5      \\
93	&     41.9	  &      6.1      \\
94	&     14.8	  &      9.6      \\
95	&     17.5	  &      8.2      \\
96	&     16.9	  &      10.6     \\
97	&     18.5	  &      7.4      \\
98	&     13.0	  &      10.7     \\
99	&     11.3	  &      14.3     \\
100	&     33.9	  &       6.1     \\
101	&     36.9	  &       6.6     \\
102	&     63.0	  &      5.3      \\
103	&     41.9	  &       6.1     \\
104	&     83.0	  &      5.3      \\
105	&     29.2	  &       7.4     \\
106	&     21.5	  &       8.3     \\
107	&     7.2	  &       16.0    \\
108	&     32.7	  &       6.4     \\
109	&     35.5	  &       6.4     \\
110	&     30.2	  &      7.8      \\
111	&     4.0	  &       19.1    \\
112	&     34.5	  &       6.4     \\
113	&     150.0	  &      5.3      \\
114	&     38.5	  &       6.4     \\
115	&     20.8	  &      6.9      \\
116	&     35.5	  &       6.4     \\
117	&     38.4	  &       7.0     \\
118	&     37.2	  &       6.8     \\
119	&     6.1	  &       16.7    \\
120	&     8.5	  &       13.7    \\
121	&     6.1	  &       17.6    \\
122	&     32.7	  &       6.4     \\
123	&     26.1	  &      8.1      \\
124	&     25.3	  &       8.7     \\
125	&     11.1	  &       11.4    \\
126	&     65.7	  &       5.6     \\
127	&     27.0	  &       7.2     \\
128	&     4.3	  &       18.6    \\
129	&     46.0	  &       6.3     \\
130	&     38.5	  &       6.4     \\
131	&     22.8	  &      6.9      \\
132	&     24.6	  &       8.0     \\
133	&     24.6	  &       8.0     \\
134	&     36.9	  &       6.6     \\
135	&     26.1	  &       8.1     \\
136	&     14.1	  &       10.9    \\
137	&     17.7	  &       10.2    \\
138	&     25.0	  &       7.9     \\
139	&     26.0	  &      7.4      \\
140	&     29.3	  &       7.1     \\
141	&     22.6	  &       8.6     \\
142	&     27.7	  &       7.7     \\
143	&     4.3	  &       25.5    \\
144	&     36.5	  &      6.1      \\
145	&     24.6	  &       8.0     \\
146	&     32.4	  &       7.0     \\
147	&     11.1	  &       9.4     \\
148	&     27.7	  &       7.7     \\
149	&     8.3	  &       10.8    \\
150	&     48.0	  &       6.2     \\
151	&     21.0	  &       7.9     \\
152	&     38.5	  &       6.4     \\
153	&     18.6	  &       9.9     \\
154	&     22.6	  &       8.6     \\
155	&     11.7	  &       10.9    \\
156	&     8.2	  &       12.0    \\
157	&     25.9	  &       8.3     \\
158	&     9.2	  &       15.2    \\
159	&     7.9	  &      9.0      \\
160	&     14.0	  &       11.1    \\
161	&     35.9	  &       7.2     \\
162	&     6.7	  &       8.4     \\
163	&     36.0	  &      5.4      \\
164	&     10.1	  &       10.4    \\
165	&     72.9	  &       5.1     \\
166	&     4.6	  &       26.0    \\
167	&     8.3	  &       18.4    \\
168	&     21.0	  &       8.9     \\
169	&     14.7	  &       11.3    \\
170	&     10.8	  &       11.0    \\
171	&     25.7	  &      10.2     \\
172	&     6.0	  &       12.5    \\
173	&     42.5	  &       6.4     \\
174	&     32.5	  &       6.4     \\
175	&     90.0	  &      5.1      \\
176	&     102.9	  &      5.3      \\
177	&     62.2	  &      5.8      \\
178	&     23.0	  &       7.1     \\
179	&     25.3	  &      8.7      \\
180	&     16.0	  &       10.5    \\
181	&     15.3	  &       7.6     \\
182	&     8.2	  &       8.9     \\
183	&     72.9	  &       5.1     \\
184	&     26.0	  &      7.4      \\
185	&     10.5	  &       10.9    \\
186	&     11.1	  &       9.9     \\
187	&     9.6	  &       13.6    \\
188	&     61.8	  &       6.0     \\
189	&     16.8	  &       11.1    \\
190	&     9.2	  &       16.2    \\
191	&     9.1	  &       11.3    \\
192	&     17.5	  &       9.7     \\
193	&     6.2	  &       9.9     \\
194	&     74.0	  &       5.2     \\
195	&     6.1	  &       18.3    \\
196	&     46.0	  &       6.3     \\
197	&     22.3	  &      7.3      \\
198	&     13.9	  &       11.7    \\
199	&     14.4	  &       10.0    \\
200	&     26.0	  &       7.9     \\
201	&     10.1	  &       10.8    \\
202	&     12.9	  &       11.0    \\
203	&     6.9	  &       21.6    \\
204	&     12.4	  &       9.8     \\
205	&     32.5	  &       6.4     \\
206	&     38.0	  &       6.9     \\
207	&     26.6	  &      8.2      \\
208	&     32.9	  &       7.3     \\
209	&     17.5	  &       8.2     \\
210	&     9.8	  &      13.5     \\
211	&     33.8	  &       7.3     \\
212	&     32.4	  &       7.0     \\
213	&     17.5	  &      8.4      \\
214	&     18.5        &       9.1     \\
215	&     5.9         &       18.4    \\
216	&     11.3	  &       12.5    \\
217	&     13.1	  &       12.1    \\
218	&     57.8	  &       6.0     \\
219	&     31.3	  &       7.9     \\
220	&     18.2	  &       9.0     \\
221	&     35.6	  &       6.3     \\
222	&     8.7	  &       11.1    \\
223	&     11.1	  &      12.9     \\
224	&     12.6	  &       13.6    \\
225	&     22.6	  &       8.6     \\
226	&     13.9	  &       11.7    \\
227	&     25.5	  &       8.3     \\
228	&     28.5	  &       6.4     \\
229	&     18.3	  &      8.1      \\
230	&     12.6	  &       13.1    \\
231	&     1.4	  &       46.7    \\
232	&     10.2	  &       10.8    \\
233	&     15.3	  &       11.3    \\
234	&     46.0	  &       6.3     \\
235	&     16.7	  &       10.8    \\
236	&     32.4	  &       7.1     \\
237	&     11.9	  &       11.5    \\
238	&     4.8	  &       20.1    \\
239	&     4.2	  &       18.7    \\
240	&     20.5	  &       9.1     \\
241	&     3.8	  &       25.0    \\
242	&     20.0	  &       8.0     \\
243	&     34.7	  &       7.2     \\
244	&     27.7	  &       7.7     \\
245	&     28.0	  &      7.8      \\
246	&     14.3	  &       10.8    \\
247	&     65.7	  &       5.6     \\
248	&     5.8	  &       12.3    \\
249	&     14.9	  &       12.9    \\
250	&     25.2	  &      7.8      \\
251	&     17.5	  &       8.2     \\
252	&     39.8	  &       6.8     \\
253	&     21.0	  &       8.6     \\
254	&     22.7	  &       9.0     \\
255	&     36.9	  &       6.6     \\
256	&     21.5	  &       8.3     \\
257	&     11.3	  &       13.2    \\
258	&     10.8	  &       14.5    \\
259	&     7.2	  &       10.8    \\
260	&     8.9	  &       11.5    \\
261	&     11.4	  &       9.1     \\
262	&     17.8	  &      10.3     \\
263	&     8.5	  &       13.7    \\
264	&     29.4	  &      7.9      \\
265	&     9.0	  &       16.0    \\
266	&     13.1	  &       11.6    \\
267	&     29.3	  &       7.1     \\
268	&     29.3	  &       7.1     \\
269	&     3.1	  &       18.6    \\
270	&     8.1	  &      9.0      \\
271	&     46.0	  &       6.3     \\
272	&     33.2	  &      6.8      \\
273	&     16.4	  &       7.9     \\
274	&     6.0	  &       11.1    \\
275	&     13.0	  &       10.6    \\
276	&     13.0	  &       10.6    \\
277	&     13.6	  &       12.0    \\
278	&     9.5	  &       15.4    \\
279	&     22.6	  &       8.6     \\
280	&     11.8	  &       9.7     \\
281	&     6.0	  &       11.7    \\
282	&     30.1	  &       7.6     \\
283	&     7.3         &       19.2    \\
284	&     17.2	  &       10.6    \\
285	&     17.5	  &       8.2     \\
286	&     46.2	  &       6.7     \\
287	&     2.0	  &       20.1    \\
288	&     4.6	  &       29.3    \\
289	&     20.1	  &      8.9      \\
290	&     10.1	  &      7.5      \\
291	&     10.0	  &       9.1     \\
292	&     3.9	  &       25.8    \\
293	&     58.5	  &       6.4     \\
294	&     14.0	  &       9.0     \\
295	&     27.7	  &       7.7     \\
296	&     41.9	  &       6.1     \\
297	&     17.5	  &       8.2     \\
298	&     6.2	  &       17.9    \\
299	&     17.5	  &       8.2     \\
300	&     24.6	  &       8.0     \\
301	&     20.4	  &       9.0     \\
302	&     6.6	  &       15.3    \\
303	&     21.5	  &       8.3     \\
304	&     11.3	  &       11.7    \\
305	&     10.4	  &       7.9     \\
306	&     6.2	  &       14.7    \\
307	&     2.0	  &       9.4     \\
308	&     64.2	  &      5.0      \\
309	&     36.9	  &       6.6     \\
310	&     5.5	  &       25.4    \\
311	&     9.6	  &       9.1     \\
312	&     3.9         &       18.1    \\
313	&     29.2	  &       7.4     \\
314	&     58.5	  &       6.4     \\
315	&     5.6	  &       22.5    \\
316	&     42.5	  &       6.4     \\
317	&     4.0	  &       13.2    \\
318	&     41.9	  &       6.1     \\
319	&     22.7	  &       8.5     \\
320	&     32.4	  &       7.0     \\

\end{supertabular}

\end{center}
%%%%%%%%%%%%%%%%%%%%%%%%%%%%%%%%%%%%%%%%%
%%%% end table ofstars
%%%%%%%%%%%%%%%%%%%%%%%%%%%%%%%%%%%%%%%%%

\end{document}